\documentclass[final]{elsarticle}
\usepackage{color}
\usepackage{appendix}

\usepackage{amssymb}
\usepackage{lineno}

\usepackage{soul}

\newcommand{\ab}[1]{{\color{black}{#1}}} %
\newcommand{\cs}[1]{{\color{black}{#1}}} %

\journal{Planetary and Space Sciences}
\begin{document}

\begin{frontmatter}

\title{Viscosity contrasts in the Venus mantle from tidal deformations}

\author[inst1]{Christelle Saliby}

\affiliation[inst1]{organization={Geoazur, CNRS, Observatoire de la Cote d'Azur, Universite Cote d'Azur},
            addressline={250 av. A. Einstein}, 
            city={Valbonne},
            postcode={06560}, 
            country={France}}

\author[inst1,inst2]{Agn\`es Fienga}
\author[inst1]{Arthur Briaud}
\author[inst1]{Anthony M\'emin}
\author[inst1]{Carianna Herrera}

\affiliation[inst2]{organization={IMCCE, Observatoire de Paris, PSL University, CNRS, Sorbonne Universite},
            addressline={77 av. Denfert-Rochereau}, 
            city={Paris},
            postcode={75014}, 
            country={France}}

\begin{abstract}
The tidal deformations of a planet are often considered as markers of its inner structure. In this work, we use the tide excitations induced by the Sun on Venus for deciphering the nature of its internal layers. In using a Monte Carlo Random Exploration of the space of parameters describing the thickness, density and viscosity of 4 or 5 layer profiles, we were able to select models that can reproduce the observed mass, total moment of inertia, $k_2$ Love number and expected quality factor $Q$. Each model is assumed to have homogeneous layers with constant density, viscosity and rigidity. These models show significant contrasts in the viscosity between the upper mantle and the lower mantle. \cs{They also rather favor a S-free core and a slightly hotter lower mantle consistent with previous expectations}. 

\end{abstract}

%

\begin{keyword}
Venus \sep Internal structure \sep Geophysics \sep Celestial Mechanics
\PACS 0000 \sep 1111
\MSC 0000 \sep 1111
\end{keyword}

\end{frontmatter}

\section{Introduction}

The terrestrial planet Venus is \cs{reminiscent} of the Earth since it \ab{is} only $5\%$ smaller and $2\%$ less dense. \cs{Therefore Venus is considered to be the twin planet of the Earth in size and density.} Despite the similarities between Venus and the Earth, these two neighbors have evolved differently as witnessed by the lack of plate tectonics and of magnetic field on Venus. In addition it has a CO$_2$-rich atmosphere $92$ times more massive than the Earth atmosphere. These discrepancies reflect differences in the internal structure, which can be constrained by Venus global properties (mass, radius and distance to the Sun) and \cs{geophysical data} such as topography and gravity field. Most prominently is the latter and its global deformation due to tidal forces from the Sun. 

The presence of hot spots on the surface of Venus has been clearly demonstrated in 2008 with the measurements obtained by the mission Venus Express (VEX) \cite{2010Sci...328..605S,2015GeoRL..42.4762S}. The question is then not if Venus is active but more about the extent of its activity. As there is no indication of plate tectonics on Venus surface \cite{Bercovici2014Plate,Crameri2010Parameters}, its volcanic activity should be driven by plumes emitted from the planet inner part to the crust. The high temperature and pressure at its surface (about 740 K for 93 bars respectively) can favor a more ductile crust than on the Earth. But how are the plumes produced? From which layer of the planet do they come from? These are some of the open questions that will be addressed by the future ESA and NASA missions to Venus \cite{EnVision,2021AGUFM.P34B..01S}.

In this paper, we use tidal deformations as a tool for exploring the internal structure of the planet and more specifically its mantle and its core.
Tidal forces on a planet cause deformations and mass redistributions in its interior leading to surface motions and variations of its gravity field that can be observed with geophysical experiments. \cite{Love1909} studied a compressible homogeneous Earth model and showed that the resulting effects could be represented by a set of dimensionless numbers, so-called Love numbers (hereafter LNs). These Love numbers reflect the internal structure of the planet as they describe the capability of the planet to resist or enhance a forcing excitation. In particular, the change in the gravitational field of a planet due to the influence of an external gravity field, more specifically its degree 2, is primarily described by the tidal Love number (hereafter TLN) $k$ of degree $2$, denoted by $k_2$. This number can be estimated from the analysis of spacecraft radio tracking data.
Indeed, Venus TLN $k_2$ has been estimated by \cite{Konopliv1996Venusian} from Doppler tracking of Magellan and Pioneer Venus orbiters (PVO) to $k_2 = 0.295 \pm 0.066$ at $2$-$\sigma$. Due to these uncertainties, the distinction between liquid and solid core cannot be done \cite{Aitta,Dumoulin2017}. Therefore constraining the internal structure of Venus is still limited for now \citep{Xiao2021Possible}. The absence of a present internal magnetic field is not a constraint since both a liquid and a solid core are compatible with this observation \cite{STEVENSON20031}. However, from the TLNs, it is possible to estimate the energy loss of the planet induced by its visco-elastic deformation at tidal frequencies. It is quantified by the quality factor, Q (as defined i.e. by \citep{Murray2000}), and can be derived by considering the real and the imaginary parts of the TLNs. Generic studies about the energy loss of the solar system planets \cite{1966Icar....5..375G} as well as works on the long term spin evolution of Venus \citep{Correia2003Long-term} provide an interval of possible values for Q for Venus ranging from 20 to 100. 


In this paper we compute the TLN $k_2$ and the quality factor $Q$ of Venus using the Fortran program, \ab{the updated version of ALMA,} ALMA$^3$ \citep{Melini2022On} which calculates the TLNs of a planet under a periodic forcing. In the first part of the paper, we present the basics of the tidal deformation modeling and the internal model of Venus. We explore the effect of two different rheologies (Andrade and Maxwell) and the influence of the thick and dense Venusian atmosphere on $k_2$ and $Q$.
In the second part of the paper, we randomly explore the space of the internal structure parameters of Venus (densities, viscosities and thicknesses) for $4$- and $5$-layer models. We use the mass, the total moment of inertia, the value of $k_2$ derived from observations and the expected limits for the quality factor $Q$  to filter out models that are not consistent with these constraints. We end up with new scenarii for the internal structure of Venus. In particular, we demonstrate that the mantle of Venus presents a clear gradient of viscosities that exists whatever the state of the deeper layers: with or without solid inner core.

\section{Model of Venus tidal deformation}

\subsection{Tidal modeling}
    \label{sec:model}
The LNs describe how a planetary body deforms in response to a surface load or an external potential and how consequently the equipotential surfaces are modified \cite{Love1909,Spada2008ALMAAF,Melini2022On}. The open-source Fortran 90 program ALMA \cite{Spada2006Using,Spada2008ALMAAF,Melini2022On} computes LNs using a semi-analytical approach and for a spherically symmetric (1 dimensional), incompressible, visco-elastic model of planet. The method used in ALMA is similar to the Visco-elastic Normal-Modes method (hereafter VNM) introduced by \cite{Peltier1974} and is based on finding the solution of the equilibrium equations in the Laplace domain. \cs{This method invokes the Correspondence Principal of linear viscoelasticity \citep{Lee1955Stress} which states that, by defining the complex rigidity (also called shear modulus) $\tilde{\mu}$, the equilibrium equations for the viscoelastic problem can be written similarly to the elastic problem. The Correspondence Principal is based on the fact that the Laplace or Fourier transforms of the equilibrium equations for a viscoelastic body are similar to an elastic body of the same geometry \citep{Lee1955Stress}. In this case the equilibrium equations are functions of $\tilde{\mu}(s)$ where $s$ is the Laplace or Fourier variable and $\tilde{\mu}$ depends on the rheology of the viscoelastic body \citep{WuandPel1982}.} 
The planet is assumed to be incompressible, in Sect. \ref{sec:dumoulin} we discuss more the assumption of incompressibility and its effect on the calculation of the TLN $k_2$. The original version of ALMA aimed at evaluating time-dependent LNs for a forcing term following a Heaviside time-history. 

In the case of the tidal excitation, the forcing is periodic and in the case of Venus, the main tides are induced by the Sun with a period of $58$ days \cite{Cottereau2011The}. We then use a modified version of the code, called ALMA$^3$, to estimate the \cs{frequency-dependent} TLNs for a periodic forcing acting on the planet \cite{2022LPICo2678.1349B,Melini2022On}. The difference between ALMA and ALMA$^3$ is that the latter accommodates the periodic perturbations which are used in this study to constrain the internal structure of Venus. This version ALMA$^3$ calculates the complex LNs for a given tidal frequency $\omega$ where the real and imaginary parts account for the amplitude and phase lag of the tidal response, respectively. The quality factor Q can then be estimated \cite{1966Icar....5..375G,Murray2000}, it is calculated as the ratio between the module of $k_2$ and its imaginary part, see Eq. \ref{eq:Q}. The theory behind ALMA$^3$ is explained in details in \cite{2022LPICo2678.1349B} and \cite{Melini2022On}.

To compute the LNs, it is required as inputs a multi-layered discretization of \ab{rheological} 1D profiles (i.e., radius, density, rigidity and viscosity) such as the PREM (Preliminary reference Earth model) model \cite{Dziewonski1981Preliminary}.
 
\begin{figure}[]
           \includegraphics[width=12cm]{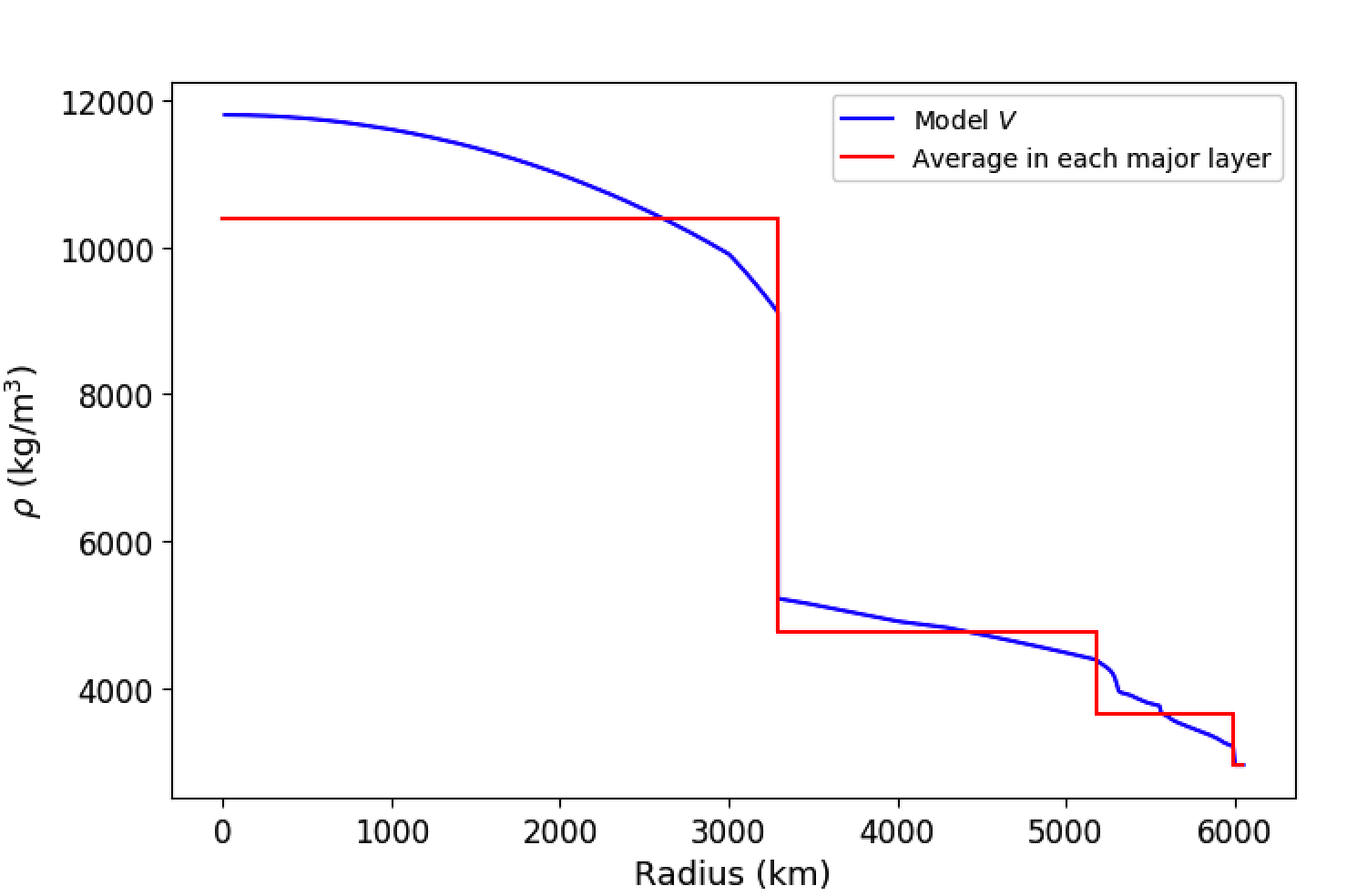}\\ 
           \includegraphics[width=12cm]{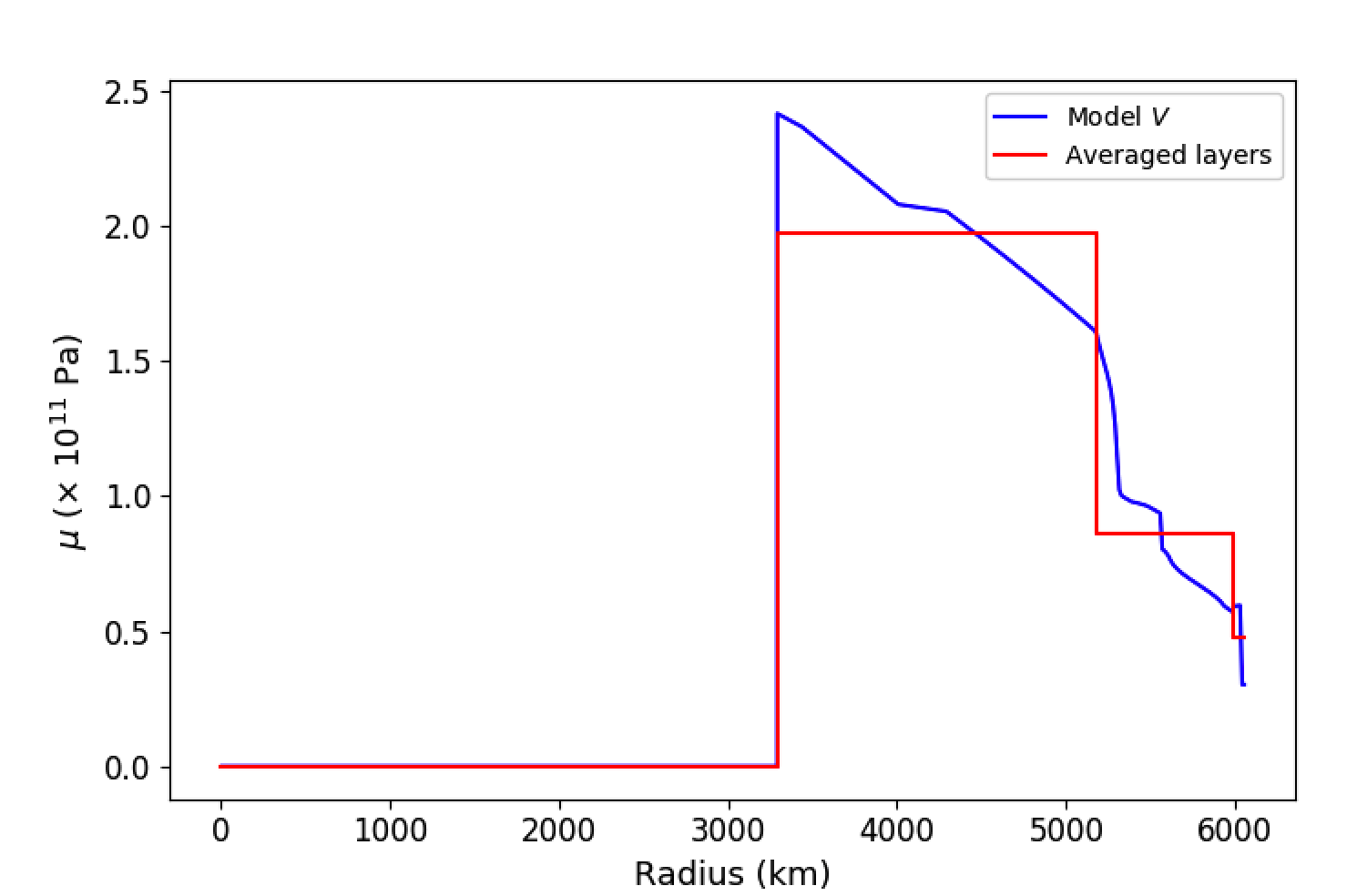}
    \caption{Density $\rho$ in $\mathrm{kg.m^{-3}}$ (top) and rigidity $\mu$ in $\mathrm{10^{11}\ \mathrm{Pa}}$ (bottom). Each major layer has been averaged for the introduction in ALMA$^3$. The model $\mathbf{V}$ refers to the \cite{Dumoulin2017} reference profile. It is built as an Earth-like Venus model with a lower Fe content ($8.1$ wt\%, FeO in the mantle and the crust) to explain the density deficit of Venus in comparison to the Earth \cite{Basaltic1981}.}
\label{Fig:DumRho}
\end{figure}

The models tested in this work have 4 or 5 layers excluding the atmosphere. A first family of models is constituted by 4 homogeneous layers : the core, the lower mantle, the upper mantle and the crust. The core could be either fluid (with a viscosity up to 10$^{-5}$ Pa.s) or solid (with a viscosity up to 10$^{31}$ Pa.s). 
The models with 5 layers are constituted as the models of 4 layer with an additional solid inner core.
The layers for the lower and the upper mantle are visco-elastic and are described with an Andrade rheology. Finally the crust is \ab{assumed} to be elastic. Fig. \ref{Fig:DumRho} shows the profiles in densities (top figure) and viscosities (bottom figure) used for the initial benchmark of the model. They will be one of the possible profiles explored with the Monte Carlo exploration (see Sect. \ref{sec:MC}).

The Andrade’s creep function used in this work was deduced from the work of both \cite{Jackson2002} and \cite{CastilloRogez2011} on the olivine mineral, a magnesium iron silicate, the primary component of the Earth upper mantle. The creep function $J(\omega)$ defining the complex rigidity is given by
\begin{equation}
    \label{eq:AndradeCreepFunction}
    J(\omega) = \frac{1}{\mu}+\beta \frac{\Gamma(\alpha + 1)}{(i \omega)^\alpha} - \frac{i}{\eta \omega}
\end{equation}
with $\Gamma$ is the Gamma function, $\mu$ is the rigidity, $\eta$ the viscosity, $\alpha$ and $\beta$ respectively determine the transient response duration in the primary creep and its amplitude. More precisely $\beta$, characterizes the intensity of anelastic friction in the material. \cite{CastilloRogez2011} approximated $\beta$ which is affixed to the density of the defects, to be $\beta = \mu^{\alpha-1}/\eta^{\alpha}$. The value of $\alpha$ has been determined for olivine-rich rocks to be within $[0.1, 0.5]$, most often within $[0.2,0.4]$ (see \cite{CastilloRogez2011}). The transient creep of this law translates in the second addend of Eq. \ref{eq:AndradeCreepFunction}.

\subsection{Validation: Comparisons to \cite{Dumoulin2017}}
\label{sec:dumoulin}

In \cite{Dumoulin2017}, the TLN $k_2$ is computed by integrating the radial functions associated with the gravitational potential (denoted as $y_5$), as defined by \cite{Takeuchi1972Seismic}, for $10$ models with different profiles for the density $\rho$ and the rigidity $\mu$ but all with a fluid core. These 10 models are based on either hot or cold temperature profiles, as well as composition and hydrostatic pressure from PREM \cite{Dziewonski1981Preliminary} extrapolation.
For comparison with our estimates, we select the model $5$ from the hot temperature models in \cite{Dumoulin2017}, denoted in their work as V$5$-$T_{hot}$, referred hereafter as $\mathbf{V}$. If the composition of Venus was the same as the Earth, its density would have been $1.9\%$ higher than that of the currently observed one \cite{Ringwood1977Earth,Lewis1972Metalsilicate,Goettel1982Density}. One reasonable explanation is that Venus and the Earth have different internal structures, and for example, Venus could have a lower Fe content than that of the Earth \cite{Basaltic1981}. This is the basis of the model $\mathbf{V}$ which was constructed in \cite{Dumoulin2017} using possible Earth-like chemical content with a lower Fe from \cite{Basaltic1981}, specifically $8.1$ wt\%, i. e. percentage by weight, FeO in the mantle and the crust.

The density and rigidity profiles corresponding to the model $\mathbf{V}$ are shown on Fig. \ref{Fig:DumRho}. 
The model $\mathbf{V}$ was also chosen by \cite{Dumoulin2017} to explore different scenarii for the state of the core other than a fluid one, assuming a solid or a partially fluid and partially solid core. The model has $500$ layers excluding the atmosphere, hence a radial discretization with a step slightly larger than $12$ km. 
The model $\mathbf{V}$ was also used by \cite{Dumoulin20} to test the effect of incompressibility. Since their code can be applied to both an incompressible and a compressible model, the TLN $k_2$ has been calculated for model $\mathbf{V}$ for both cases. For this test the mantle is assumed to follow an Andrade rheology with $\alpha=0.3$ with an homogeneous viscosity of $10^{20}$ Pa.s. The real part of $k_2$ was found to be equal to $0.2948$ ($4.6\%$ smaller than the compressible case). \cs{This difference is not big enough to affect our study considering the current large uncertainty of the observed $k_2$ (Table \ref{Tab:TableVenus})}. The imaginary part of $k_2$ was found to be the same for both cases with a value of $0.0087$. In what follows the models are assumed to be incompressible resulting from the limitations of ALMA$^3$.

\begin{figure}
 \includegraphics[width=15cm]{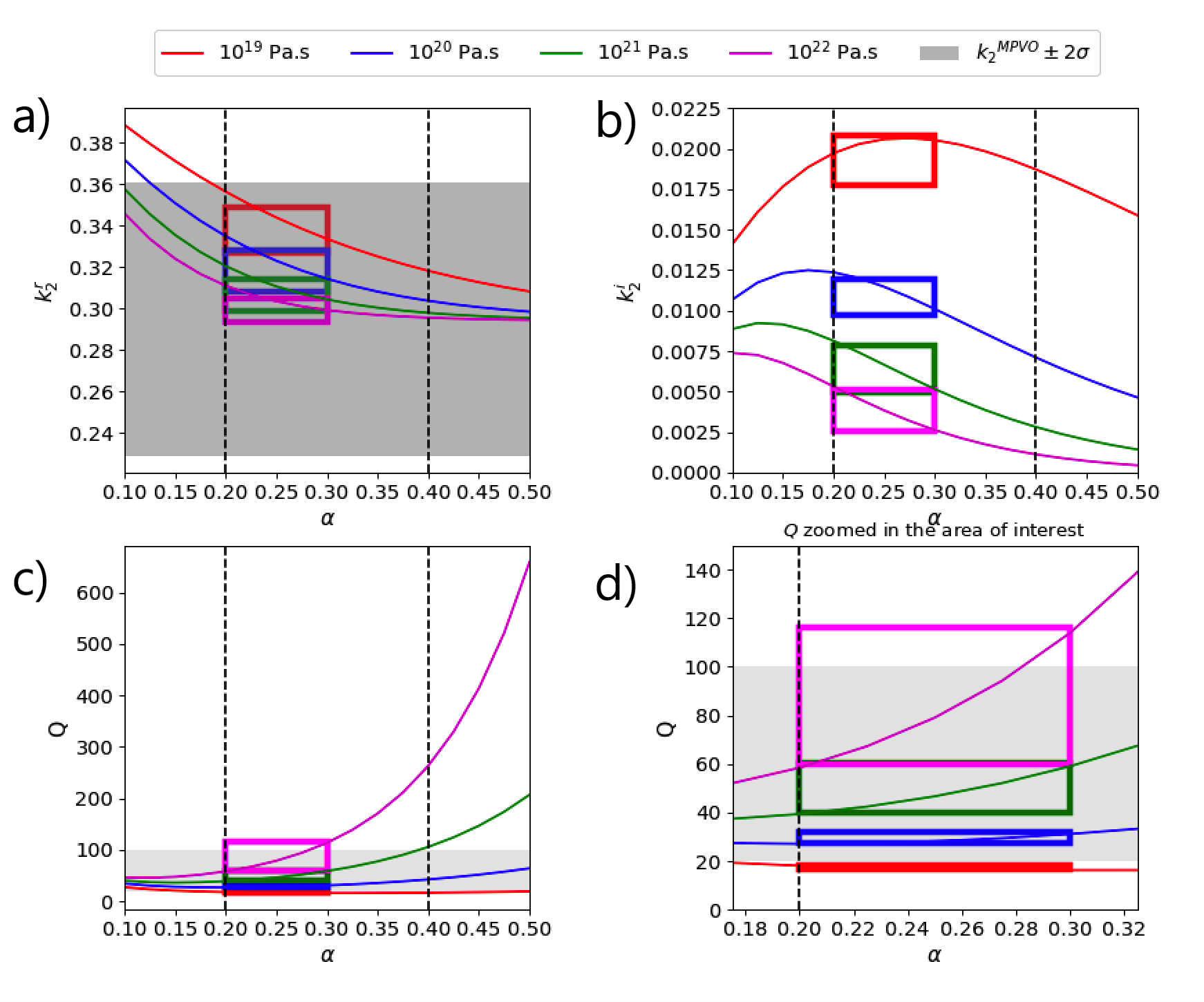}
 \caption{Real part of $k_2$ (a), Imaginary part of $k_2$ (b), Quality factor $Q$ (c) and and its zoom (d) as a function of $\alpha$ (x-axis). The mantle follows an Andrade rheology with different viscosities specified in the legend, from 10$^{19}$ Pa.s to 10$^{22}$ Pa.s.} The rectangles represent the intervals obtained by \cite{Dumoulin2017} for $\alpha \in [0.2,0.3]$. The black dashed lines represent the range of $\alpha$ for olivine-rich rocks. The gray delimitation shows the most recently observed value range according to an uncertainty of $2$-$\sigma$ from \cite{Konopliv1996Venusian}.
 \label{Fig:Dum_Comp}
\end{figure}

We average sub-layers corresponding to each major Venus layer as a single homogeneous layer, reducing our initial $500$ layers to $4$ layers without the atmosphere. To compare with model $\mathbf{V}$, we used for the mantle the Andrade rheology and four viscosities $\eta$ from 10$^{19}$ to 10$^{22}$ Pa.s. 
Fig. \ref{Fig:Dum_Comp} (a) and Fig. \ref{Fig:Dum_Comp} (b) show the real part (i.e. $k_2^r$) and the imaginary part (i.e. $k_2^i$) of $k_2$, respectively. Their associated quality factor $Q$ is calculated as 

\begin{equation}
    \label{eq:Q}
        Q^{-1}=\displaystyle{\frac{k_2^i}{\| k_2 \|}},
\end{equation}
with $\| k_2 \|=\sqrt{{k_2^r}^2+{k_2^i}^2}$ and is shown on Fig. \ref{Fig:Dum_Comp} (c) and (d). \cs{Additionally in their work, the mantle viscosity is assumed to be constant throughout the mantle which is similar to our study where we assume a lower mantle a an upper mantle with the same viscosity.}. The variation of $\alpha$ in \cite{Dumoulin2017} is between $0.2$ and $0.3$. The range of values obtained in their work is represented as vertical lines on Fig. \ref{Fig:Dum_Comp}. 
For the real part $k_2^r$ and for $\alpha$ between $0.2$ and $0.3$, the maximum difference between our results and those of \cite{Dumoulin2017} are between $1.8\%$ to $2\%$ depending on the mantle viscosity. These differences are consistent with the one obtained by \cite{Spada2011The} when comparing different methods to calculate the LN for a Heaviside step function.

Furthermore, as one can see on Fig. \ref{Fig:Dum_Comp} (a), the results for $k_2^r$ for $\alpha \in [0.2,0.4]$ \citep{CastilloRogez2011} (corresponding to olivine-rich rocks) fall into the range of the most recently estimated value from the data of Magellan and PVO, therefore denoted by $k_2^{\mathrm{MPVO}}$, with a $\pm 2$-$\sigma$ uncertainty. For each mantle viscosity, the maximum difference in the values of $k_2^r$ we obtain for this range of $\alpha$ is decreasing with increasing mantle viscosity. 

The imaginary part $k_2^i$ (see Fig. \ref{Fig:Dum_Comp} (b)), for $\eta \geq 10^{20}$ Pa.s, is different between $1\%$ and $2.16\%$ from our estimates and the ones of \cite{Dumoulin2017} depending on $\alpha$. Nonetheless, for $\eta=10^{19}$ Pa.s, the peak of the curve falls in the range of $\alpha \in [0.2,0.3]$. This is the main difference between the two results, since the range of variations between the minimum and maximum for the considered $\alpha$ range values is smaller than that of \cite{Dumoulin2017}. 
The quality factor $Q$ is illustrated on Fig. \ref{Fig:Dum_Comp} (c) and (d). One can see on these figures that its span (upper and lower boundaries) for $\alpha \in [0.2,0.3]$ is almost the same for each viscosity.

\begin{figure}[h!]
 \centering
 \includegraphics[width=13cm]{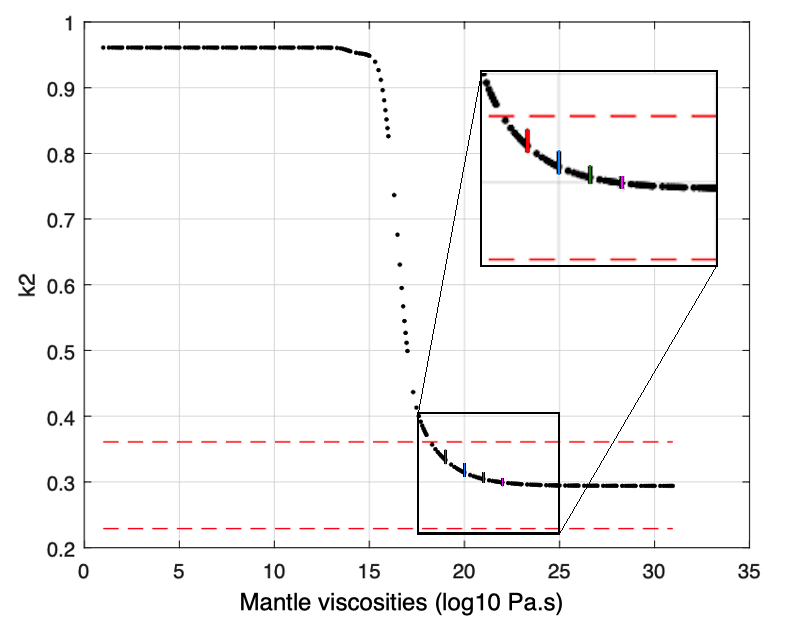}
 \caption{The real $k_2$ as a function of the mantle viscosity and for $\alpha=0.3$. The model used is $\mathbf{V}$. The dashed red lines indicate the interval of $k_2$ as observed by Magellan and PVO \cite{Konopliv1996Venusian}. The colored vertical lines represent the intervals obtained by \cite{Dumoulin2017} for $\alpha$ between $0.2$ and $0.3$ for different mantle viscosities. The color code of the vertical lines is similar to Fig. \ref{Fig:Dum_Comp} which indicate the four different mantle viscosities.}
 \label{Fig:Diff_mantle_viscosities}
\end{figure}

Finally, we expand the viscosity range of the mantle from the previous range of $10^{19}$, $10^{20}$, $10^{21}$ and $10^{22}$ Pa.s to a complete variation from the elastic limit ($\eta \rightarrow 10^{31}$ Pa.s) to the fluid one ($\eta \rightarrow 0$ Pa.s) for $\alpha = 0.3$. Fig. \ref{Fig:Diff_mantle_viscosities} shows the real $k_2$ as a function of the mantle viscosity. The red dashed line illustrates the range of the observed $k_2$ with the Magellan and PVO $2$-$\sigma$ uncertainty. One can see that for $\eta > 10^{18}$ Pa.s the value of $k_2$ fits well into the observed range. This is consistent with the choice of the mantle viscosity range of \cite{Dumoulin2017}, also used in our study for the comparison.

\subsection{Sensitivity to rheologies}

\begin{figure}[h!]
 \centering
 \includegraphics[width=14cm]{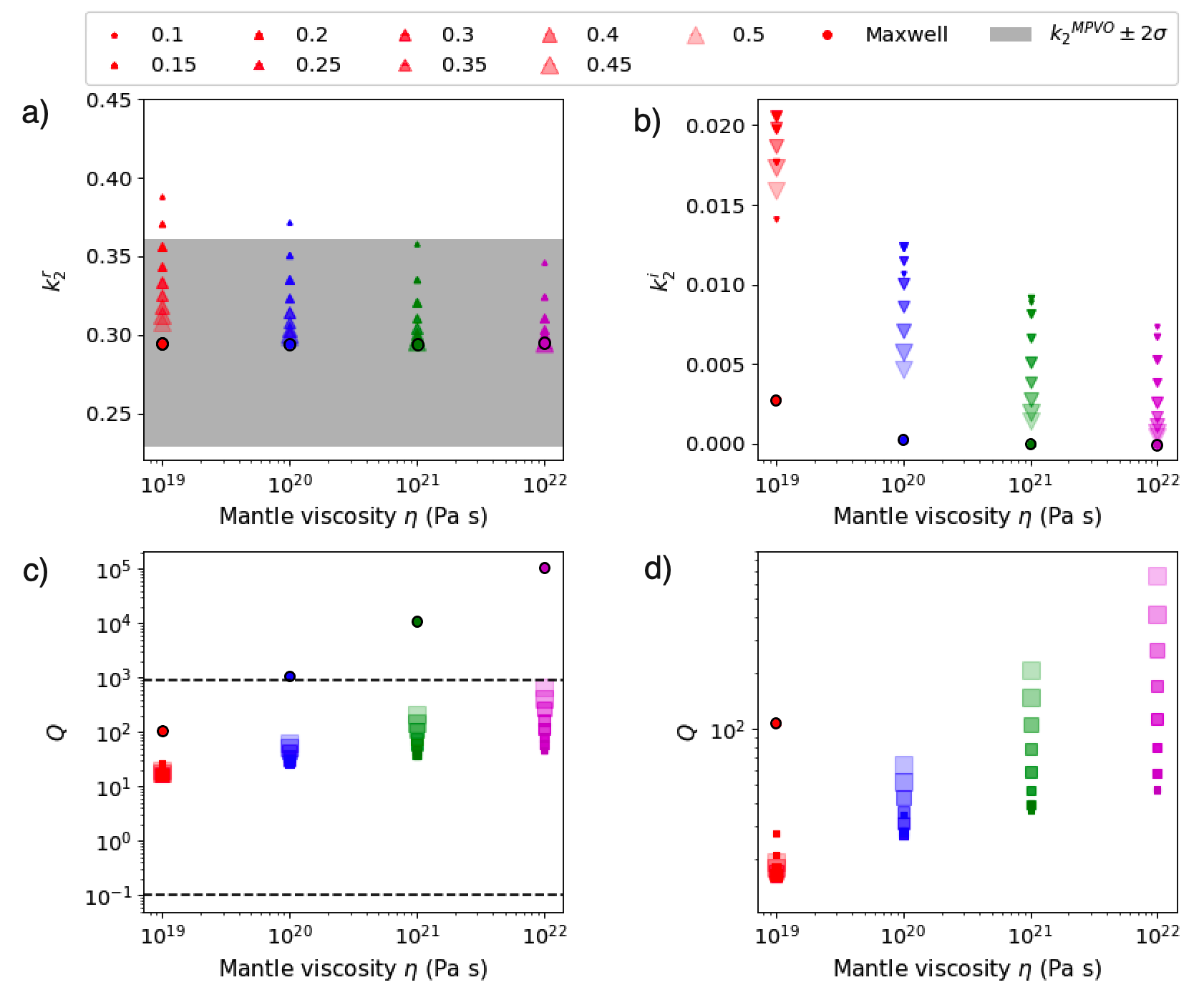}
 \caption{Real part of $k_2$ (a), Imaginary part (b), Quality factor $Q$ (c) and its zoom (d) as a function of mantle viscosities $\eta$ in Pa.s (x-axis) for a mantle with an Andrade rheology for different values of $\alpha \in [0.1,0.5]$. The black dashed lines in (c) delimits the zoomed area in (d). The gray delimitation shows the most recently observed value range according to an uncertainty of $2$-$\sigma$ from \cite{Konopliv1996Venusian}.}
 \label{Fig:MaxVsAnd}
\end{figure}

A comparison between the Andrade and the Maxwell rheologies is performed in order to assess the model (and more specifically the quality factor) sensitivity to the rheologies.
Fig. \ref{Fig:MaxVsAnd} shows the results of $k_2^r$, $k_2^i$ and $Q$ for different mantle viscosities $\eta$ in Pa.s. 

\cs{Fig. \ref{Fig:Dum_Comp} (a)} shows that $k_2^r$ is decreasing with increasing $\alpha$, for each of the explored mantle viscosities. \cs{More specifically Fig. \ref{Fig:MaxVsAnd} (a)} shows that these values approach the results for a Maxwell mantle with higher $\alpha$ values, which is also the case for $k_2^i$ (see Fig. \ref{Fig:MaxVsAnd} (b)).
The quality factor, plotted on Fig. \ref{Fig:MaxVsAnd} (c) and (d), is sensitive to the mantle viscosity $\eta$ for both Maxwell and Andrade rheologies. However, when Q computed with the Andrade rheology remains in the expected interval of $20<Q<100$ \cite{Correia2003Long-term}, the value of Q obtained with the Maxwell rheology reaches far bigger values (from 100 with a low viscosity of 10$^{19}$ Pa.s to 100 000 for a viscosity of 10$^{22}$ Pa.s). Moreover, regarding the Andrade rheology, only Q estimated with $\alpha < 0.3$ are smaller than 100 for all considered viscosities.
These results are in agreement with the other studies \cite{Dumoulin2017,CastilloRogez2011} which suggest that an Andrade rheological law is a better choice to mimic the attenuation behavior of planetary rocks at tidal periods \citep{Dumoulin2017,Bagheri2019Tidal}.

\subsection{Influence of the atmosphere}

\begin{figure}[h!]
 \centering
 \includegraphics[width=15cm]{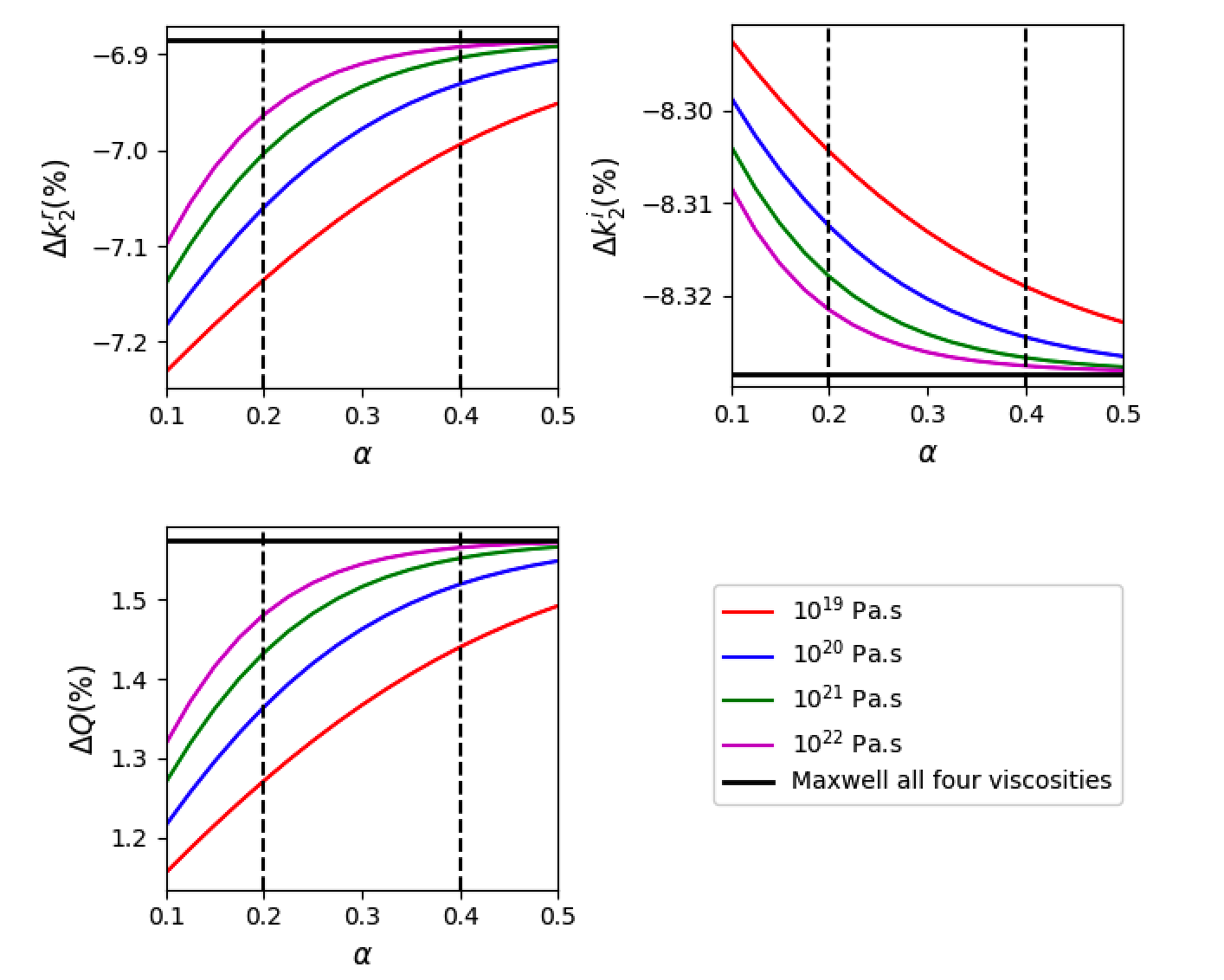}
 \caption{Representation of the percentage variation in ($\%$) of $k_2^r$, $k_2^i$ and $Q$ (y-axis) of Venus without and with an atmosphere as a function of $\alpha \in [0.1,0.5]$ (x-axis). The values correspond to a mantle with either an Andrade and Maxwell rheologies for different values of mantle viscosities $\eta$ in Pa.s. The black dashed lines represent the range of $\alpha$ for olivine-rich rocks.}
 \label{Fig:Diff_Atmosphere_p}
\end{figure}

Finally, we test the effect of the dense Venus atmosphere on the global tidal deformation of the planet. A model of the atmosphere is added as a viscous layer on top of the surface. The TLN $k_2$ with the atmosphere is calculated with ALMA$^3$ \cs{by adding the atmosphere as an additional layer above the crust}. The model of the atmosphere is taken from the Venus International Reference Atmosphere \cite{Seiff1985Models}. The atmosphere has a thickness of $100\ \mathrm{km}$, a density $\rho_{atmo} = 36.5\ \mathrm{kg.m^{-3}}$ and no rigidity ($\mu_{atmo} = 0\ \mathrm{Pa} $) . The viscosity of the atmosphere is fixed to $10^{-5}$ Pa.s for each computation. Fig. \ref{Fig:Diff_Atmosphere_p} shows the variations (in \%) of the $k_2^r$, $k_2^i$ and $Q$ when we include the effect of the atmosphere.
We can see that the atmosphere induces a decrease of the real and imaginary parts of $k_2$ at a maximum level of respectively $7.2\%$ and $8.34\%$ (Fig. \ref{Fig:Diff_Atmosphere_p} (a) and (b)). The former percentage of $7.2\%$ is equivalent to a decrease in $k_2^r$ of a maximum of $0.026$ which is lower than the $1$-$\sigma$ uncertainty of PVO. The variation depends slightly on the value of $\alpha$ and the mantle viscosity. The effect on quality factor $Q$ (see Fig. \ref{Fig:Diff_Atmosphere_p} (c)) is only of a maximum of $+1.65 \%$. We then conclude that the atmosphere does affect the studied parameters but not outside the $\pm 2$-$\sigma$ of the observed $k_2$, despite its high density and low viscosity. \cs{In what follows the models of Venus are only of the solid part of the planet, hence do not include its atmosphere}.

\section{Monte Carlo exploration}
\label{sec:MC}
Based on the previous comparisons, we extend the space of parameters to explore (thicknesses of the layers but also their densities and viscosities) in order to build profiles for the internal structure of Venus that match with the present geophysical constraints. These constraints are the total mass of Venus, its moment of inertia, the TLN $k_2$ and the planet quality factor $Q$.

\subsection{geophysical constraints}
    \label{sec:obs}
The mean surface radius of Venus is set to $R_V= 6051.8$ km \cite{Rosenblatt1994Comparative}. The total mass with its atmosphere is denoted by $M_{V+a}$. It is determined with its uncertainty from the gravitational constants $G$ and $GM_{V+a}$. Using $G=(6.67430 \pm 0.00015) \times 10^{-11} \mathrm{m^3 kg^{-1} s^{-2}}$ \cite{Tiesinga2018CODATA} and $GM_{V+a}=324858.592 \pm 0.006\ \mathrm{km^3 s^{-2}}$ \cite{Konopliv1999Venus} we deduce $M_{V+a}$. The mass of the atmosphere equals to $4.77 \times 10^{20}\ \mathrm{kg}$ \cite{Taylor1985The} is therefore subtracted to obtain the mass $M_V$ without the atmosphere as given on Table \ref{Tab:TableVenus}. Several parameters of Venus are used to constrain its interior in addition to its mass without the atmosphere. These parameters are the normalized moment of inertia $\tilde{C}=C/M_V R_V^{2}$ (hereafter $\mathrm{MoI}$) \cite{Margot2021Spin}, such that $C$ is its polar moment of inertia and its observed TLN $k_2$ shown on Table \ref{Tab:TableVenus}. Finally we also consider the possible values for the quality factor $Q$ at $58$ days as given by \cite{Correia2003Long-term}. Table \ref{Tab:TableVenus} gathers these \cs{literature values} that are used as constraints for this work.

\begin{table}
\caption{Venus state-of-the-art geophysical constraints. The mass $M_V$ is without the atmosphere.}
    \centering
        \begin{tabular}{c c c c}
\hline
Constant & Value & $\pm 1$-$\sigma$ & References\\
\hline
         $R_V\ \mathrm{(km)}$ & $6051.8$ & 1  & \cite{Rosenblatt1994Comparative} \\
         $M_V\ \mathrm{(\times 10^{24} kg)}$  &  $4.8673$ & $1.1 \times 10^{-4}$ & - \\
         $\mathrm{MoI}$  & $0.337$ & $0.024$ & \cite{Margot2021Spin}\\
         $k_2$& $0.295$ & $0.033$ & \cite{Konopliv1996Venusian}\\
         $Q$ & $20< Q <100$ &  &  \cite{Correia2003Long-term}\\
         \hline
          & & & \\
          \end{tabular}
\label{Tab:TableVenus}
\end{table}

\subsection{Method}
\label{sec:MC}

As explained in Sect. \ref{sec:model}, to compute the tidal deformation of the planet and then to compare the TLN and quality factor to the state-of-the-art values, a discretized description of the Venus internal structure in terms of profiles of density, rigidity and viscosity is requested, considering different possible rheologies (Newton, Andrade or Maxwell). The aim of this work is to explore the space of these internal structure parameters (ISP) by using the geophysical constraints given in Sect. \ref{sec:obs} as references for filtering acceptable combinations of ISP. 

Three types of profiles are considered: the \textbf{Class 1} is constituted with an elastic crust, two visco-elastic layers for the mantle and an inviscid fluid core, the \textbf{Class 2} has a solid core instead of an inviscid fluid core and the \textbf{Class 3} has both a solid inner core and a fluid outer core. We also impose no density inversion in the profiles but we allow equal densities for successive layers. This leaves the algorithm free to propose 3-layer models with either the same characteristics for the upper and the lower mantle or for the crust and upper mantle layer. 

Finally the total mass of the planet is conserved in each model. To do so, the density of the innermost layer of each class is not randomly selected, but instead calculated from its random thickness and random densities and thicknesses of the other layers. Consequently, the densities of the fluid core for \textbf{Class 1}, of the solid core for \textbf{Class 2} and of the solid inner core for \textbf{Class 3} are not randomly sampled but deduced from the other layers. 

For each class, the crustal thickness and density are both fixed to $60\ \mathrm{km}$ and $2950$ $\mathrm{kg.m^{-3}}$ \cite{Steinberger2010Deep}, respectively. As a consequence the upper mantle boundary is fixed to $5991.8\ \mathrm{km}$. The thicknesses that vary are the ones of the lower mantle and the core. In contrast, the third class gets three layer radial boundaries that vary. The crustal thickness is constrained in \cite{James2013Crustal} to be from $8$ to $25$ km. Testing the effect of the crustal thickness, we replace the original crustal thickness of $60$ km in model V to $8$ km. The effect on the real and imaginary parts of $k_2$ are $0.6\%$ and $0.7\%$, respectively. 

In this work we uniformly explore the structural and rheological parameters in intervals given by Table \ref{Tab:PRIOR}. The fluid core of \textbf{Class 1} is assumed to be an inviscid fluid therefore its viscosity is fixed to $\eta = 0$ Pa.s. The viscosity of the fluid outer core of \textbf{Class 3} can not be set to be an inviscid fluid (zero viscosity). \cs{It is one limitation of the code ALMA$^{3}$ where its boundary conditions allow only the first layer from the center to be an inviscid fluid. Therefore the code does not converge if another layer, in this case the outer core which is the second layer from the center, to have a zero viscosity. Therefore the rheology of the outer core is set to be fluid} with an arbitrary low viscosity (here 10$^{-5}$ Pa.s) to mimic the behavior of an inviscid fluid.
For the solid layers (the mantle layers for all classes, the core for \textbf{Class 2} and the inner core for \textbf{Class 3}), we consider an Andrade rheology with $\alpha = 1/3$ \citep{Louchet2009Andrade}. The rigidities are fixed for each layer and are equal to the values
corresponding to the rigidity profile given on Fig. \ref{Fig:DumRho} and on Table \ref{Tab:PRIOR}.
Ultimately, we select models according to the constraints mentioned in Sect. \ref{sec:obs} considering a $3$-$\sigma$ interval for the mass and TLN $k_{2}$, a 1-$\sigma$ interval for the MoI and the range specified in Tab. \ref{Tab:TableVenus} for the 58-day quality factor $Q$.


\begin{table}[ht]
\caption{Venus internal parameters, both fixed and simulated with random Monte-Carlo within their respective range.Values indicated with a star are fixed values and values marked with a dagger are deduced as explained in Sect. \ref{sec:MC}.}
\centering
\renewcommand{\arraystretch}{1.5}
\begin{tabular}{c c c c c }
\hline
& $R\ \mathrm{(km)}$ & $\rho\ \mathrm{(kg.m^{-3})}$ & $\eta$  ($\log10$({Pa.s})) & $\mu$ (GPa)\\
\hline
Crust & 6051.8$^{*}$ & 2950$^{*}$ & $\infty$ & 47.65$^{*}$ \\
Upper mantle & 5991.9$^{*}$ & 1000-15000 & 15-25 & 85.7 \\
Lower mantle & 2000-5900 & 3000-15000 & 15-25& 196.94\\
Fluid core (\textbf{Class 1}) & 1000-5000 & 7000-22000{$^{\dagger}$} & $-\infty^{*}$ & 0$^{*}$\\
Solid core (\textbf{Class 2}) & 1000-5000 & 6000-22000{$^{\dagger}$} & 11-22& 125.63 \\
Outer core (\textbf{Class 3}) & 1000-5000 & 1000-15000 & -5$^{*}$ & 0$^{*}$ \\
Inner core (\textbf{Class 3}) & 1-5000 & 5000-30000{$^{\dagger}$} & 10-20& 273.91 \\
\hline
\end{tabular}
\label{Tab:PRIOR}
\end{table}

\section{Results}
\label{sec:res}
$65000$ models have been produced by varying the thickness, density and viscosity for the different layers. After a first filtering with $\mathrm{MoI} \pm 1$-$\sigma$, we retain between $54.5$ and $68\%$ of the $65000$ models which correspond to $35472$, $35443$ and $44390$ models for \textbf{Class 1}, \textbf{Class 2} and \textbf{Class 3}, respectively. A second filtering considering the TLN results in $13077$, $16172$ and $9944$ models for \textbf{Class 1}, \textbf{Class 2} and \textbf{Class 3}, respectively. Finally the quality factor Q filter is performed resulting in $4703$, $4536$ and $4160$ selected models. \cs{In \ref{sec:Appendix_Q}, are given the results when the quality factor $Q$ is not used as a constraint.} To test whether the number of models simulated are enough, we tested subsets of the original $65000$ models. The randomly chosen subsets of models consist increasingly of $650$ to $65000$ models. After filtering with the MoI, $k_2$ and $Q$ filters we illustrate (see Fig. \ref{Fig:Subsets_percentage}) in \ref{sec:Appendix_rigidity} the percentage of selected models after several filters for \textbf{Class 1} as an example.
Table \ref{tab:results} gives the statistics of the selected models namely the mean and the first and third quartiles of the parameters that have been randomly sampled and selected according to our method. 

\begin{table}
\centering
    \caption{Results of the selection process over $65000$ randomly sampled profiles. Are given in Column 1, the type of models considered and on Column 2 the layers. Column 3 gives the mean and first and third quartiles (25$\%$ and 75$\%$) of the layer thicknesses ($\mathrm{km}$), Column 4 the densities $\mathrm{(kg.m^{-3})}$ and Column 5 the viscosities in $\log10$(Pa.s).}
\begin{tabular}{ c  c  c  c  c }
\hline
Models & Layers & thickness & density & viscosity \\
 &  & (km) & (kg.m$^{-3}$) & $\log10$(Pa.s) \\
\hline
Fluid (\textbf{Class 1}) & upper mantle & $963_{600}^{1417}$ & $3765_{3446}^{4123}$ & $19.9_{18.3}^{22.3}$ \\
\\
            & lower mantle & $1839_{1418}^{2232}$ & $4890_{4484}^{5360}$ & $20.78_{19.85}^{21.85}$ \\
            \\
            & core         & $3166_{2898}^{3372}$ & $10899_{9892}^{11909}$ & $-5$ \\
\hline 
Solid (\textbf{Class 2A}) & upper mantle & $1432_{883}^{1996}$ & $3987_{3619}^{4306}$ & $20.95_{19}^{23.48}$ \\
\\
            & lower mantle  & $1313_{715}^{2007}$ & $5057_{4612}^{5561}$ & $21_{19.95}^{22.85}$ \\
            \\
            & core          & $3240_{2944}^{3460}$ & $10527_{9373}^{11713}$ & $14.95_{13.4}^{19.48}$ \\
\hline
Solid (\textbf{Class 2B}) & upper mantle & $1052_{773}^{1275}$ & $3562_{3275}^{3844}$ & $20.85_{18.95}^{22.9}$ \\
\\
            & lower mantle  & $410_{220}^{689}$ & $4719_{4188}^{5277}$ & $20.9_{18.95}^{23.48}$ \\
            \\
            & core          & $4502_{4257}^{4702}$ & $7209_{6917}^{7597}$ & $20.7_{19.85}^{21}$ \\
\hline
Fluid/Solid (\textbf{Class 3}) & upper mantle & $925_{585}^{2099}$ & $3722_{3377}^{4010}$ & $20_{18.3}^{22.6}$ \\
\\
            & lower mantle & $1718_{1330}^{2099}$ & $4932_{4431}^{5369}$ & $21.48_{20.54}^{22.78}$ \\
            \\
            & outer core   & $381_{159}^{712}$ & $8204_{6723}^{9900}$ & $-5$\\
            \\
            & inner core   & $2825_{2402}^{3141}$ & $11450_{10425}^{12220}$ & $15.7_{12.95}^{17.81}$\\
           \hline
\end{tabular}
\label{tab:results}
\end{table}

\begin{figure}[h!]
 \centering
 \includegraphics[width=11cm]{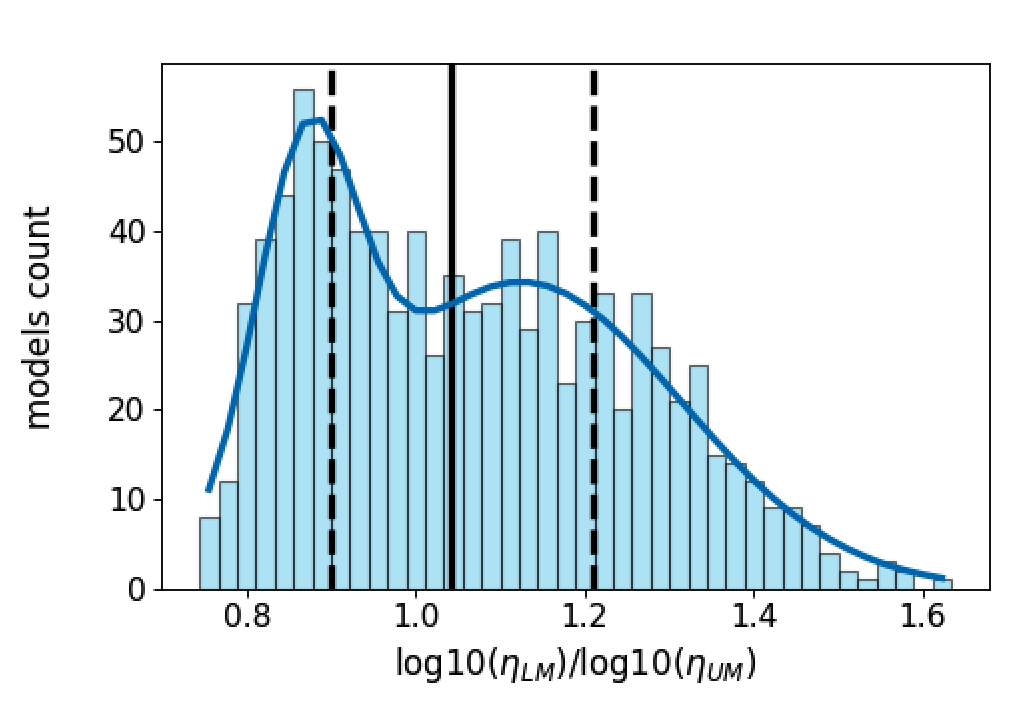}
 \caption{\textbf{Class 1}: Histogram of the ratio between the lower mantle viscosity versus the upper mantle viscosities. The black curve is the bi-modal fit of the ratio distribution. Black plain and dash lines correspond to the median and the first and third quartiles, respectively.}
 \label{Fig:CS_visco_C1}
\end{figure}

\begin{figure}[h!]
 \centering
 \includegraphics[width=13.5cm]{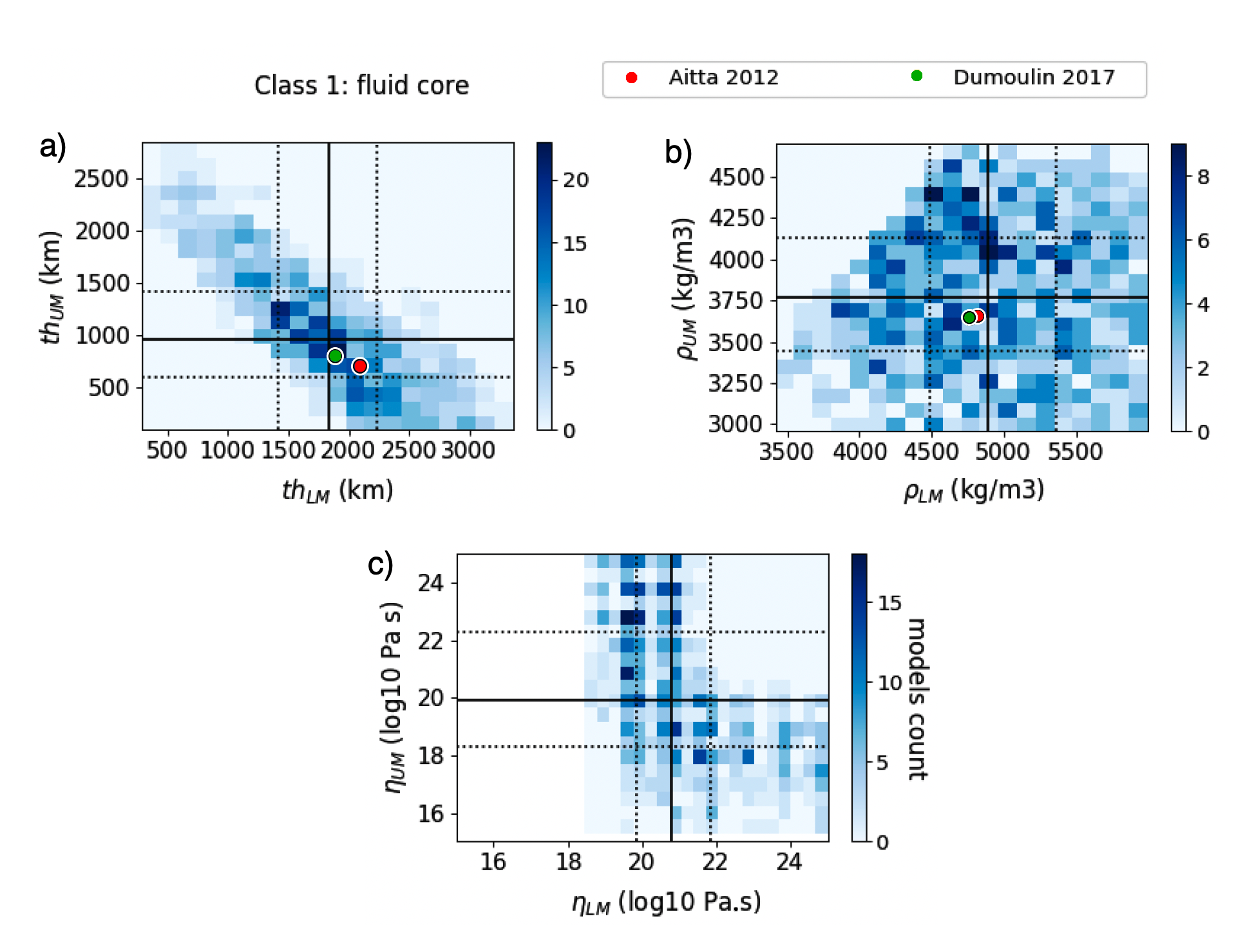}
 \caption{\textbf{Class 1}: 2-D Histograms for the thicknesses (top left-hand side), the densities (top right-hand side) and the viscosities (bottom side) of the lower mantle versus that of the upper mantle. The red and green dots show the values considered by \cite{Aitta} and \cite{Dumoulin2017}, respectively.}
 \label{Fig:CS_2D_C1}
 \end{figure}

\subsection{\textbf{Class 1}: only a fluid core}
\label{sec:c1}
The models of \textbf{Class 1} include a mantle with two separated layers (upper and lower), to reproduce the Earth structure, and an inviscid fluid core. From our simulations, it appears that when only a fluid core is present, the lower mantle (1839$_{1418}^{2232}$~km) is significantly thicker than the upper mantle (963$_{600}^{1417}$~km) with a higher density (4890$_{4484}^{5360}$ versus 3765$_{3446}^{4123}$~kg.m$^{-3}$) and a significantly higher viscosity (20.78$_{19.85}^{21.85}$ and 19.9$_{18.3}^{22.3}$~$\log10$(Pa.s)). 
The higher lower mantle density is a result of the assumption of no density inversion in each model. This assumption is also driven by the fact that an increased density from the surface to the center of the planet can be obtained by integrating its pressure equations. 
These significant differences between the lower and the upper mantles stress the dichotomy of state and nature of these two layers. 
Furthermore, the distribution of the ratio between the lower and the upper mantle viscosities (Table \ref{Tab:Gaussiens} and Fig. \ref{Fig:CS_visco_C1}) show two trends of models: the first trend has a peak of distribution for \cs{$\log10 (\eta_{\mathrm{LM}})/\log10 (\eta_{\mathrm{UM}}) \approx 0.9$} and the second trend has a peak of distribution for \cs{$\log10 (\eta_{\mathrm{LM}})/\log10 (\eta_{\mathrm{UM}}) \approx 1.1 $}. Fig. \ref{Fig:CS_visco_C1} also shows that we have slightly more models with a more viscous lower mantle since the mean (second quartile) of the histogram is for $\eta_{\mathrm{LM}}/\eta_{\mathrm{UM}}>1$. It is unexpected since the lower mantle is expected to be less viscous than the upper mantle by the Arrhenius law \citep{Roller1986Rheology}. The result is based on the selection of models with geophysical constraints and statistical study with minimal initial assumptions on the chemical content or temperature profile of Venus.
Table \ref{Tab:Gaussiens} gives the results in terms of $\chi^{2}$ for two adjustments of the viscosity ratio distribution considering a bi-modal and a Gaussian profiles. The bi-modal model gives a better $\chi^{2}$ than the Gaussian profile (0.91 versus 1.5), favoring a double distribution of the upper and lower viscosity ratios: one centered around $0.87 \pm 0.058$ (with a more viscous upper mantle) and one around $1.13 \pm 0.19$ (with a more viscous lower mantle).
The possible entanglement of the lower and upper mantle viscosities is even more visible on the 2-D histograms shown on Fig. \ref{Fig:CS_2D_C1}. On this Figure, one can see that a more viscous lower mantle relates to a more fluid upper mantle and vice versa, unless the two layers have similar viscosities. Models with the same viscosity (between $10^{19}$ and $10^{21}$ Pa.s) for both the lower and the upper mantles represent about 1$\%$ of the models. Fig. \ref{Fig:CS_2D_C1} (c) also shows that the lower and upper mantles can not be both more fluid ($\eta_{\mathrm{LM}} < 10^{19.5}$ Pa.s and $\eta_{\mathrm{UM}} < 10^{19}$ Pa.s) or more viscous ($\eta_{\mathrm{LM}}> 10^{21.8}$ Pa.s and $\eta_{\mathrm{UM}}> 10^{21}$ Pa.s). 

Finally, the distribution of the thicknesses of the lower and upper mantle (Fig. \ref{Fig:CS_2D_C1} (a)) shows a direct correlation, expected for a terrestrial planet as Venus. Moreover, the density of the fluid core that we obtain ($10899_{9892}^{11909}$ kg.m$^{-3}$) is consistent with what is expected for a planet of the size of Venus composed by iron alloys \cite{Aitta}.

\subsection{\textbf{Class 2}: only a solid core}
\label{sec:c2}

 \begin{figure}[h!]
 \centering
\includegraphics[width=14cm]{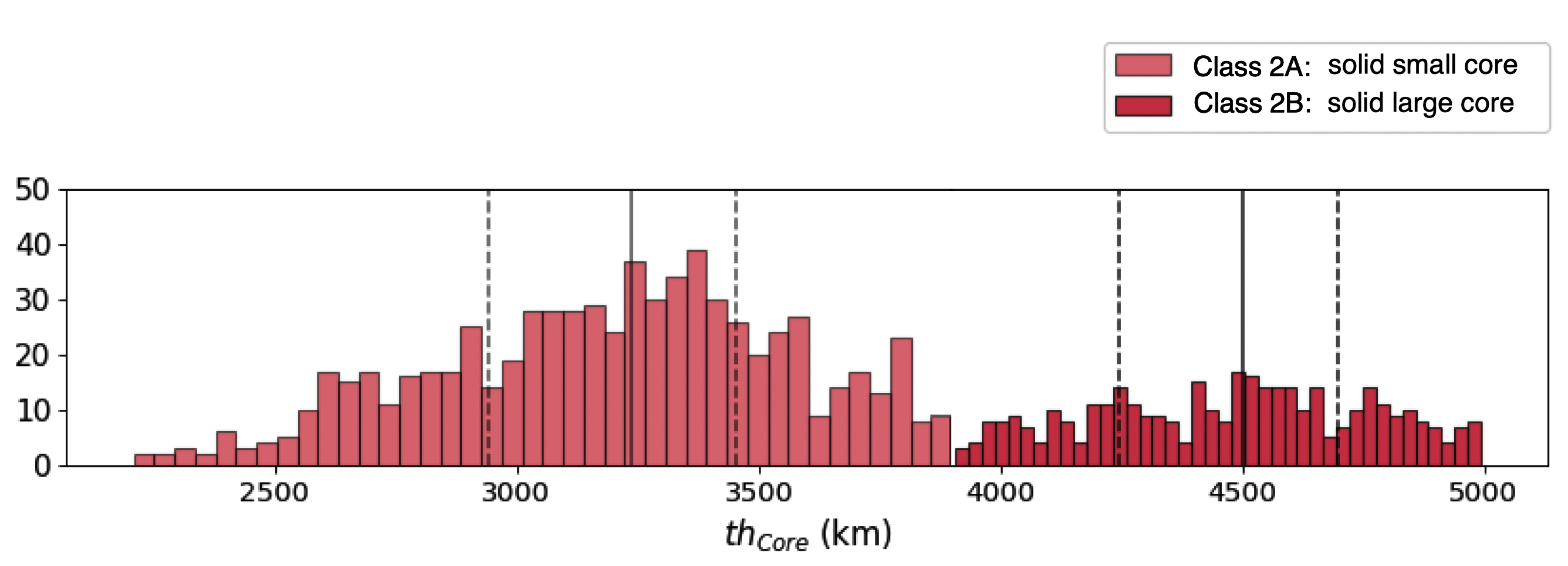}
 \caption{\textbf{Class 2}: Histograms of the thicknesses of the core. Black plain and dash lines correspond to the median and the first and third quartiles, respectively for each sub-class.}
 \label{Fig:CS_visco_C2}
\end{figure}

The models of \textbf{Class 2} include a mantle with two separated layers, as for the models of \textbf{Class 1}, and a solid visco-elastic core following an Andrade rheology, similar to the one of the mantle, with a rigidity of $125.63$~GPa (see \ref{sec:Appendix_rigidity} for the impact of the fixed rigidity on the results).

Fig. \ref{Fig:CS_visco_C2} shows the histogram of the selected \cs{sizes for the solid inner core}. Two families of models \cs{can be distinguished}: one with a large core of about $4500$ km and one with a small core of about $3235$ km. Fitting the \cs{size of the core} with a Gaussian and with a bi-modal distributions gives values of $\chi^2$ of $2.47$ and $0.94$, respectively. Therefore the two families have been defined in fitting a bi-modal distribution of the thicknesses and in separating the two distributions at 92$\%$ of the two populations.
The two families, labelled \textbf{Class 2A} and \textbf{Class 2B} on Table \ref{tab:results} and Fig. \ref{Fig:CS_visco_C2}, are considered separately in the analysis. As expected, the bigger core (\textbf{Class 2B}) which is 4499$_{4244}^{4699}$~km thick presents a lower density $7215_{6924}^{7611}$ kg.m$^{-3}$, favoring a scenario of a core enriched in light elements. This low density core is also associated with a significantly higher viscosity ($20.7_{19.85}^{21}$ $\log10$(Pa.s)) in comparison with models of \textbf{Class 2A} that 
have a smaller core (3235$_{2940}^{3453}$~km thick) and a lower viscosity ($14.95_{13.35}^{19.44}$ $\log10$(Pa.s)). 

Thicknesses and densities ($10533_{9376}^{11723}$ kg.m$^{-3}$) of the core for models of \textbf{Class 2A} are consistent with those of \textbf{Class 1} and are in the expected range for a planet of the size of Venus. On Table \ref{Tab:Gaussiens}, one can see that, also for \textbf{Class 2}, the bi-modal distribution of the upper and lower mantle viscosity contrast is also validated ($\chi^{2}$ of 0.97 for bi-modal versus 1.42 for Gaussian).

However, it is not so clearly the case when we separately consider the two sub-classes \textbf{2A} and \textbf{2B}. In particular, for the family with the biggest core (\textbf{Class 2B}), the contrast of viscosities between the upper and the lower mantles as presented on Table \ref{Tab:Gaussiens} is not present and the single Gaussian distribution centered on the equal viscosity for upper and lower layers gives a better $\chi^{2}$ ($1.2$) than the bi-modal distribution ($2.58$). At the opposite, as one can see on Table \ref{Tab:Gaussiens}, the models of the \textbf{Class 2A} favor a bi-modal distributions of the viscosities ($\chi^{2}$ bi-modal equals to $1.22$ where $\chi^{2}$ Gaussian is equal to $1.7$).

\begin{table}[ht]
\caption{Gaussian and bi-modal distributions of the ratio between viscosities of the lower mantle and the viscosities of the upper mantle (viscosity contrast) for the different classes of models. Are given in Columns 2, 5 and 7, the centroids $M$, $M_1$ and $M_2$, in Columns 3, 6 and 8, the uncertainties $\sigma$, $\sigma_1$ and $\sigma_2$ and in Columns 4 and 9, the $\chi^2$ of the each fit.}
\centering
\renewcommand{\arraystretch}{1.6}
\begin{tabular}{c | c c c | c c c c c}
\hline
& \multicolumn{3}{c}{\bfseries Gaussian distribution} & \multicolumn{5}{|c}{\bfseries Bi-modal distribution}\\
\hline
 & $M$ & $\sigma$ & $\chi^2$ & $M_1$ & $\sigma_1$ & $M_2$ & $\sigma_2$ & $\chi^2$ \\
\textbf{Class 1} & $0.99$ & $0.26$ & $1.5$ & $0.87$ & $0.058$ & $1.13$ & $0.19$ & $0.91$\\
\textbf{Class 2} & $1$ & $0.19$ &$1.42$ & $0.9$ & $0.11$ & $1.1$ & $0.16$ & $0.97$ \\
\textbf{Class 2A} & $1$ & $0.18$ & $1.7$ & $0.88$ & $0.09$ & $1.07$ & $0.17$ & $1.22$\\
\textbf{Class 2B} & $1$ & $0.2$ & $1.2$ & $0.97$ & $0.16$ & $1.26$ & $0.09$ & $2.58$\\
\textbf{Class 3} & $1.03$ & $0.21$ & $1.42$ & $0.96$ & $0.12$ & $1.25$ & $0.13$ & $1.2$\\
\hline
\end{tabular}
\label{Tab:Gaussiens}
\end{table}

\subsection{\textbf{Class 3}: fluid outer core and solid inner core}

 On Table \ref{Tab:Gaussiens}, \textbf{Class 3} also shows a bi-modal distribution of the viscosity ratio of upper and lower layers of the mantle ($\chi^{2}$ for bi-modal distribution of 1.2 versus 1.42 for Gaussian) but with a shift of the centroids towards higher ratios. Indeed, where for \textbf{Class 1}, the modes were centered on $0.87 \pm 0.058$ and $1.13 \pm 0.19$ (marginally compatible with $1$), the first mode of \textbf{Class 3} is marginally compatible with a center at 1 ($0.96 \pm 0.12$) and the second mode is centered on $1.25 \pm 0.13$. In addition, Table \ref{tab:results} shows that the viscosities of the lower mantle increased in comparison to the viscosity of the same layer for models of \textbf{Class 1} ($20.78_{19.85}^{21.85}$ versus $21.48_{20.6}^{22.78} \log10$ (Pa.s)) when the viscosities of \textbf{Class 1} and \textbf{Class 3} remain identical for the upper layer.
The densities of the two layers increase marginally for \textbf{Class 3} relatively to \textbf{Class 1} but not significantly. 
The mechanism of the increase of the lower mantle viscosity induced by the introduction of the solid inner core is then confirmed by the results presented on Tables \ref{tab:results} and \ref{Tab:Gaussiens}.
In terms of core densities, they are high both for the solid inner core ($11450_{10450}^{12220}$ kg.m$^{-3}$) and for the fluid outer core ($8276_{6723}^{9912}$ kg.m$^{-3}$). They are in average compatible with the densities of \textbf{Class 1} and \textbf{Class 2A}.
Finally, when for the fluid core, the viscosity remains close to a low value, the viscosity of the inner core is obtained to be also quite small ($15.6_{12.95}^{17.78} \log10$(Pa.s)) for the \textbf{Class 3} but compatible with the value found for \textbf{Class 2A}.

\section{Discussion}
\label{sec:discussion} 
\begin{figure}[h!] 
        \centering
         {\includegraphics[width=14cm]{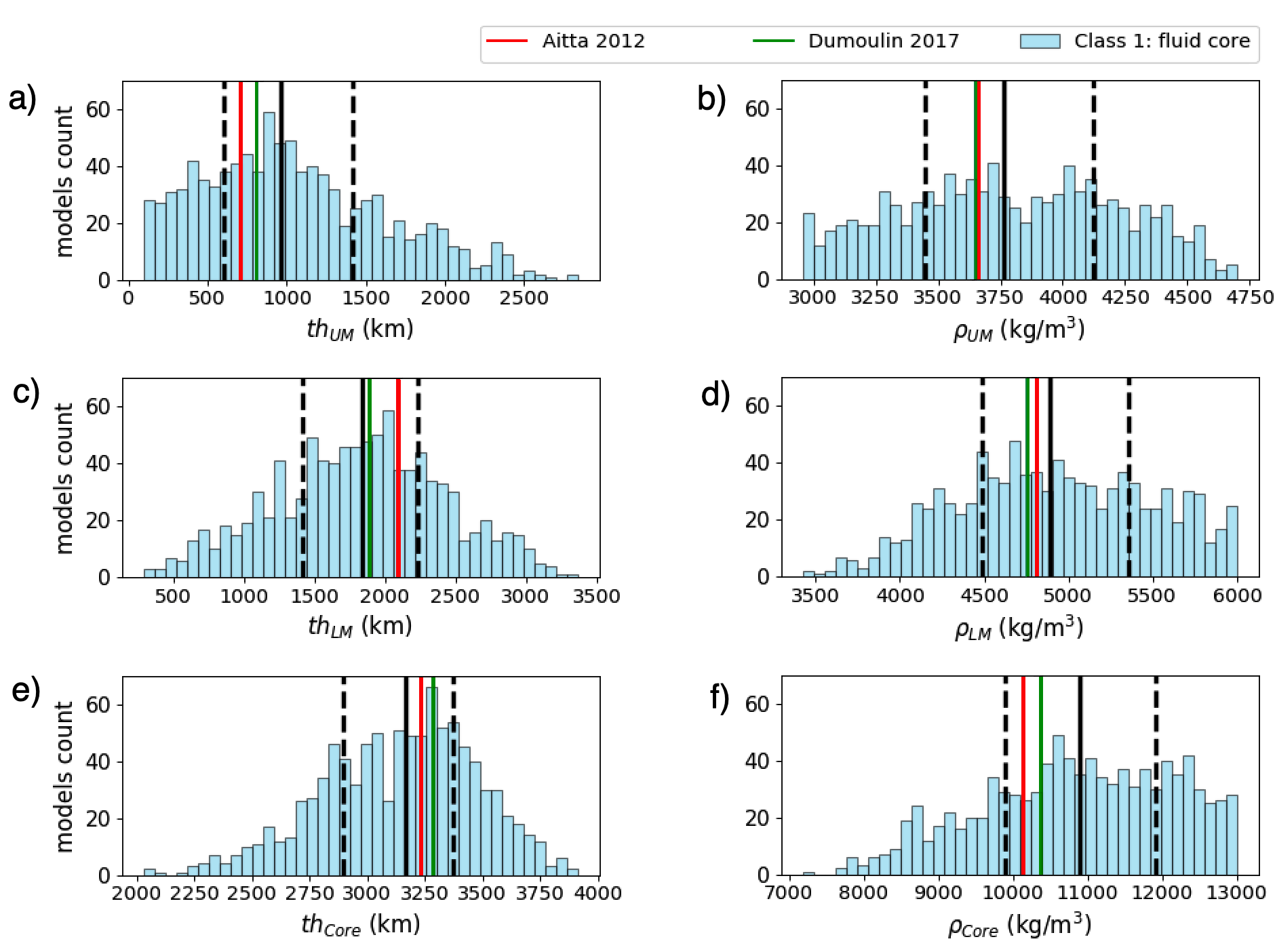}} 
         \caption{Comparisons between the models of \textbf{Class 1} and those from \cite{Dumoulin2017} and \cite{Aitta}: Histograms in thicknesses in $\mathrm{(km)}$ (left-hand side column) and densities in $\mathrm{(kg.m^{-3})}$ (right-hand side column). The solid black line corresponds to the mean and the dashed black lines correspond respectively to the first and third quartiles. The red and green vertical lines indicate the limits of the models proposed by \cite{Dumoulin2017} and \cite{Aitta} respectively.}
         \label{Fig:Histograms_FC12}
\end{figure}

\begin{figure}[h!]
    \centering
    \includegraphics[scale=0.33]{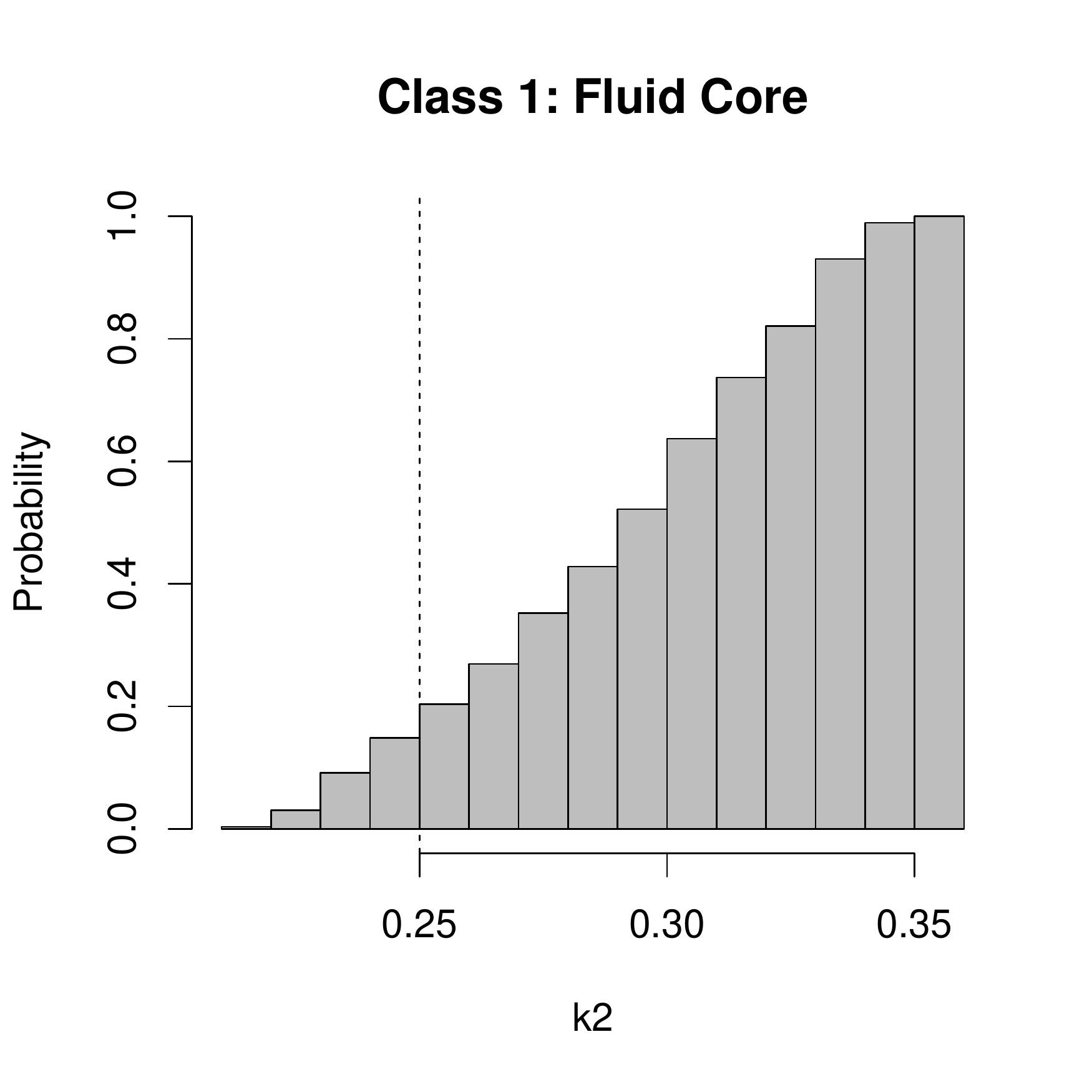}\includegraphics[scale=0.33]{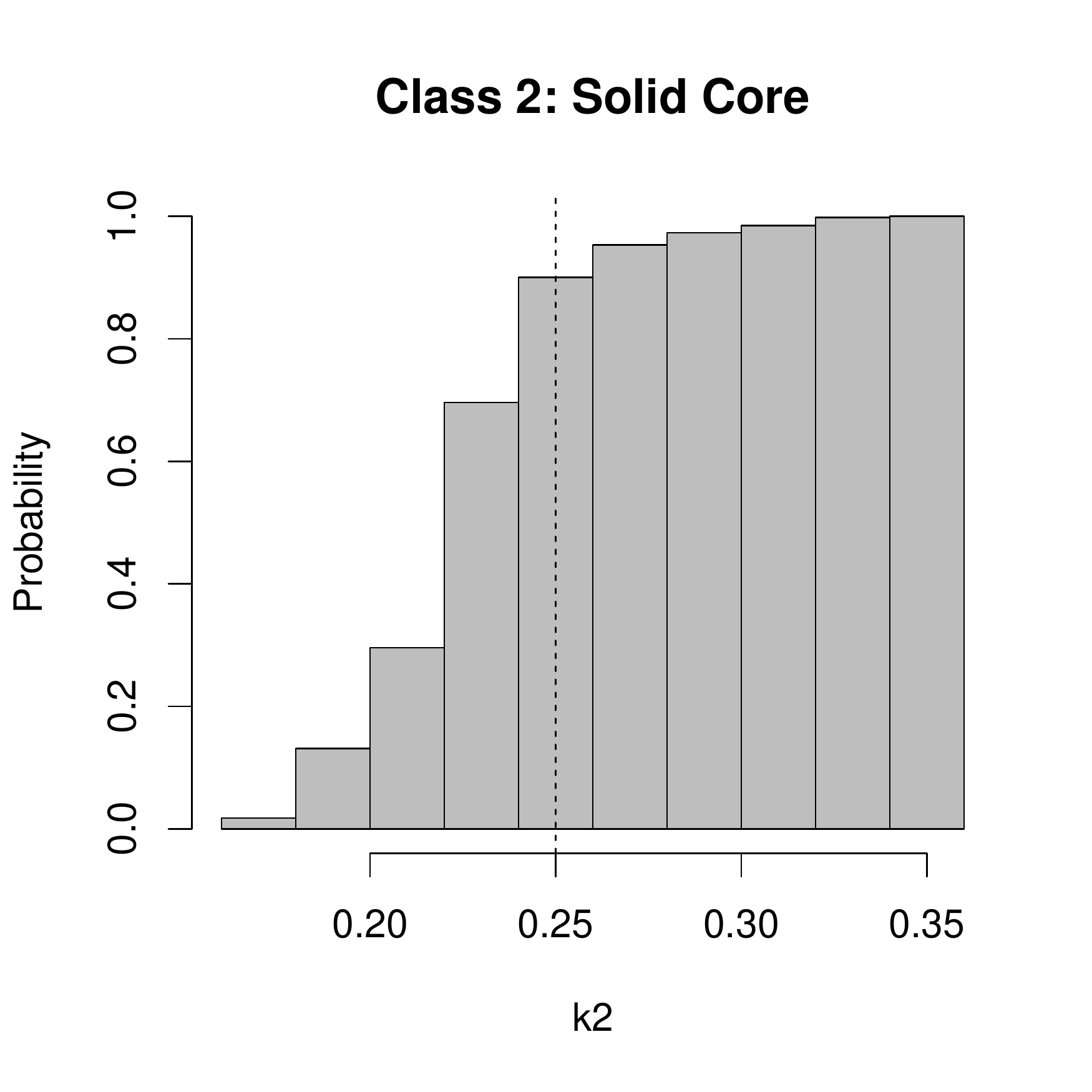}\\
    \includegraphics[scale=0.33]{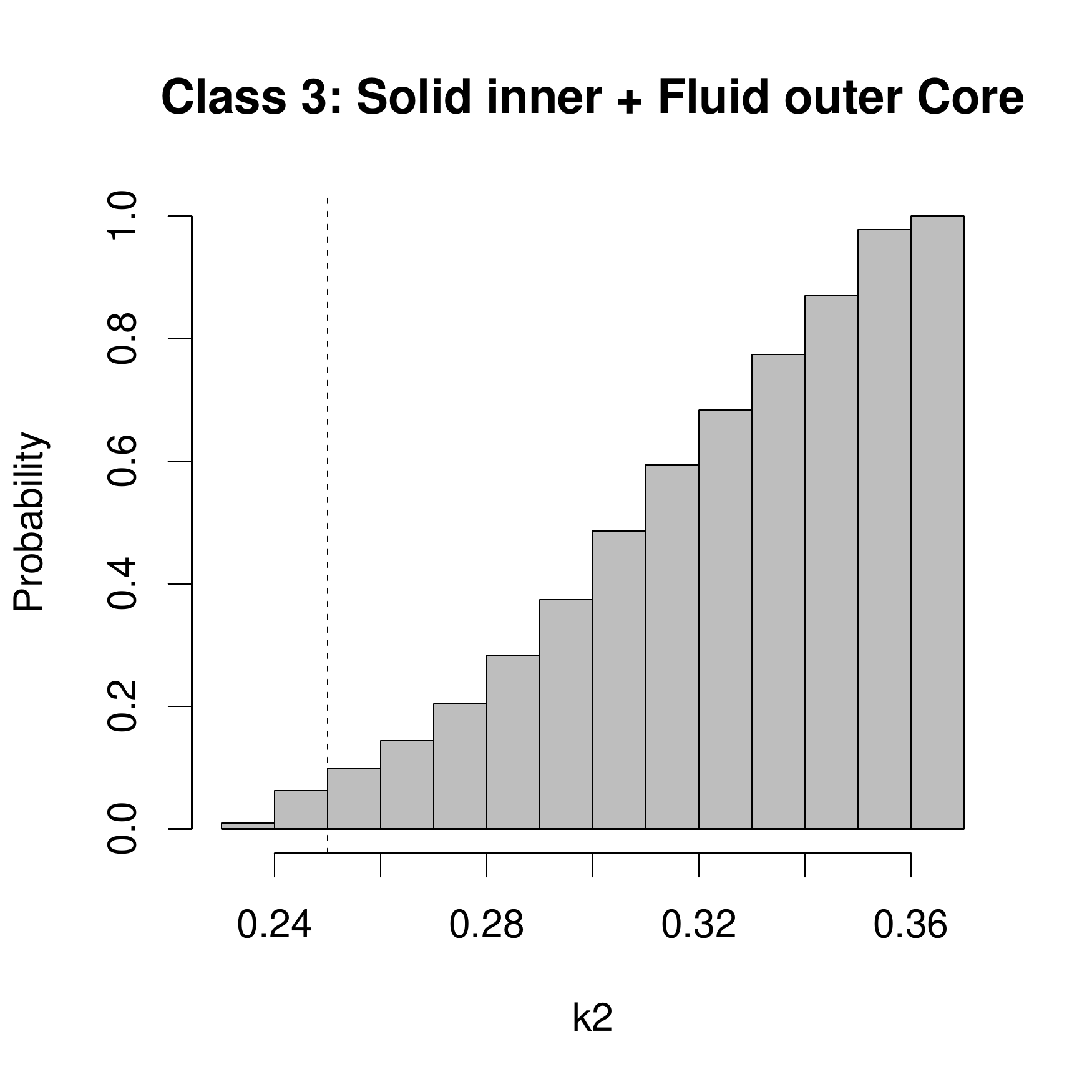}
    \caption{Cumulative histograms of the $k_2$ values obtained with the models of \textbf{Class 1} (fluid core), \textbf{Class 2} (solid core) and \textbf{Class 3} (solid inner and fluid outer core).}
    \label{fig:histo$k_2$}
\end{figure}

\begin{figure}[h!]
    \centering
    \includegraphics[width=13cm]{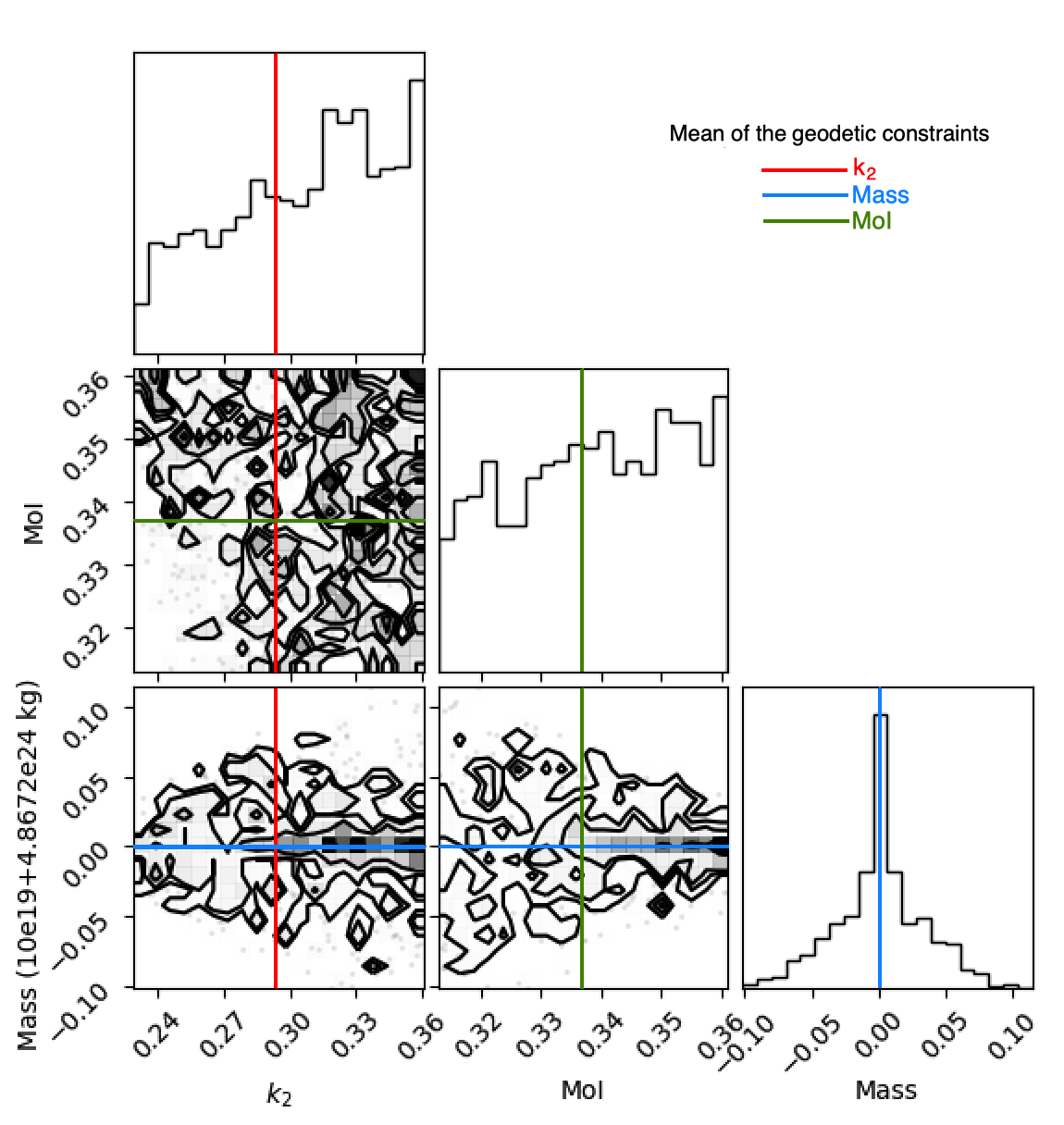}
    \caption{Histograms and corner plots of the real $k_2$, MoI and mass values obtained with the models of \textbf{Class 1} (fluid core). The solid red, black and green lines corresponds to the mean values of the geophysical constraints $k_2$, mass and MoI from Table \ref{Tab:TableVenus}.}
    \label{fig:Corner_FC}
\end{figure}

\begin{figure}[h!]
    \centering
    \includegraphics[width=13cm]{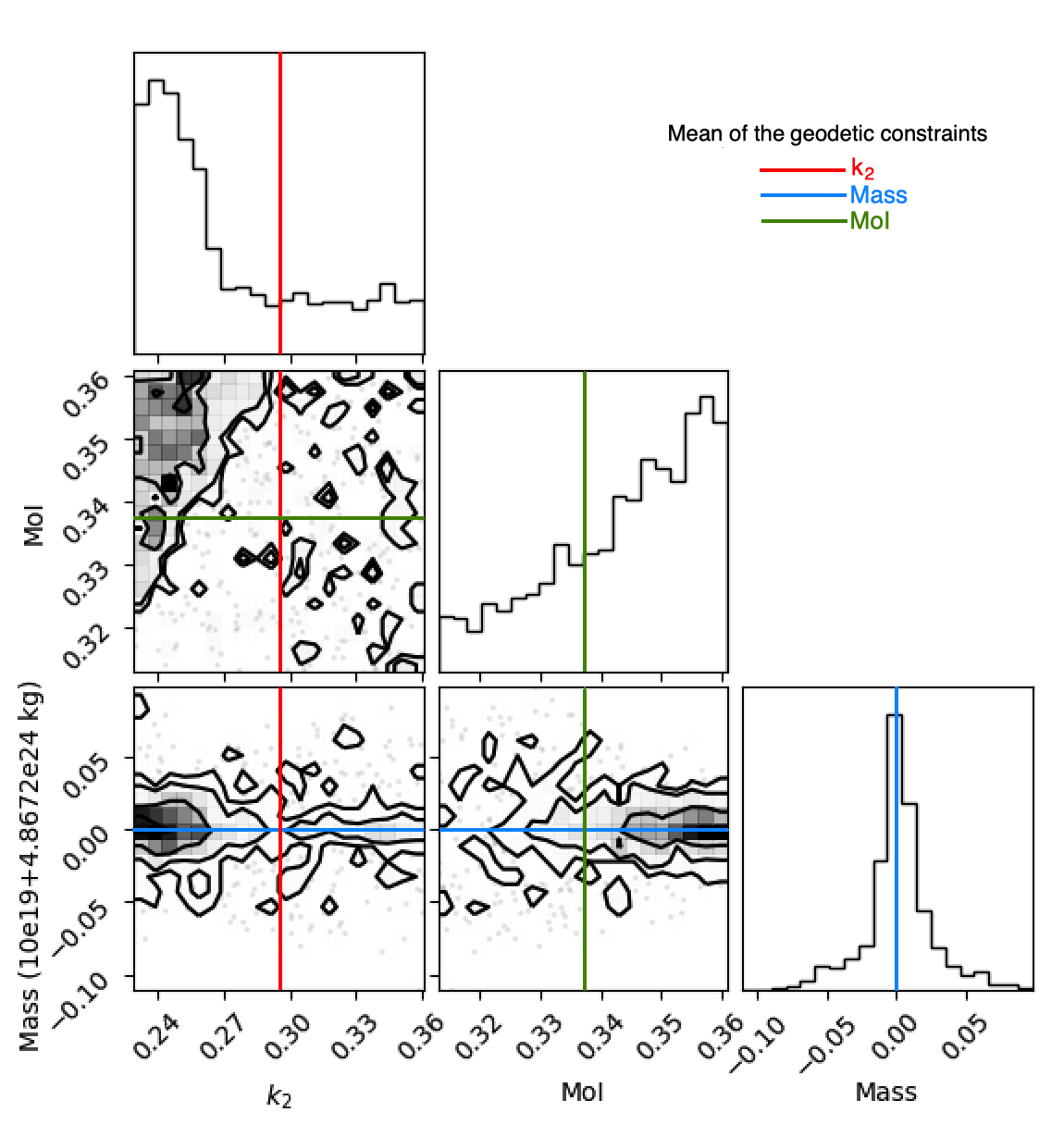}
    \caption{Histograms and corner plots of the real $k_2$, MoI and mass values obtained with the models of \textbf{Class 2} (fluid core). The solid red, black and green lines corresponds to the mean values of the geophysical constraints $k_2$, mass and MoI from Table \ref{Tab:TableVenus}.}
    \label{fig:Corner_SC}
\end{figure}

\begin{figure}[h!]
    \centering
    \includegraphics[width=13cm]{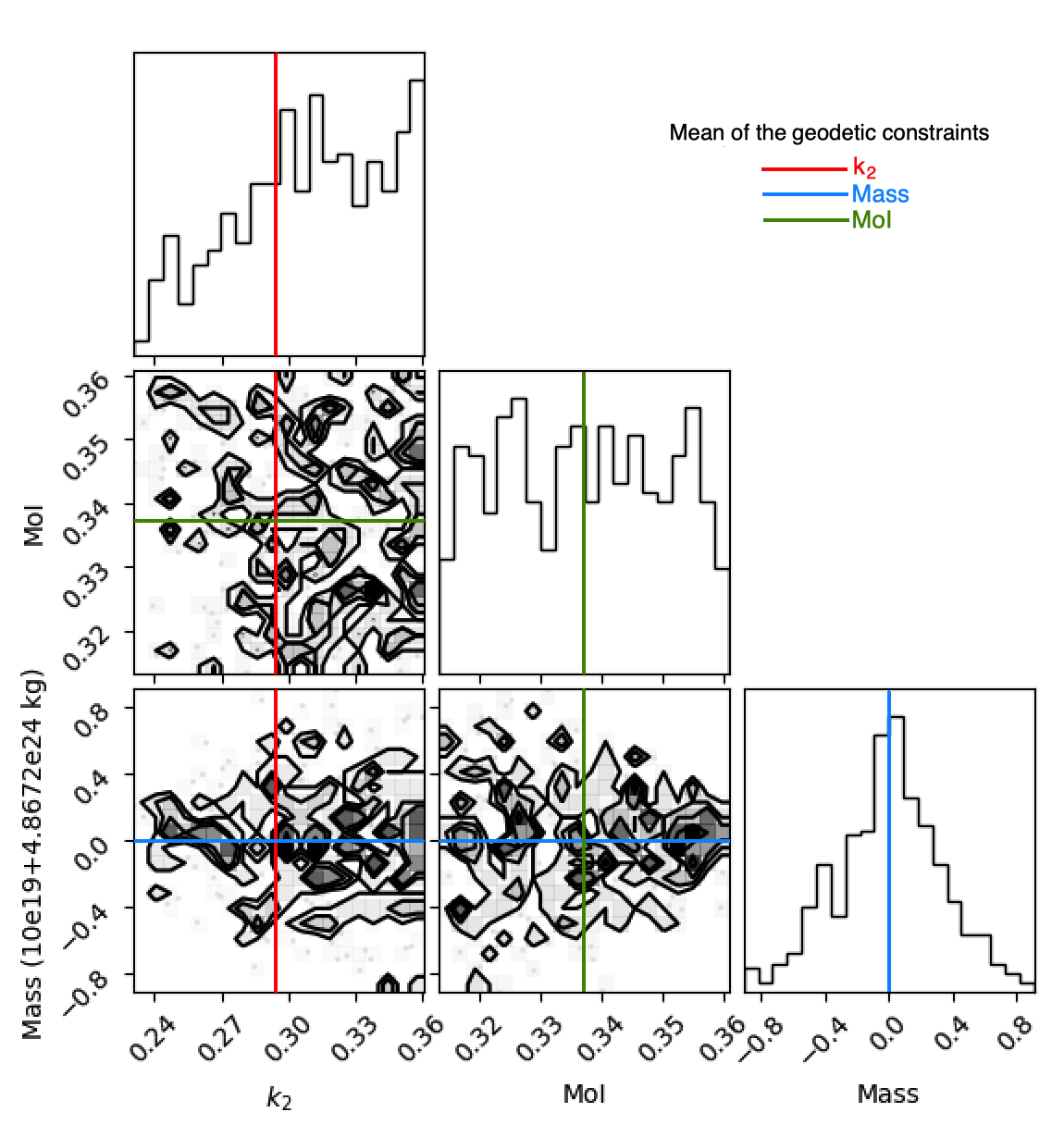}
    \caption{Histograms and corner plots of the real $k_2$, MoI and mass values obtained with the models of \textbf{Class 3} (fluid core). The solid red, black and green lines corresponds to the mean values of the geophysical constraints $k_2$, mass and MoI from Table \ref{Tab:TableVenus}.}
    \label{fig:Corner_SFC}
\end{figure}

We compare our results with previous studies such as the one of \cite{Dumoulin2017} and \cite{Aitta} which constructed a scaled model of the density of Venus as a function of depth using the density profile of PREM \cite{Dziewonski1981Preliminary}. Both studies consider a Venus model with a fluid core and a mantle divided into a lower and an upper layer, as the Earth. These profiles are then comparable with models of \textbf{Class 1} as defined in Sect. \ref{sec:c1}, except that there is no viscosity contrast in between the two mantle layers in \cite{Dumoulin2017}. As one can see from Fig. \ref{Fig:Histograms_FC12}, our \textbf{Class 1} models are in good agreement with the limits extracted from \cite{Dumoulin2017} model V and \cite{Aitta} (illustrated respectively with red and green vertical lines). Moreover, we agree on the \cite{Dumoulin2017} conclusion that a solid core with a high density is mostly likely to be associated with a low viscosity. This case corresponds to the models of our \textbf{Class 2A} with a density for the solid core not smaller than 9376 kg.m$^{-3}$ and a viscosity not greater than 10$^{19.5}$ Pa.s. 
We also agree that the probability of having a $k_2$ $<$ 0.25 is of about 90 $\%$ with a solid core (\textbf{Class 2}) but only of 18 $\%$ and 6 $\%$ with a fluid core or a solid inner core and a fluid outer core, respectively (see Fig. \ref{fig:histo$k_2$}). This result stresses that the $k_2$ value is indeed a good marker of the core state. Figs. \ref{fig:Corner_FC}, \ref{fig:Corner_SC} and \ref{fig:Corner_SFC}, obtained with \cite{corner}, show the relation between the real part of $k_2$, MoI and mass for each of the \textbf{Classes 1}, \textbf{2} and \textbf{3}, respectively. \cs{Our results depend on the presently known observations as the mass and MoI. In the future, the results of this work can be recreated, be better constrained and updated after future observations with smaller uncertainties. Therefore further exploration could prove beneficial in verifying the results obtained in this study.}
The mean value of the geophysical constraints of $k_2$, MoI and mass are $0.295$, $0.337$ and $4.8673 \times 10^{24}$ kg respectively (see Table \ref{Tab:TableVenus}) and are illustrated in red, black and green. These Figures show that the model distribution is centered around the mean mass. The models of \textbf{Class 1} are $55.8\%$ higher and $44.17\%$ lower than the MoI mean (see \ref{fig:Corner_SFC}. These values are respectively $72.39\%$ and $27.6\%$ for \textbf{Class 2} (see Fig. \ref{fig:Corner_SC}) and $49.16\%$ and $50.83\%$ for \textbf{Class 3} (see Fig. \ref{fig:Corner_SFC}). Additionally $59.57\%$ of the models of \textbf{Class 2} have a MoI higher and a $k_2$ respectively higher and lower than the mean estimated value from Table \ref{Tab:TableVenus} (see Fig. \ref{fig:Corner_SC}). Therefore a better estimation of the MoI of Venus, additionally to $k_2$, will better constrain the core structure between a totally solid state and a partially or totally fluid one as the conclusion made with $k_2$ (see Fig. \ref{fig:histo$k_2$}).

\begin{figure}[h!]
    \centering
    \includegraphics[scale=0.4]{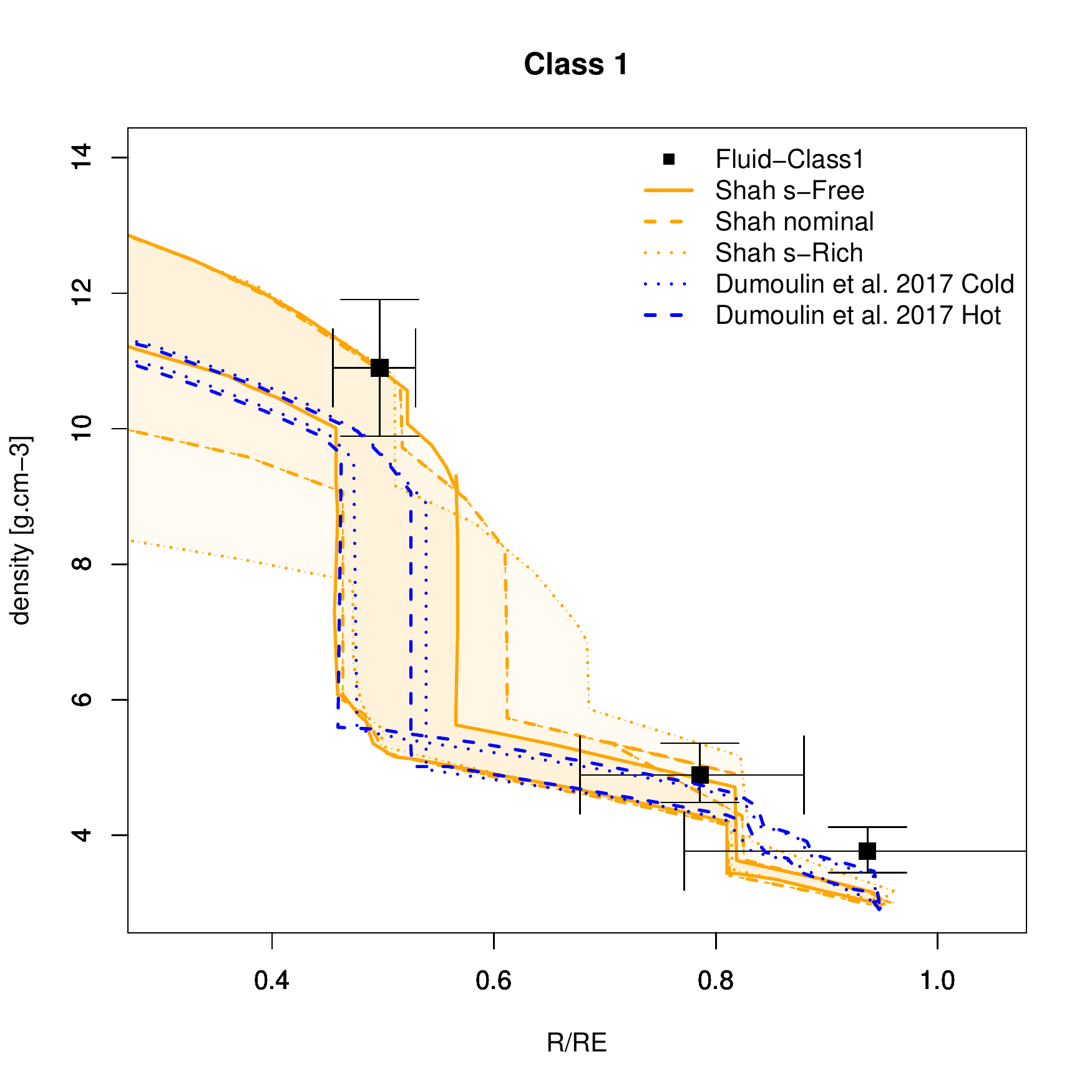}\includegraphics[scale=0.4]{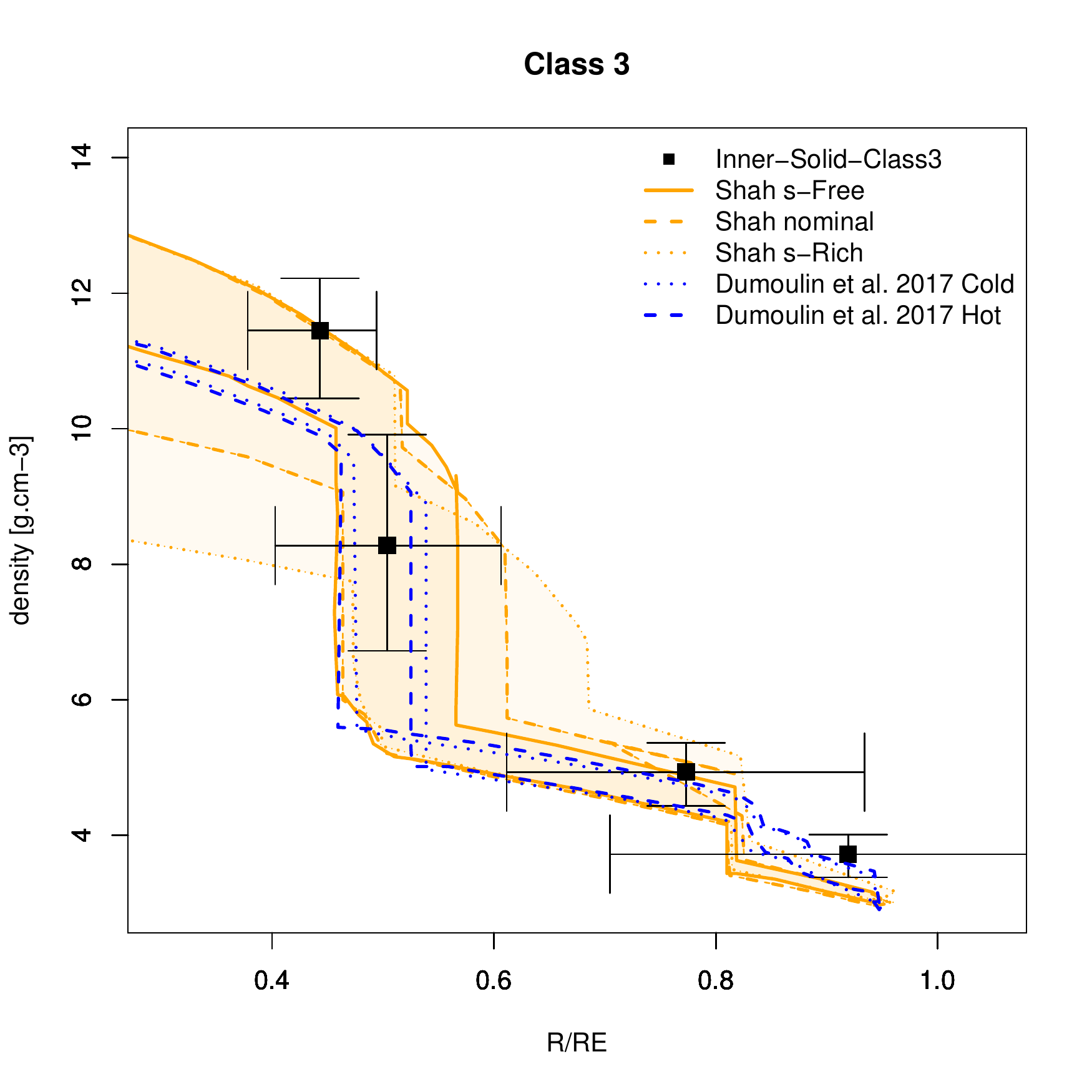}\\
    \includegraphics[scale=0.4]{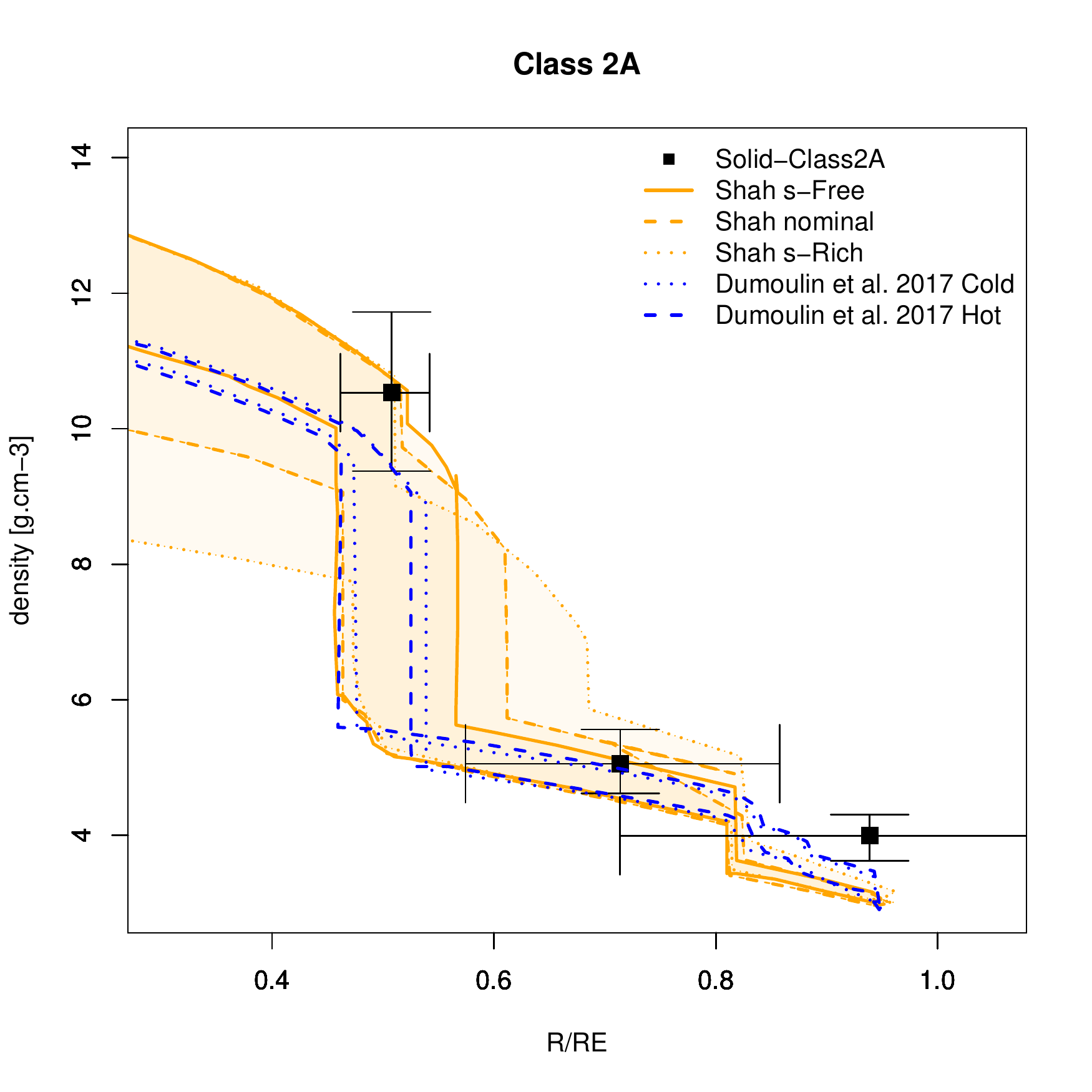}\includegraphics[scale=0.4]{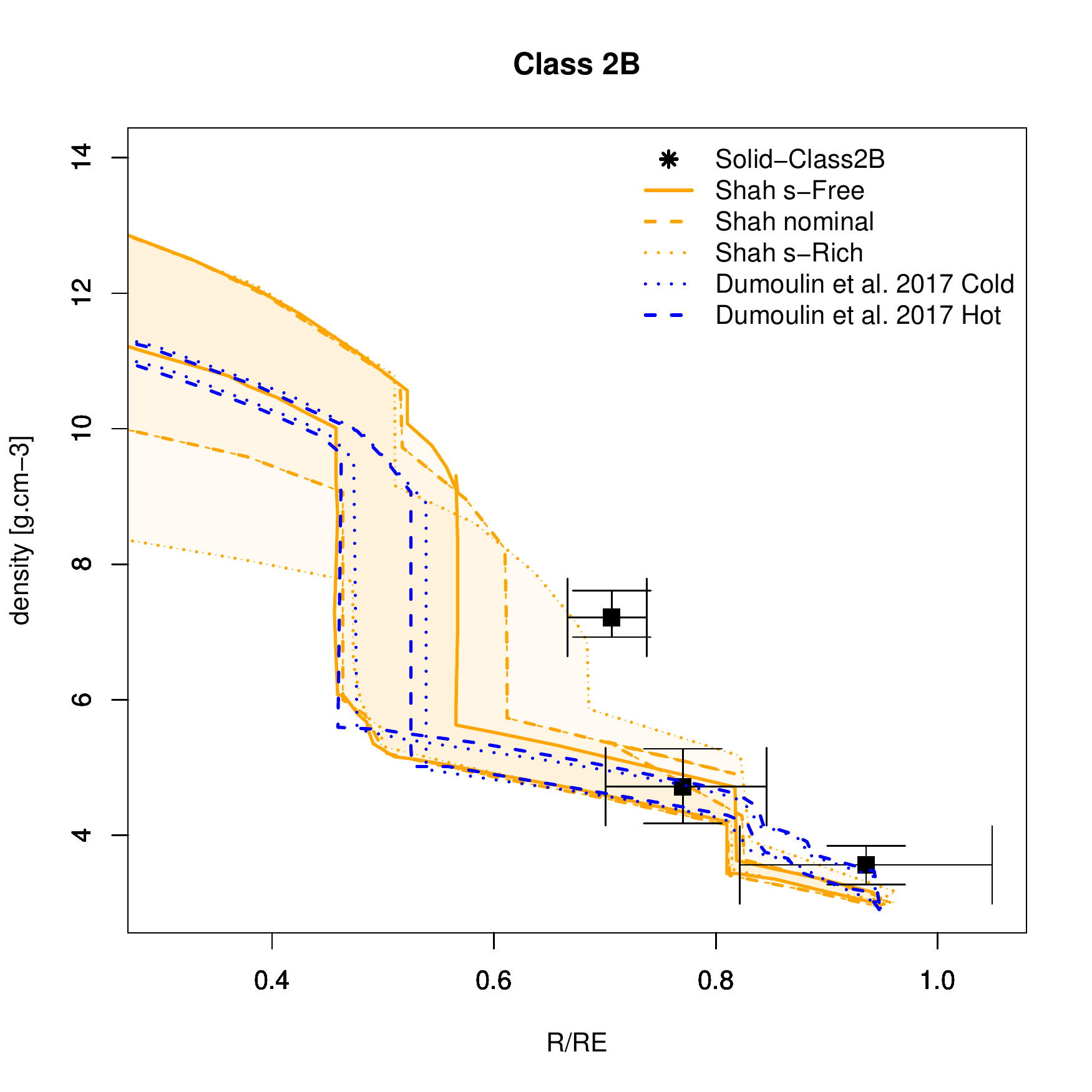}
    \caption{Comparison of the densities obtained with this work for the different classes of models with the profiles from \cite{Shah2021Interior} without subcategories according to the MoI values and \cite{Dumoulin2017} for the two temperature profiles considered in this study (hot and cold). The x-axis gives the ratio between the radius R of each layer and the Earth radius. The error-bars are given at 2-$\sigma$.}
    \label{fig:shah_density}
\end{figure}

We finally compare the density estimates from our classes of models with the end-members of the \cite{Dumoulin2017} density profiles for hot and cold temperature mantle hypothesis (Fig. \ref{fig:shah_density}). We obtain that the density of the lower and the upper mantles match with \cite{Dumoulin2017} profiles within the 2-$\sigma$ error bars, except for models of \textbf{Class 2B} which have a core density completely out the range from \cite{Dumoulin2017} profiles \cs{(black triangle)}. Regarding the density of the cores, our estimates appear to be slightly higher than the one from \cite{Dumoulin2017}, except for the fluid outer core of models of \textbf{Class 3} (including also a solid inner core), which seems to match well with the profiles of \cite{Dumoulin2017}. So despite the fact that our models favor a viscosity contrast between the two mantle layers (as discussed in Sect. \ref{sec:res}), upper and lower mantle densities from all our models match well with that of \cite{Dumoulin2017}. It is not the case for the core densities which are significantly higher than the one of \cite{Dumoulin2017} for the solid inner core, the fluid core and the solid core. Additionally, our densities \cs{show a better compatibility} with the \cite{Shah2021Interior} S-free density profiles (Fig. \ref{fig:shah_density}). Fig. \ref{fig:shah_density} shows the densities obtained in this work as a function of the relative radius (R) with respect to the Earth radius (RE) and compared with the density profiles from \cite{Shah2021Interior} for the three core compositions (S-free, Nominal-S and S-rich). 


\cite{Shah2021Interior} studied also different structure models of Venus based on the equations of state evolution for different hypothesis of core compositions: without sulfur (S-free, with 0 wt\%), with the same amount of sulfur as the Earth (Nominal-S, with 4.6-7.6 wt\%) and with more sulfur than the Earth (S-Rich, with 9.1-22 wt\%). 
Their estimated MoI values are encompassed in the 1-$\sigma$ uncertainty of \cite{Margot2021Spin} and they considered two different patterns of model: those with low MoI values (generally smaller than 0.323) representing models within the 1$\%$ lowest possible values of MoI and those with high MoI (generally greater than $0.323$) gathering models within the $1\%$ highest possible values of MoI. 
\cs{In the goal of better comparing with their study and results, we consider two subcategories} of our models according to their MoI values using the same MoI intervals as in \cite{Shah2021Interior}. More than $57\%$ of our models included in the \cite{Shah2021Interior} Low and High MoI intervals belong to the High MoI models when only $13\%$ of our models belong to the Low MoI type of models. $30\%$ of the rest of our results have either smaller or higher values for the MoI.
From Fig. \ref{fig:shah_density}, we can see that there is a good consistency between our densities and \cite{Shah2021Interior} results for the three compositions, except for the \textbf{Class 2B} set of models which appears to be outside \cite{Shah2021Interior} ranges. The inner core, the outer core and the lower mantle densities obtained for 5 or 4 layers with solid or fluid cores are encompassed in the intervals proposed by \cite{Shah2021Interior} without considering MoI discrimination. For the upper mantle (above 0.8 Earth radius), our estimations appear to be slightly larger. However, \cite{Shah2021Interior} consider the upper mantle and the crust as a single layer while we consider two different layers. 

\begin{figure}[h!]
    \centering
    \includegraphics[scale=0.4]{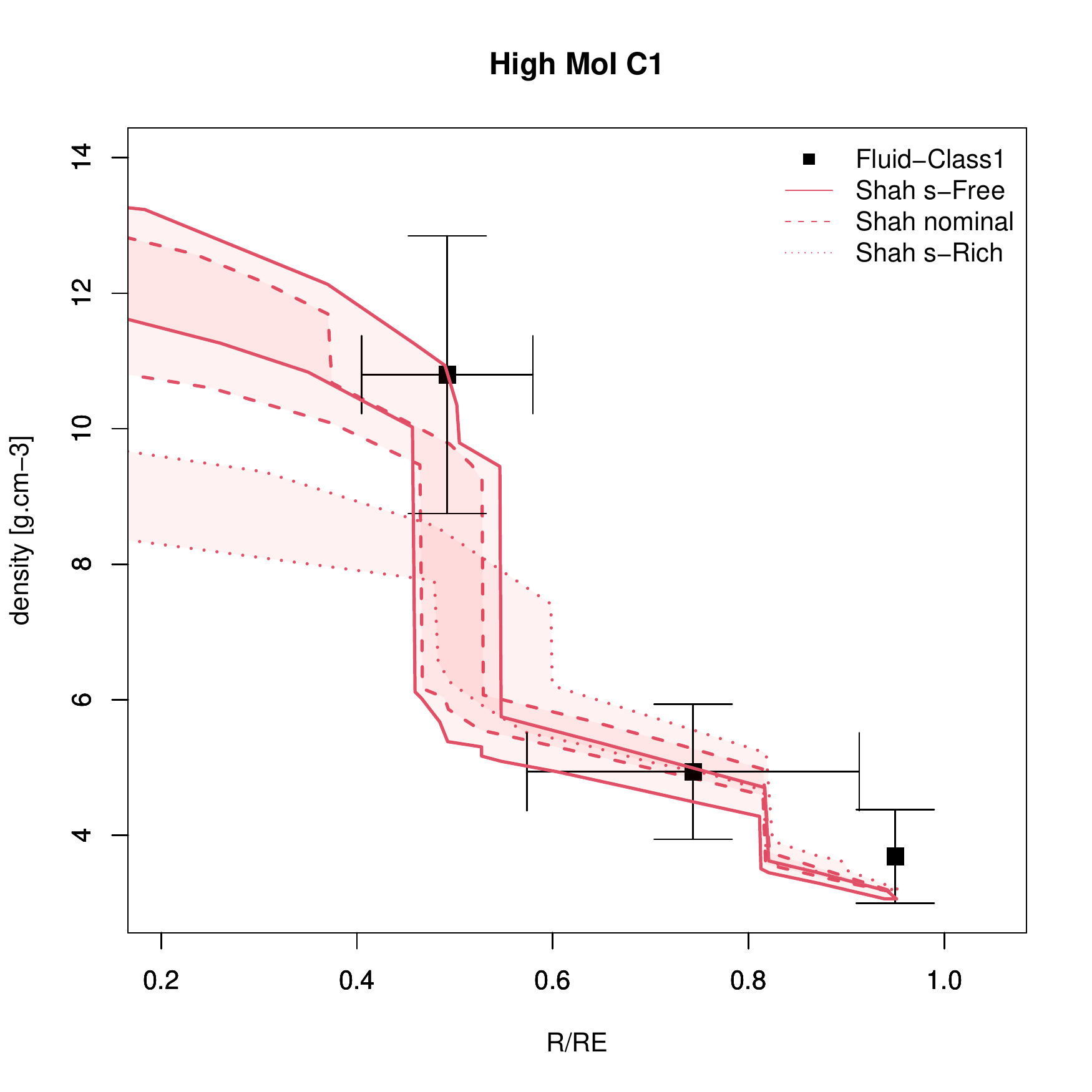}\includegraphics[scale=0.4]{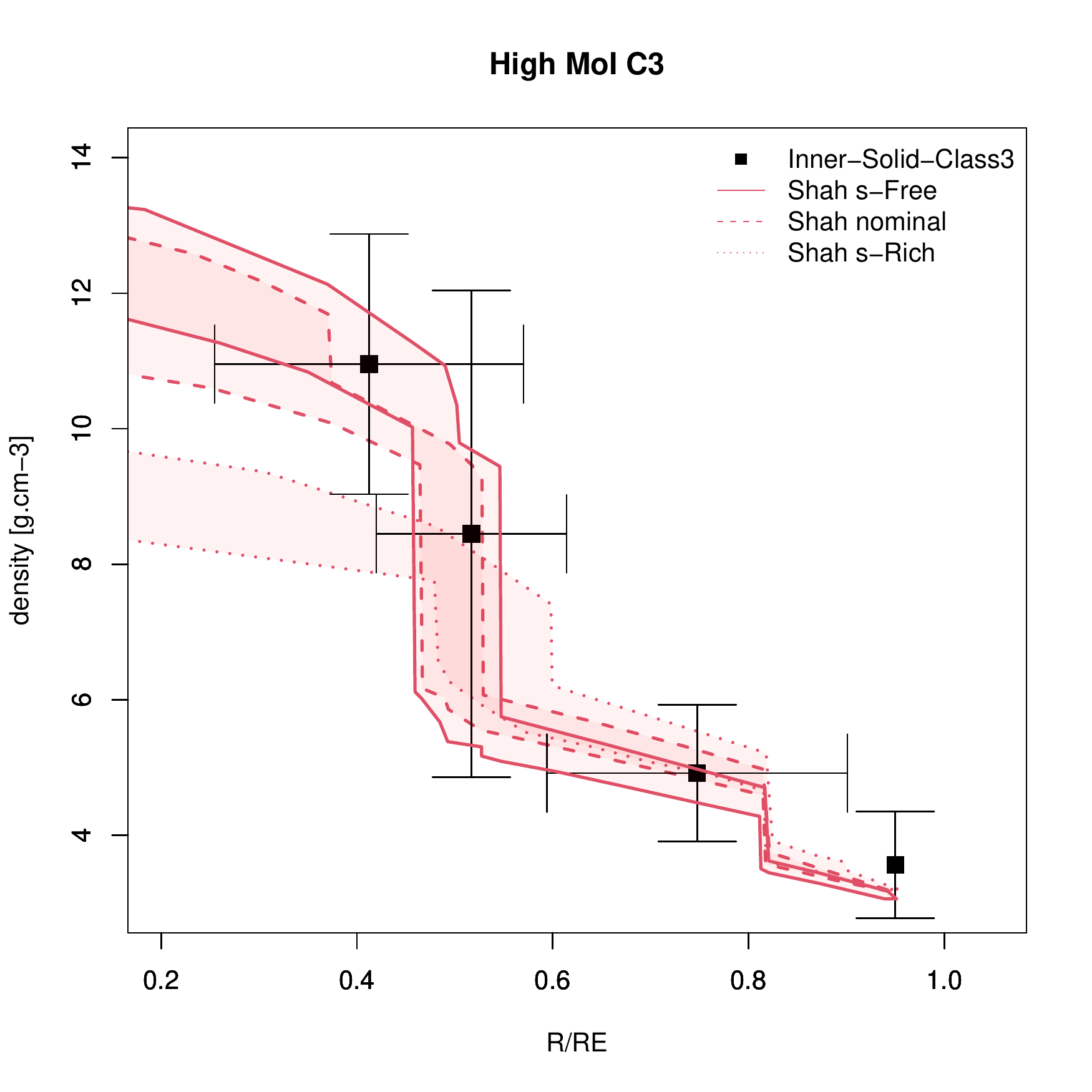}\\
    \includegraphics[scale=0.4]{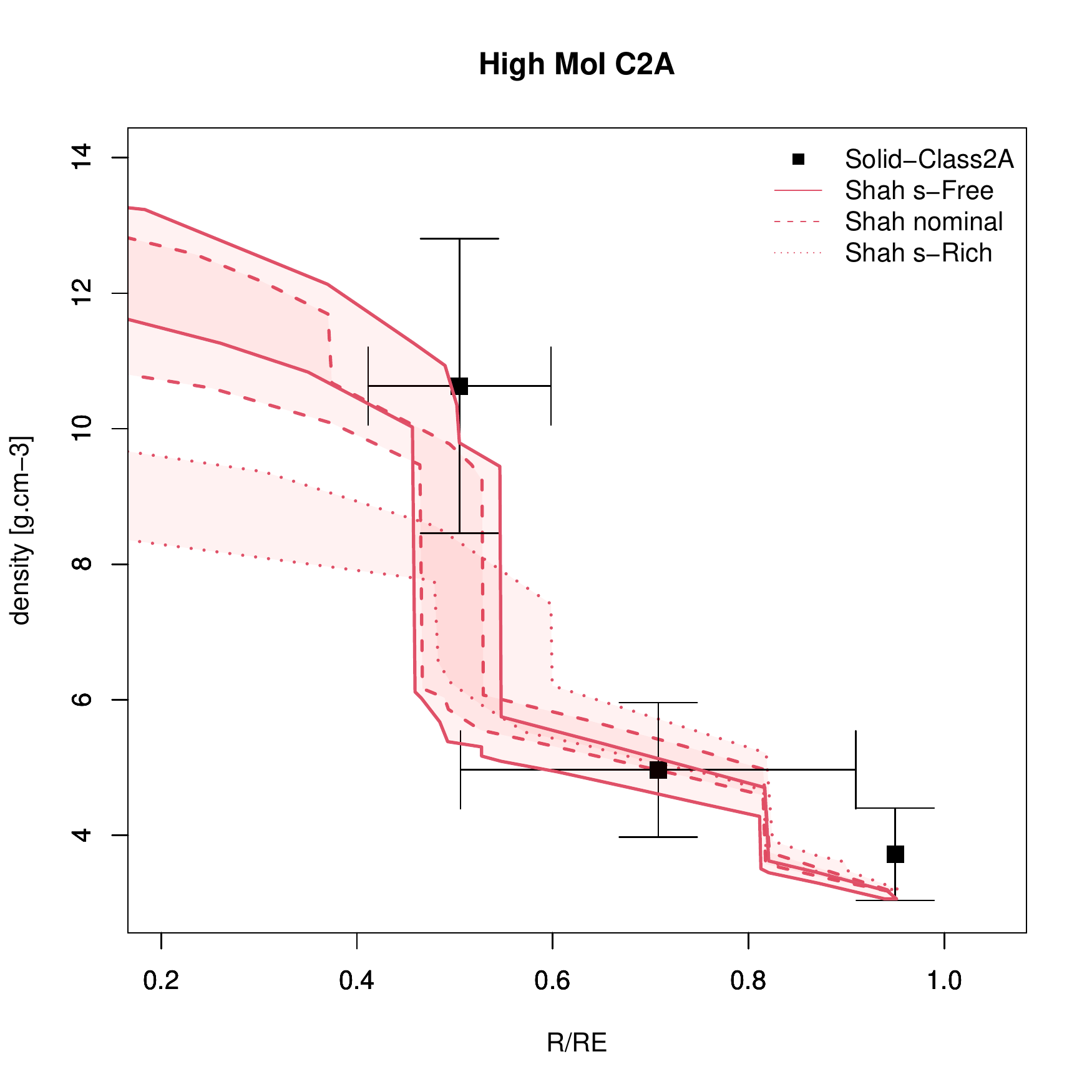}\includegraphics[scale=0.4]{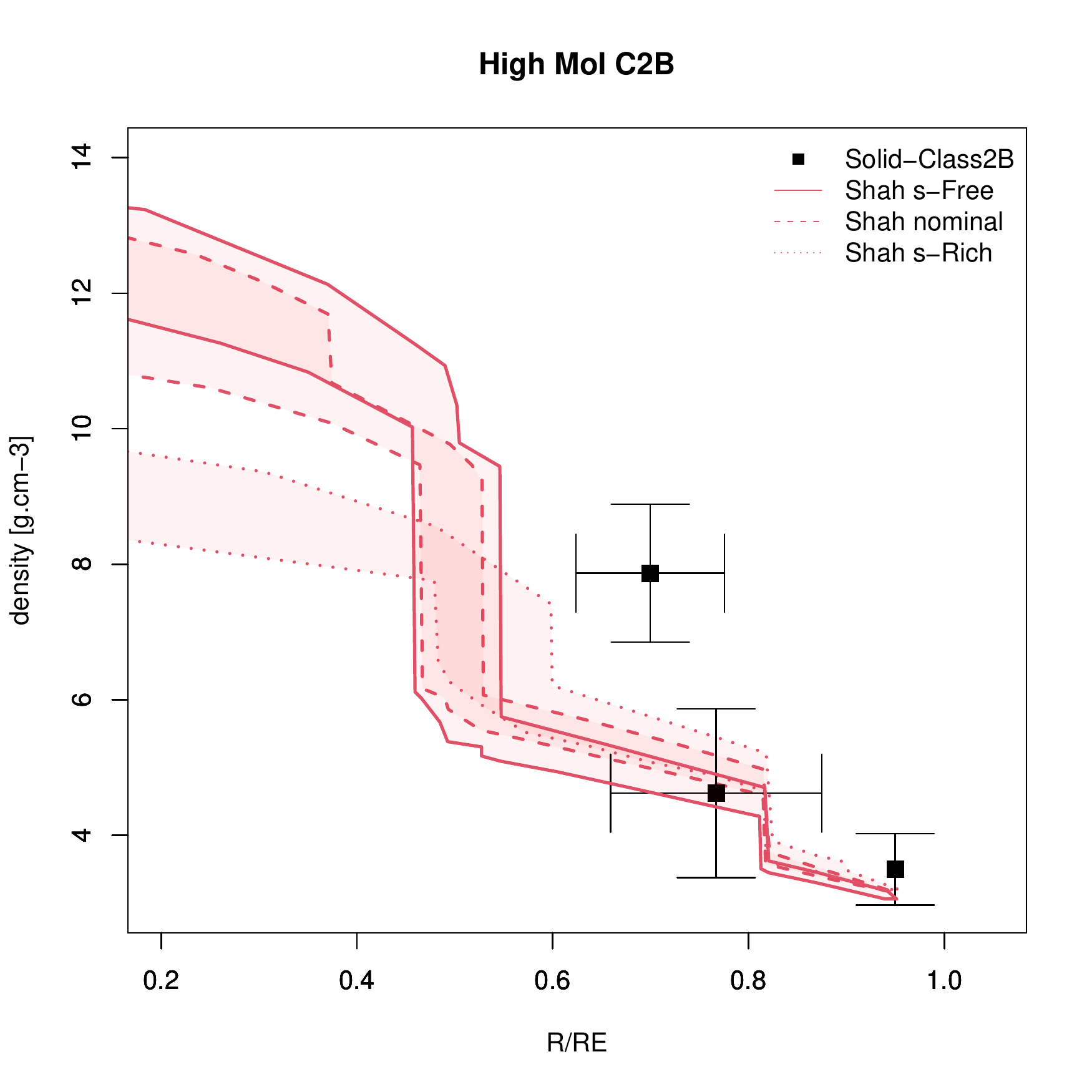}
    \caption{Comparison of the densities obtained with this work for the different classes of models  with the profiles from \cite{Shah2021Interior} considering High MoI as defined in \cite{Shah2021Interior}. The x-axis gives the ratio between the radius R of each layer and the Earth radius. The error-bars are given at 2-$\sigma$.}
    \label{fig:shah_density2h}
\end{figure}

\begin{figure}[h!]
    \centering
    \includegraphics[scale=0.4]{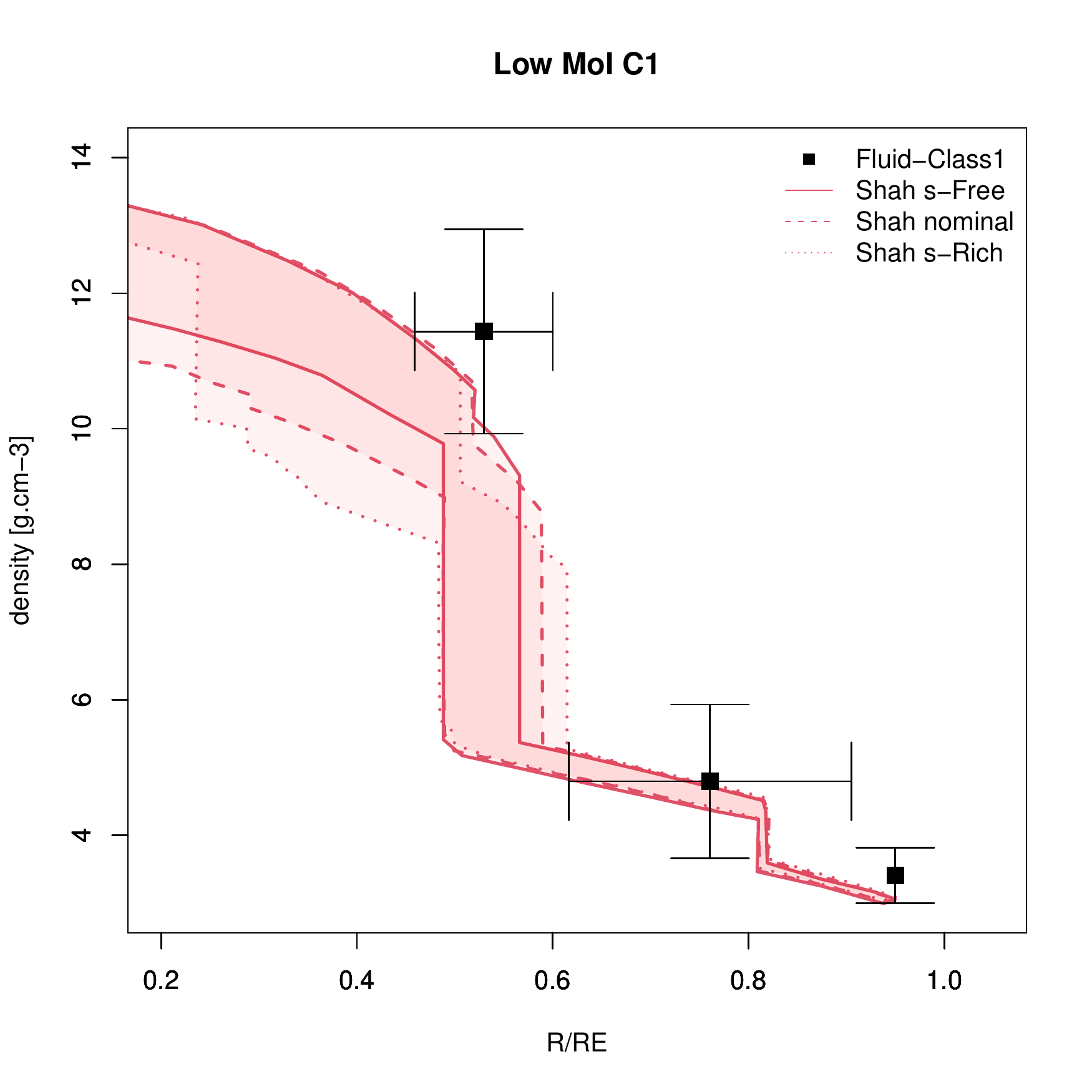}\includegraphics[scale=0.4]{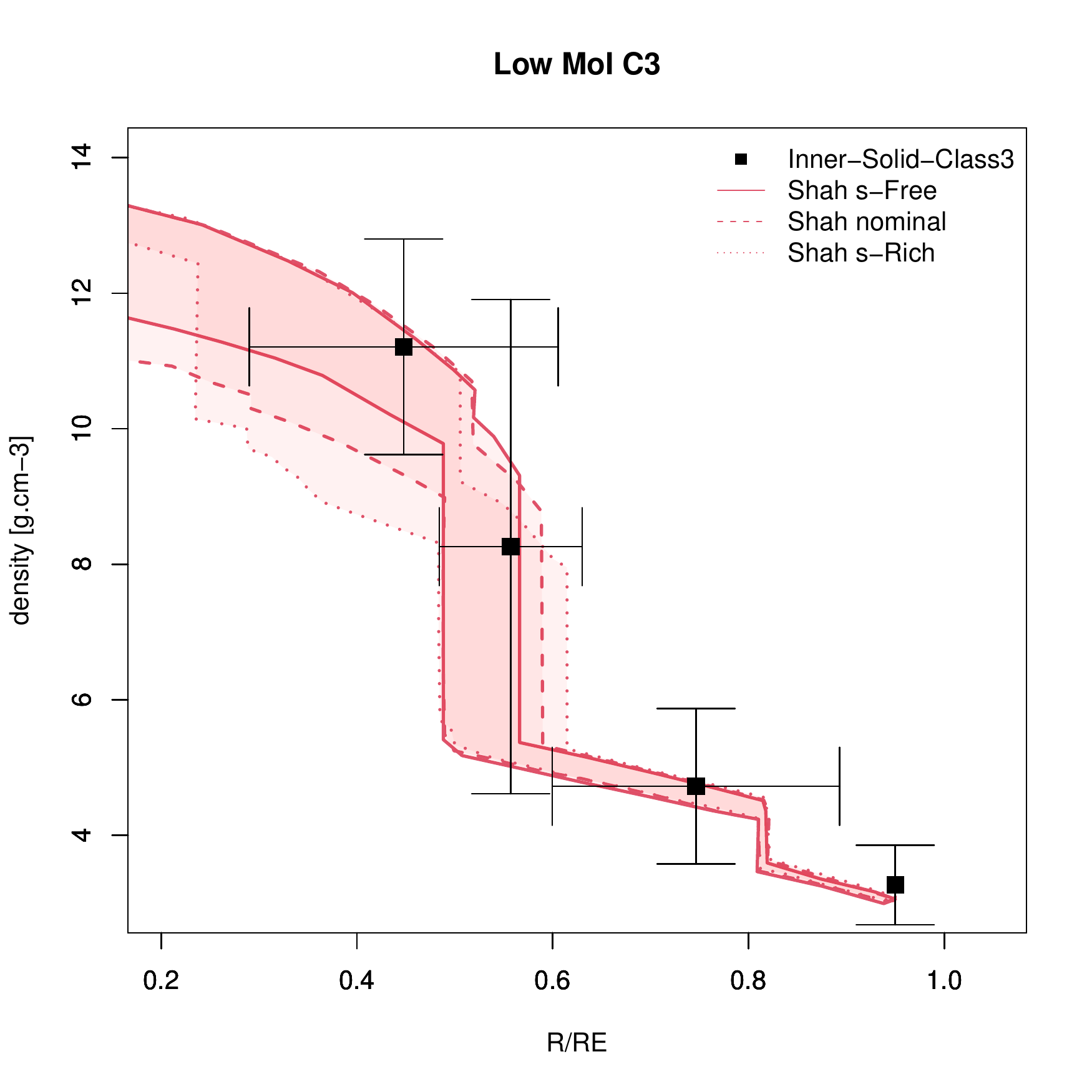}\\
    \includegraphics[scale=0.4]{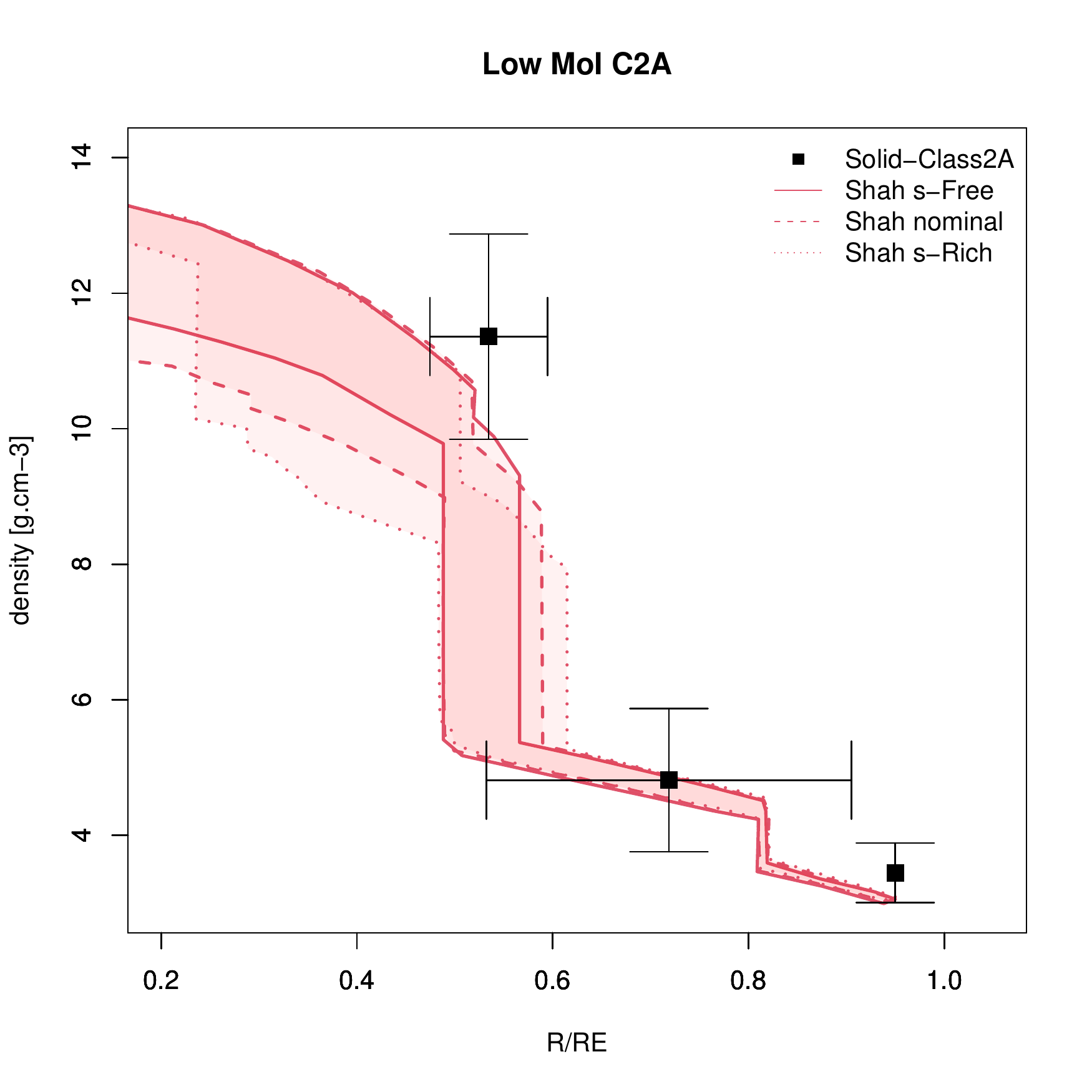}\includegraphics[scale=0.4]{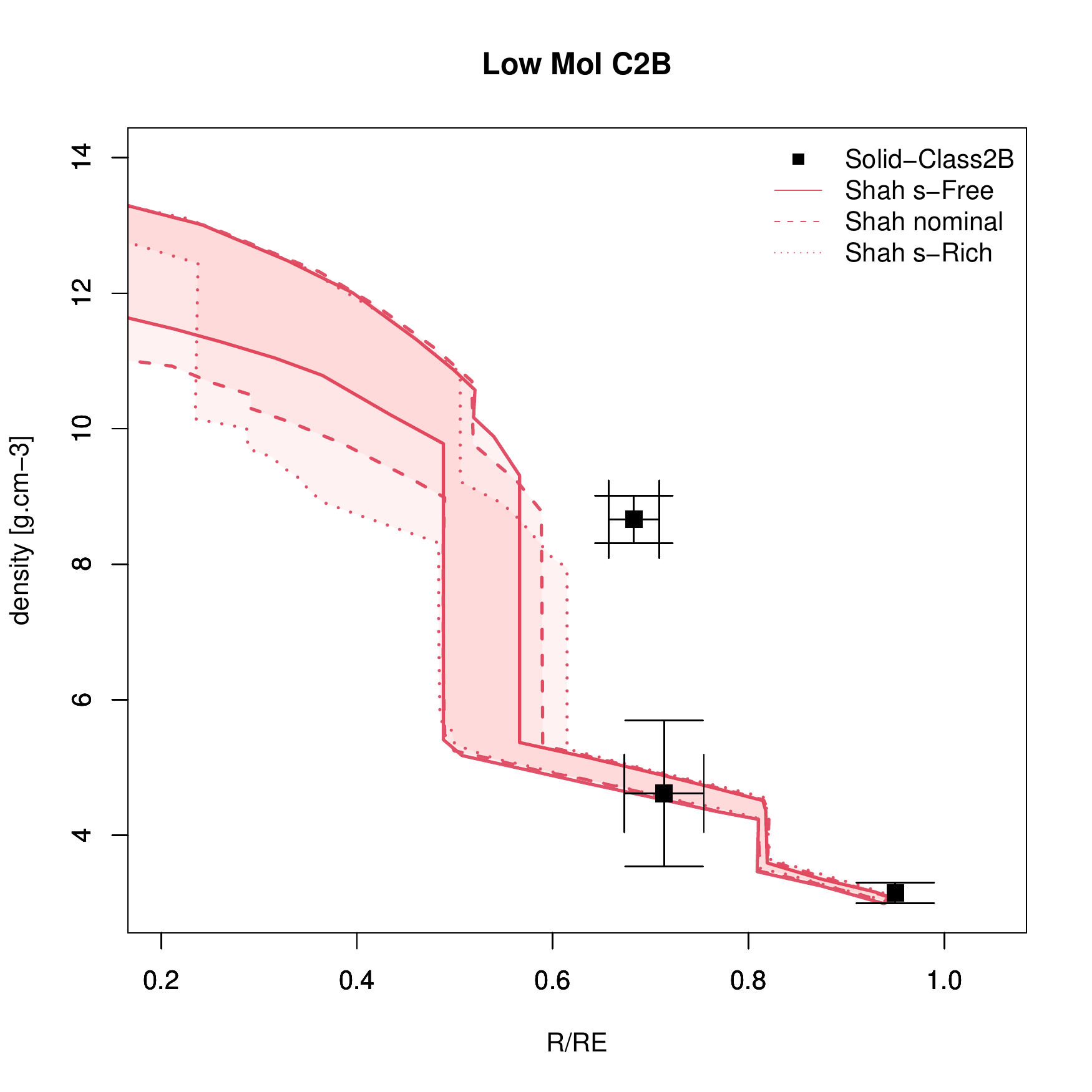}
    \caption{Comparison of the densities obtained with this work for the different classes of models  with the profiles from \cite{Shah2021Interior} considering Low MoI as defined in \cite{Shah2021Interior}. The x-axis gives the ratio between the radius R of each layer and the Earth radius. The error-bars are given at 2-$\sigma$.}
    \label{fig:shah_density2l}
\end{figure}

The radius and density of the upper mantle obtained in this work are about $5968$ km and $3765$ kg.m$^{-3}$, respectively. The radius of the lower mantle is about $5005$ km. In averaging the upper mantle with the crust which has a fixed density of 2950 kg.m$^{-3}$, we obtain a crust+upper mantle density of about 3688 kg.m$^{-3}$, closer to the value expected by \cite{Shah2021Interior}.
We consider the same MoI subcategories as the one proposed in \cite{Shah2021Interior}, and as one can see on \cs{Figs. \ref{fig:shah_density2h} and \ref{fig:shah_density2l}} where the densities are plotted versus the relative depth for models presenting High MoI and Low MoI, respectively. The same conclusions remain for the Low MoI case (see Fig. \ref{fig:shah_density2l}) as the total MoI range (see Fig. \ref{fig:shah_density}). For models with High MoI (gathering more than 57$\%$ of our models), see Fig. \ref{fig:shah_density2h}, we see a \cs{rather better consistency} between our estimates and the S-free and the Nominal S profiles. 
In particular in Fig. \ref{fig:shah_density2l} the {\bf{Class 3}} (top-right) inner core density and the {\bf{Class 1}} (top-left) and {\bf{Class 2A}} (bottom-left) core densities are significantly different from the one expected with a S-rich profile, whereas they are statistically consistent at 2-$\sigma$ with the nominal-S profile and totally encompassed in the S-free profile. 

One way to model the temperature dependence of the viscosity is to use the Arrhenius
law.  \cite{Roller1986Rheology} shows that the viscosity of a material can be expressed as an exponential function of temperature, in other words as an Arrhenius-type function. Based on this fact, \cite{Karato1993Rheology,Karato2008Deformation}  deduced an expression of the temperature which is highly dependent on the viscosity. Eq. \ref{eq: temp4} is reformulated from equation 2 of \cite{nakada2012viscosity} which is deduced from the work of \cite{Karato1993Rheology,Karato2008Deformation}. As explained in \cite{nakada2012viscosity}, it is possible if one assumes the temperature $T_{u}$ and the viscosity $\eta_{u}$ of the upper mantle layer to deduce the temperature of the lower layer, $T_{l}$, by considering the following relation:
\begin{equation}
    \label{eq: temp4}
        T_{l} = \frac{H^{*}_l T_{u}}{H^{*}_u + T_{u} R_g ln(\frac{\eta_{l}}{\eta_{u}})}
\end{equation}
where $\eta_l$ is the lower mantle viscosity, $\eta_u$ is the upper mantle viscosity, $H^*_l$ and $H^{*}_u$ are the activation enthalpy for the lower and the upper mantle respectively and $R_g$ is the gas constant. Therefore we calculate the temperature of the lower mantle from the deduced viscosities.

\begin{figure}[h!]
    \centering
    \includegraphics[scale=0.7]{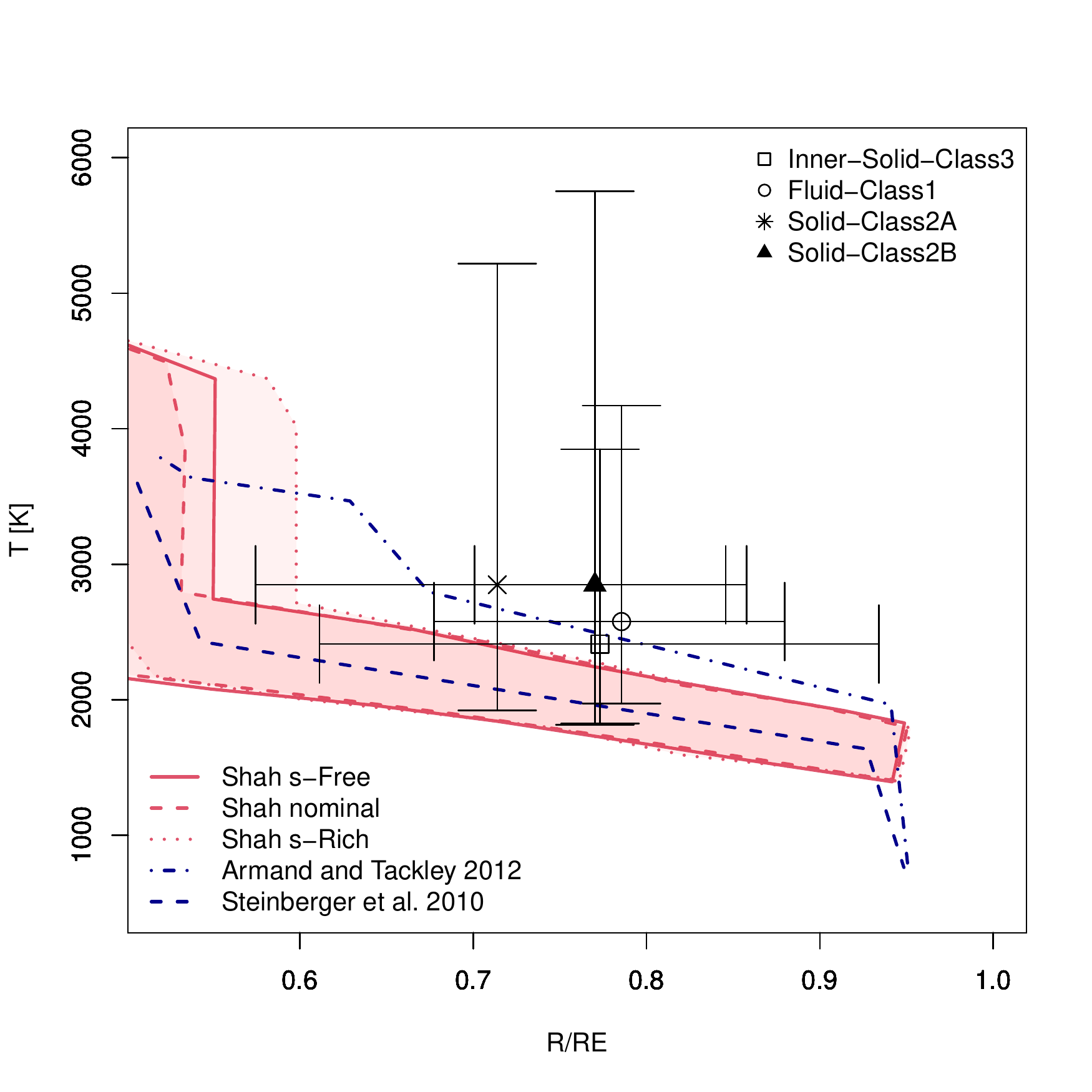}
    \caption{Comparison of the temperatures for the lower mantle for the different classes of models with profiles from \cite{Shah2021Interior}, \cite{Armann2012Simulating} and \cite{Steinberger2010Deep}. The temperature of the upper mantle is fixed to $1600$ K. Same x-axis as on Fig. \ref{fig:shah_density}. The error-bars are given at 2-$\sigma$}
    \label{fig:temprofile}
\end{figure}

\begin{figure}
    \centering
    \includegraphics[scale=0.5]{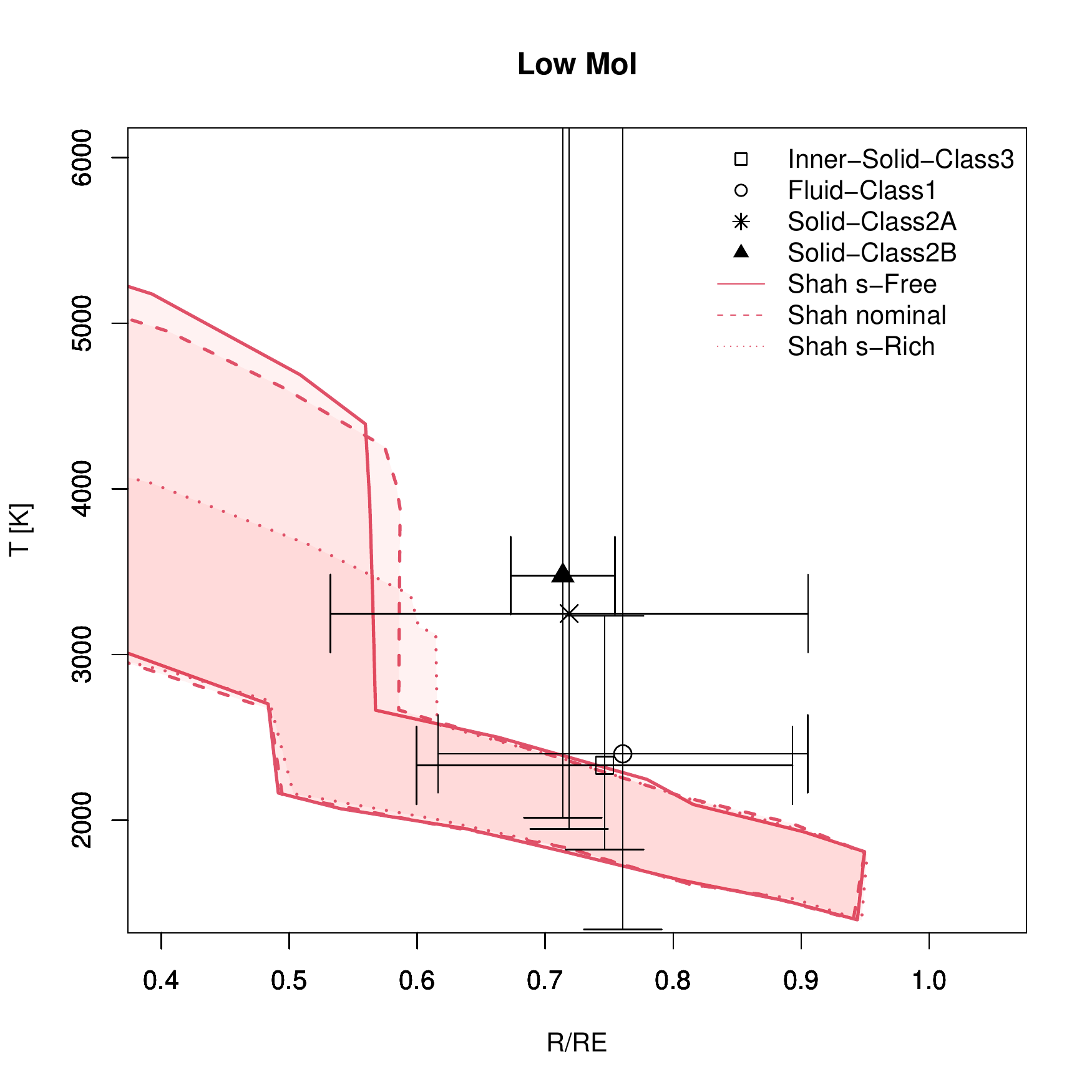}\\
    \includegraphics[scale=0.5]{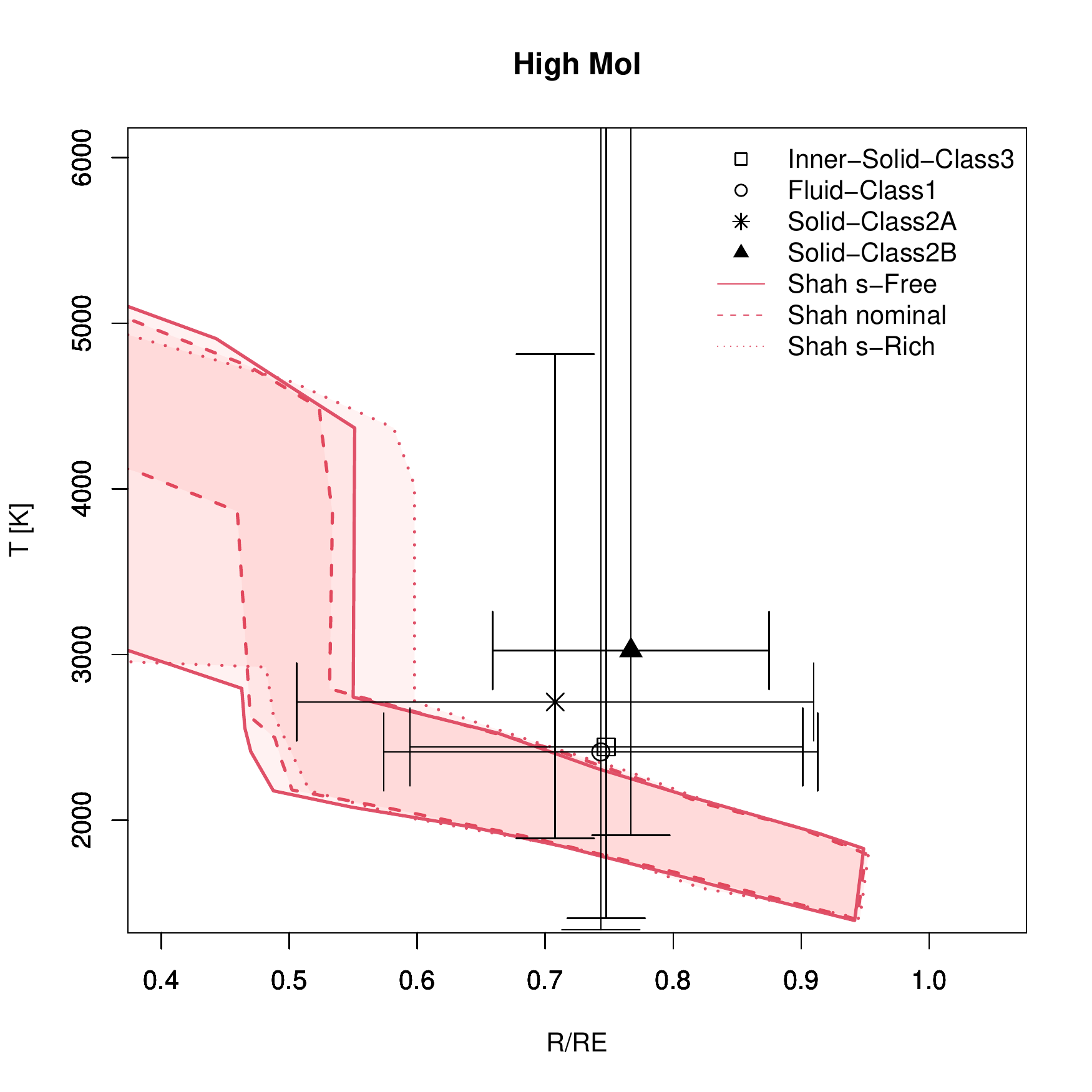}
    \caption{Comparison of the temperatures for the lower mantle obtained with this work for the different classes of models with the profiles from \cite{Shah2021Interior} considering Low and High MoI as defined in \cite{Shah2021Interior}. Same x-axis as on Fig. \ref{fig:shah_density}. The error-bars are given at 2-$\sigma$}
    \label{fig:temprofile2}
\end{figure}

From the mantle parameters (temperature, density thickness and viscosity) we obtain a Rayleigh number much higher than the critical value, therefore the mantle in convective. This result justifies the use of Eq. \ref{eq: temp4} for convective layers \citep{nakada2012viscosity}.
Using Eq. \ref{eq: temp4}, we estimate the temperature of the lower mantle from the viscosity contrast between the upper and the lower mantle layers for each class of models. They are shown on Fig. \ref{fig:temprofile}. 
We assume for this an upper mantle temperature of $1600$ K as given by the \cite{Shah2021Interior} temperature profile reproduced on Fig. \ref{fig:temprofile}. The activation enthalpy values ($H^{*}$) are taken equivalent to those given for the Earth (240 kJ.mol$^{-1}$ for the upper mantle and $430$ kJ.mol$^{-1}$ for the lower mantle) as in \cite{NAKAKUKI2010309}. We consider values of the upper mantle viscosities given by Table \ref{tab:results} for the different models. We then obtain values plotted on Fig. \ref{fig:temprofile}. The error-bars in x-axis correspond to 2-$\sigma$ uncertainties given in Table \ref{tab:results} and the error-bars in y-axis correspond to uncertainties deduced from Table \ref{tab:results} upper mantle viscosities. From these estimations, we see that our models seem to propose a slightly hotter, but still statistically consistent $T_{l}$ in comparison with \cite{Shah2021Interior} with or without MoI subcategories (see Figs. \ref{fig:temprofile} and \ref{fig:temprofile2}). 
The lower mantle temperatures deduced from our approach are also consistent with the temperature profiles from \cite{Steinberger2010Deep} and \cite{Armann2012Simulating} as one can see on Fig. \ref{fig:temprofile}. These two reference profiles give two possible extrema of temperature evolution with depth: a cold one from \cite{Steinberger2010Deep} (Venus scaled adiabatic profile) and a hot one from \cite{Armann2012Simulating} (thermo-chemical Venus evolution). Our temperature for the lower mantle is compatible at 2-$\sigma$ with the two hot and cold profiles but suggests an even hotter temperature than \cite{Armann2012Simulating}.

\begin{figure}[h!]
    \centering
    \includegraphics[scale=0.4]{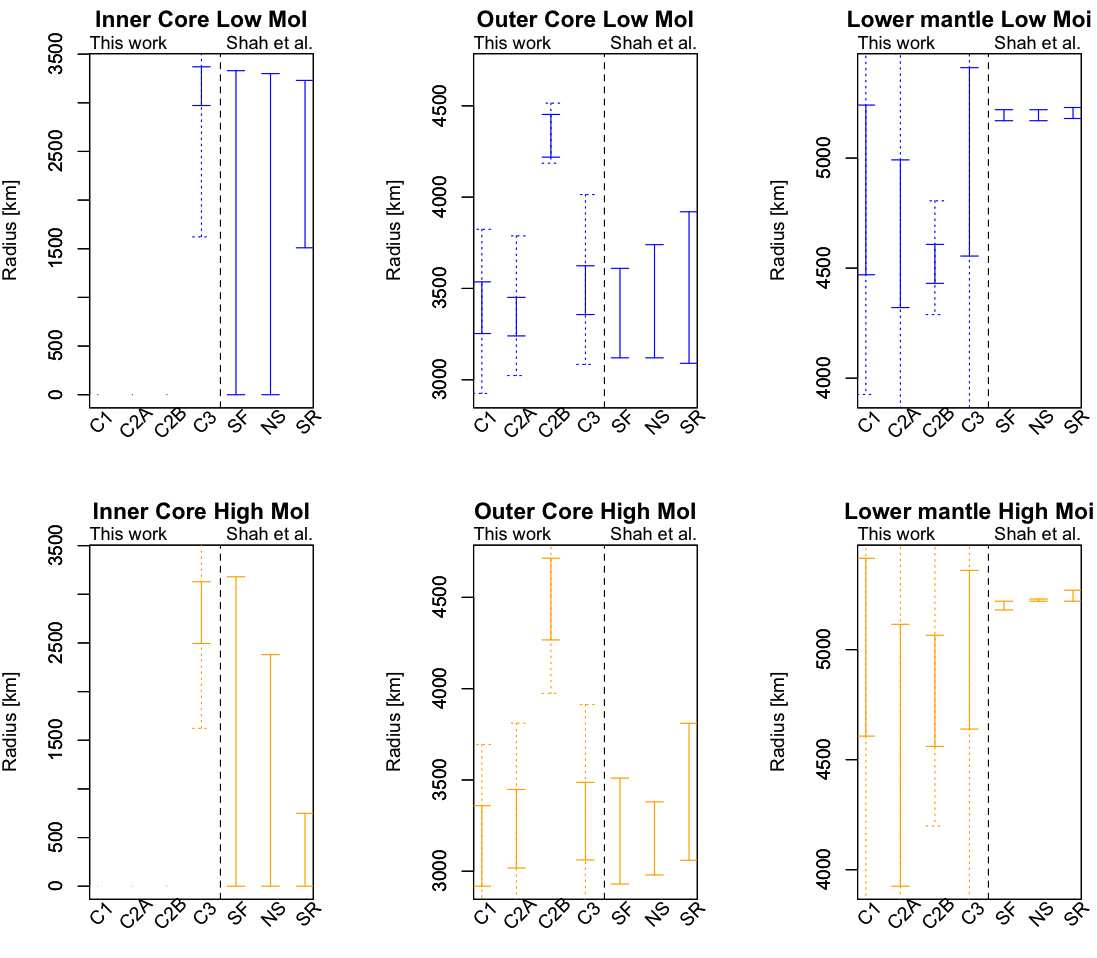}
    \caption{Comparisons between layer boundaries (radii) from \cite{Shah2021Interior} and those obtained for the different classes considering MoI subcategories as presented in \cite{Shah2021Interior}. C1, C2A, C2B, C3 stand for  \textbf{Class 1}, \textbf{Class 2A}, \textbf{Class 2B} and \textbf{Class 3} respectively and SF, NS and SR stand for S-free, Nominal-S and S-rich core models as defined in  \cite{Shah2021Interior}, respectively. The error-bars are given at 2-$\sigma$}
    \label{fig:shah}
\end{figure}

The radii of the different layers considered in this work and in \cite{Shah2021Interior} are shown on Fig. \ref{fig:shah}. As in \cite{Shah2021Interior}, we consider two regimes of models according to MoI values: high MoI (greater than 0.323) and with low MoI (smaller than 0.323). The upper mantle is not consider here as it is supposed to be fixed in \cite{Shah2021Interior}. With these comparisons, it appears that the \textbf{Class 2B} (models with a big core and a low density) is not consistent with \cite{Shah2021Interior}. This class was already pointed as an outlier of our selection (see discussion of Sect. \ref{sec:c2}) and gathers models with no viscosity contrast between the upper mantle and the lower mantle. For models with high MoI (lower row of Fig. \ref{fig:shah}), \cs{the results are to some-degree more compatible with the} S-free models from \cite{Shah2021Interior} than with Nominal-S or S-rich models, especially for \textbf{Class 3}. It is particularly \cs{more explicit} with the radius of the inner core for which \textbf{Class 3} value ($2841_{2494}^{3129}$ km) is consistent with that of S-free (from $0$ to $3180$ km) but not with that of Nominal-S (from $0$ to $2380$ km) or S-rich (from $0$ to $750$ km). For the lower mantle and the outer core radii for models of \textbf{Class 3} but also of \textbf{Class 1} and \textbf{Class 2B}, our values match well with all the models of \cite{Shah2021Interior}. For models with low MoI (upper row of Fig. \ref{fig:shah}), the totality of our models except those of \textbf{Class 2B} are consistent with the models of \cite{Shah2021Interior}. The comparisons between the densities estimated in this former work and ours, \cs{done with error-bars at $2-\sigma$}, especially for high MoI models (representing more than $57\%$ of our results) \cs{to be more compatible generally} with S-free profiles than with S-rich. \\

Finally, regarding the core, the \textbf{Class 2B} models presenting a very big core (of about $73\%$ of the total size of the planet) but with a low density ($7215$ kg.m$^{-3}$) and presenting no contrast in mantle layer viscosities, is not compatible with  \cite{Shah2021Interior} and \cite{Dumoulin2017}.




\section{Conclusion}

In this work, we have used state-of-the-art geophysical constraints of Venus (mass, total MoI, Love number $k_2$ and quality factor Q) to infer possible internal structure of the planet. Therefore we aim at constraining the internal structure of Venus with minimal assumptions about its chemical content. We adapted the deformation semi-analytical modeling of the ALMA$^3$ open-source Fortran 90 program \cite{Melini2022On} originally designed for studying the loading deformations of the Earth \cite{Spada2008ALMAAF}, to the case of the tidal deformation of Venus.

For one given set of parameters extracted from \cite{Dumoulin2017} (model V), we first demonstrate that our model leads to similar results in terms of real and imaginary Love numbers, consistent values of the Andrade rheology $\alpha$ parameter of the Venus mantle and consistent intervals of the mantle viscosities when this latter is supposed to be homogeneous. We then randomly sample the parameter space of the possible internal structure profiles, in varying the thicknesses, the densities and the viscosities of 4 or 5 layer profiles. Each layer is assumed to be homogeneous therefore having averaged values of parameters (density, rigidity and viscosity). We only consider models that induce geophysical quantities consistent with the state-of-the-art constraints given on Table \ref{Tab:TableVenus}. Over $65000$ models produced randomly, remain about $18000$ models with 23$\%$ of them being 5 layer models (with a solid inner core and a fluid outer core) and 77$\%$ having either a fluid (38 $\%$) or a solid core (39 $\%$). We assume incompressible models, which is in the theoretical basis of ALMA$^3$, with layer-fixed rigidity. The existence of the Venus inner core is not clearly demonstrated  from our results but we show that the existence of a solid core cannot be ruled out by only considering geophysical constraints. Moreover, an interesting pattern in our models is the contrast of viscosities in the mantle. Indeed as it has been discussed in Sect. \ref{sec:discussion}, only 1 $\%$ of our 4 and 5-layer models have the same viscosity for the lower and the upper mantle, inducing  significant viscosity contrasts between the two layers. Significant differences in densities and thicknesses for these two layers also stress the non-homogeneity of the Venus mantle. 

Furthermore, as one can see on Fig. \ref{fig:temprofile}, the viscosity contrasts \cs{can also be expressed as temperature differences which would than result in a lower mantle of a higher temperature (with a minimum of $1800$ K)} than the upper mantle (fixed at $1600$ K). These lower mantle temperatures are also hotter (but still in agreement at 2-$\sigma$) than the temperatures proposed by \cite{Shah2021Interior}, \cite{Steinberger2010Deep} and \cite{Armann2012Simulating}. 

The comparisons with \cite{Shah2021Interior} also \cs{indicate the our results are somewhat more compatible with their results for an S-free core} (considering the density or the radius comparisons). Such types of models with a very low percentage of sulfur are in agreement with the past literature  \cite{Lewis1972Metalsilicate, TRONNES2019165}. \cs{They also are not compatible with} the only class of models (\textbf{Class 2B}) proposing mantle layers without viscosity contrast.


Finally, for the future missions towards Venus, we confirm that the determination of a very accurate $k_2$ TLN will be a key for deciphering the state of the Venus core with 90$\%$ probability that a low $k_2$ ($k_2 < 0.25$)  will indicate a solid unique core with a density compatible with an iron alloy (not less than 9000~kg.m$^{-3}$) and a low viscosity (of about 10$^{15}$ Pa.s).
We stress, at last, that these results rely on an interval for the quality factor $Q$. The one used in this work is based on the range deduced from previous models and realistic assumptions as no direct measurement of the tidal dissipation has been done so far for Venus. Such an estimation will be an important outcome of future space missions.


\section*{Acknowledgments} 
The authors thank Giorgio Spada and Daniele Melini for providing the ALMA$^3$ program as well as many constructive discussions. The authors also gratefully acknowledge the cooperation of Caroline Dumoulin in this work and the engaging conversations with Nicolas Coltice.
This work was supported by the "Programme National de Gravitation, Références, Astronomie, Métrologie" (PN-GRAM) of CNRS/INSU co-funded by CEA and CNES.


\bibliographystyle{elsarticle-num}
\bibliography{references}


\newpage
\appendix


\section{Sensitivity to the value of dissipation}
\label{sec:Appendix_Q}
\renewcommand{\thefigure}{A\arabic{figure}}
\setcounter{figure}{0}
\renewcommand{\thetable}{A\arabic{table}}
\setcounter{table}{0}

\begin{figure}[h!]
        \centering
            \includegraphics[width=15cm]{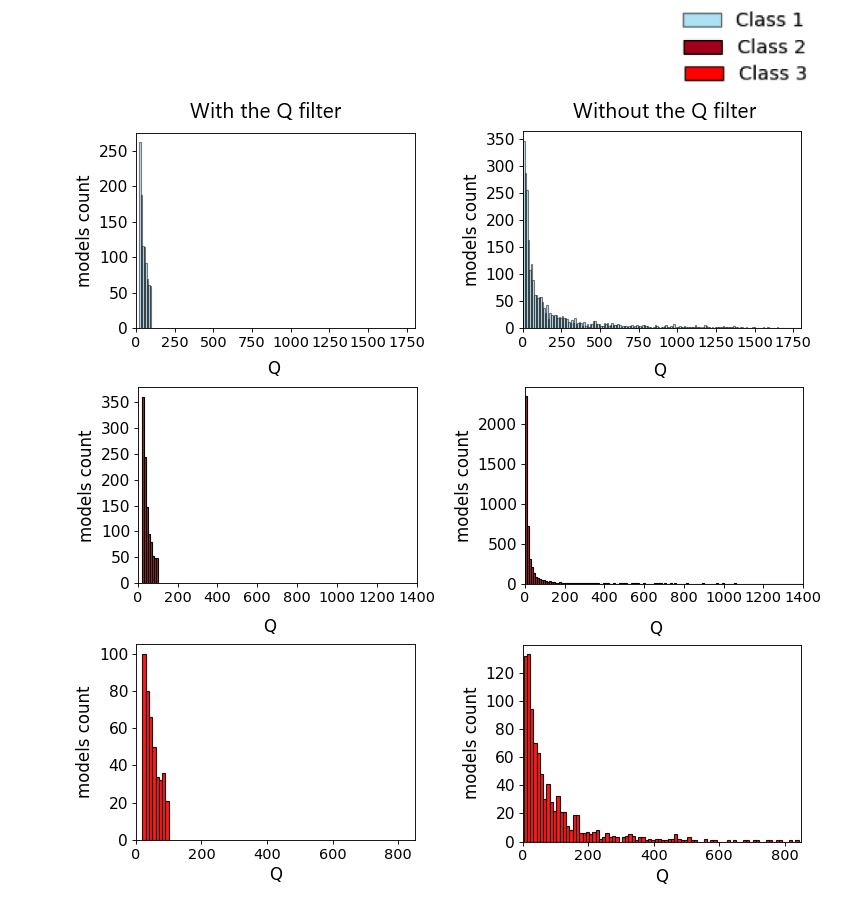}
    \caption{Histograms of the quality factor $Q$ distribution from \textbf{Classes 1}, \textbf{2} and \textbf{3} with (left) and without (right) the quality factor $Q$ filter.}
\label{Fig:Appendix_Q}
\end{figure}

The results obtained in this work are driven by the interval of possible values for the quality factor $Q$. In this section, we show the selection of models without considering the $Q$ filter. 
The original $65000$ models of each class have been filtered with the mass, MoI and observed $k_2$ as used in this work. Therefore the difference between the original results (Table \ref{tab:results}) and this section is the lack of the $Q$ filter. The following Table \ref{tab:results2} presents the results with this filter. Fig \ref{Fig:Appendix_Q} shows the histograms of $Q$ for each class with and without this filter. Without the quality factor $Q$ filter set between $20$ and $100$, the $Q$ values for \textbf{Classes 1}, \textbf{2} and \textbf{3} range in $2$-$1987$, $2$-$1707$, $2$-$1017$ respectively (see Fig \ref{Fig:Appendix_Q}).

\cs{In what follows we compare the original study (with the $Q$ filter) with the new one in this Appendix (without the $Q$ filter). Therefore we compare the difference in percentage between the quartiles of Table \ref{tab:results} of the original study and Table \ref{tab:results2} of this Appendix.} Table \ref{tab:results2} shows that for \textbf{Class 1} the thicknesses $Th_{\mathrm{Core}}$, $Th_{\mathrm{LM}}$ and $Th_{\mathrm{UM}}$ quartiles ($25\%$, $50\%$ and $75\%,$) vary from $-0.82\%$ to $-0.52\%$, $2.64\%$ to $3.5\%$ and $-4.56\%$ to $-1.72\%$ respectively. The three layers respective densities quartiles ($\rho_{\mathrm{Core}}$, $\rho_{\mathrm{LM}}$ and $\rho_{\mathrm{UM}}$) vary from $-0.52\%$ to $0.17\%$, $0.25\%$ to $0.58\%$ and $-0.81\%$ to $-0.23\%$. As for the viscosities of the lower mantle $\eta_{\mathrm{LM}}$ and the upper mantle $\eta_{\mathrm{UM}}$ vary from $0\%$ to $4.81\%$ and $1.63\%$ to $4.77$ respectively. Table \ref{tab:results2} also shows that for \textbf{Class 2} the thicknesses quartiles of the layers, $Th_{\mathrm{Core}}$, $Th_{\mathrm{LM}}$ and $Th_{\mathrm{UM}}$ vary respectively from $-2.32\%$ to $2.7\%$, $-0.46\%$ to $27.39\%$ and $-3.79\%$ to $-4.27\%$ respectively. The three layers respective densities ($\rho_{\mathrm{Core}}$, $\rho_{\mathrm{LM}}$ and $\rho_{\mathrm{UM}}$) quartiles vary from $-1.1\%$ to $3.35\%$, $-1.63\%$ to $-0.56\%$ and $-2.36\%$ to $-1.6\%$. As for the viscosities of the core $\eta_{\mathrm{Core}}$, the lower mantle $\eta_{\mathrm{LM}}$ and the upper mantle $\eta_{\mathrm{UM}}$ vary from $-6.99\%$ to $11.88\%$, $-9.25\%$ to $-0.43\%$ and $-5.78\%$ to $-0.65\%$ respectively. The same table (Table \ref{tab:results2}) shows that for \textbf{Class 3} the thicknesses $Th_{\mathrm{IC}}$, $Th_{\mathrm{OC}}$, $Th_{\mathrm{LM}}$ and $Th_{\mathrm{UM}}$ quartiles ($25\%$, $50\%$ and $75\%,$) vary from $-3.22\%$ to $-1.56\%$, $1.31\%$ to $8.8\%$, $3.15\%$ to $5.3\%$ and $-2.32\%$ to $3.15\%$. Their respective densities ($\rho_{\mathrm{IC}}$, $\rho_{\mathrm{OC}}$, $\rho_{\mathrm{LM}}$ and $\rho_{\mathrm{UM}}$) quartiles vary from $-0.21\%$ to $0.08\%$, $-0.58\%$ to $1.33\%$, $0.87\%$ to $1.15\%$ and $-0.08\%$ to $1.63\%$. The viscosities of the inner core $\eta_{\mathrm{IC}}$, the lower mantle $\eta_{\mathrm{LM}}$ and the upper mantle $\eta_{\mathrm{UM}}$ vary from $-0.38\%$ to $0.5\%$, $-2.62\%$ to $3.07\%$ and $1.32\%$ to $4.25\%$.

\begin{table}
\centering
    \caption{Results of the selection process over $65000$ randomly sampled profiles. Are given in Column 1, the type of models considered and on Column 2 the layers. Column 3 gives the mean and first and third quartiles (25$\%$ and 75$\%$) of the layer thicknesses ($\mathrm{km}$), Column 4 the densities $\mathrm{(kg.m^{-3})}$ and Column 5 the viscosities in $\log10$(Pa.s).}
\begin{tabular}{ c  c  c  c  c }
\hline
Models & Layers & thickness & density & viscosity \\
 &  & (km) & (kg.m$^{-3}$) & $\log10$(Pa.s) \\
\hline
Fluid (\textbf{Class 1}) & upper mantle & $919_{576}^{1392}$ & $3749_{3418}^{4114}$ & $20.85_{18.6}^{22.9}$ \\
\\
            & lower mantle & $1904_{1453}^{2295}$ & $4915_{4496}^{5392}$ & $21.78_{19.85}^{23.6}$ \\
            \\
            & core         & $3142_{2874}^{3355}$ & $10956_{9909}^{11938}$ & $-5$ \\
\hline
Solid (\textbf{Class 2}) & upper mantle & $1166_{777}^{1645}$ & $3739_{3385}^{4100}$ & $20.48_{17.9}^{22.85}$ \\
\\
            & lower mantle  & $971_{472}^{1661}$ & $4889_{4402}^{5441}$ & $20.48_{17.95}^{22.85}$ \\
            \\
            & core          & $3571_{3157}^{4170}$ & $9278_{7789}^{10939}$ & $17.95_{16}^{19.9}$ \\
\hline
Fluid/Solid (\textbf{Class 3}) & upper mantle & $903_{580}^{2165}$ & $3738_{3374}^{4076}$ & $20.85_{18.6}^{22.9}$ \\
\\
            & lower mantle & $1780_{1400}^{2165}$ & $4975_{4475}^{5431}$ & $21.85_{20}^{23.48}$ \\
            \\
            & outer core   & $386_{173}^{740}$ & $8313_{6795}^{9843}$ & $-5$\\
            \\
            & inner core   & $2734_{2344}^{3091}$ & $11425_{10430}^{12230}$ & $15.7_{12.9}^{17.9}$\\
\hline
\end{tabular}
\label{tab:results2}
\end{table}

\newpage

\section{Testing the effect of the number of simulated models and the sensitivity to the rigidity}
\label{sec:Appendix_rigidity}
\renewcommand{\thefigure}{B\arabic{figure}}
\setcounter{figure}{0}
\renewcommand{\thetable}{B\arabic{table}}
The rigidities of the different layers have been fixed so far. In this appendix we consider the effects of changing these parameters on the models selection. We also show the fact that the $65000$ originally simulated models are enough for the statistical analysis of this work. We take \textbf{Class 1} (fluid core) as an example. 

\begin{figure}[h!]
    \centering
    \includegraphics[width=15cm]{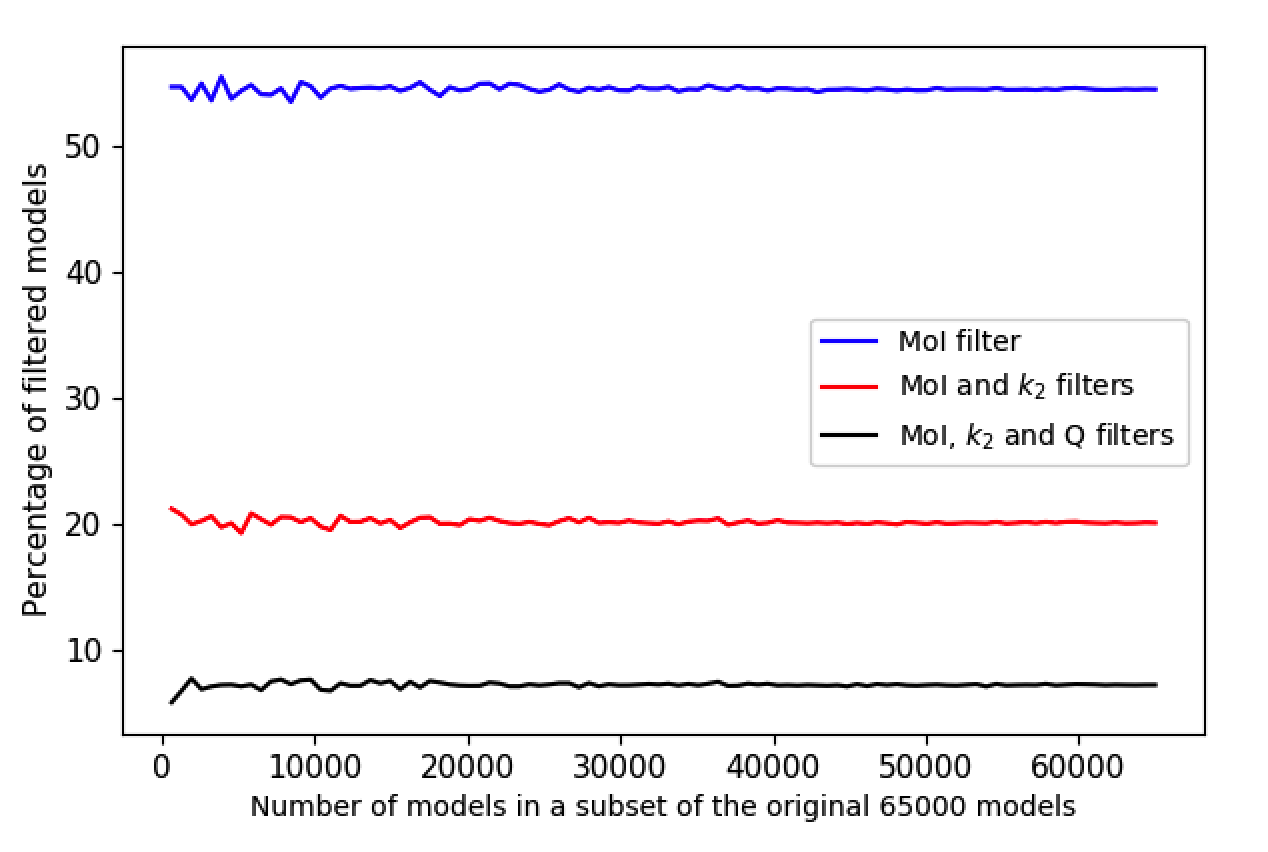}
    \caption{The number of filtered models after each additional filter: MoI ($\pm 1$-$\sigma$), $k_2$ ($\pm 2$-$\sigma$) and Q (from $20$ to $100$). The x-axis illustrates the number of models in each randomly selected subset of the original $65000$ models.}
\label{Fig:Subsets_percentage}
\end{figure}

\begin{figure}[h!]
    \centering
    \includegraphics[width=15cm]{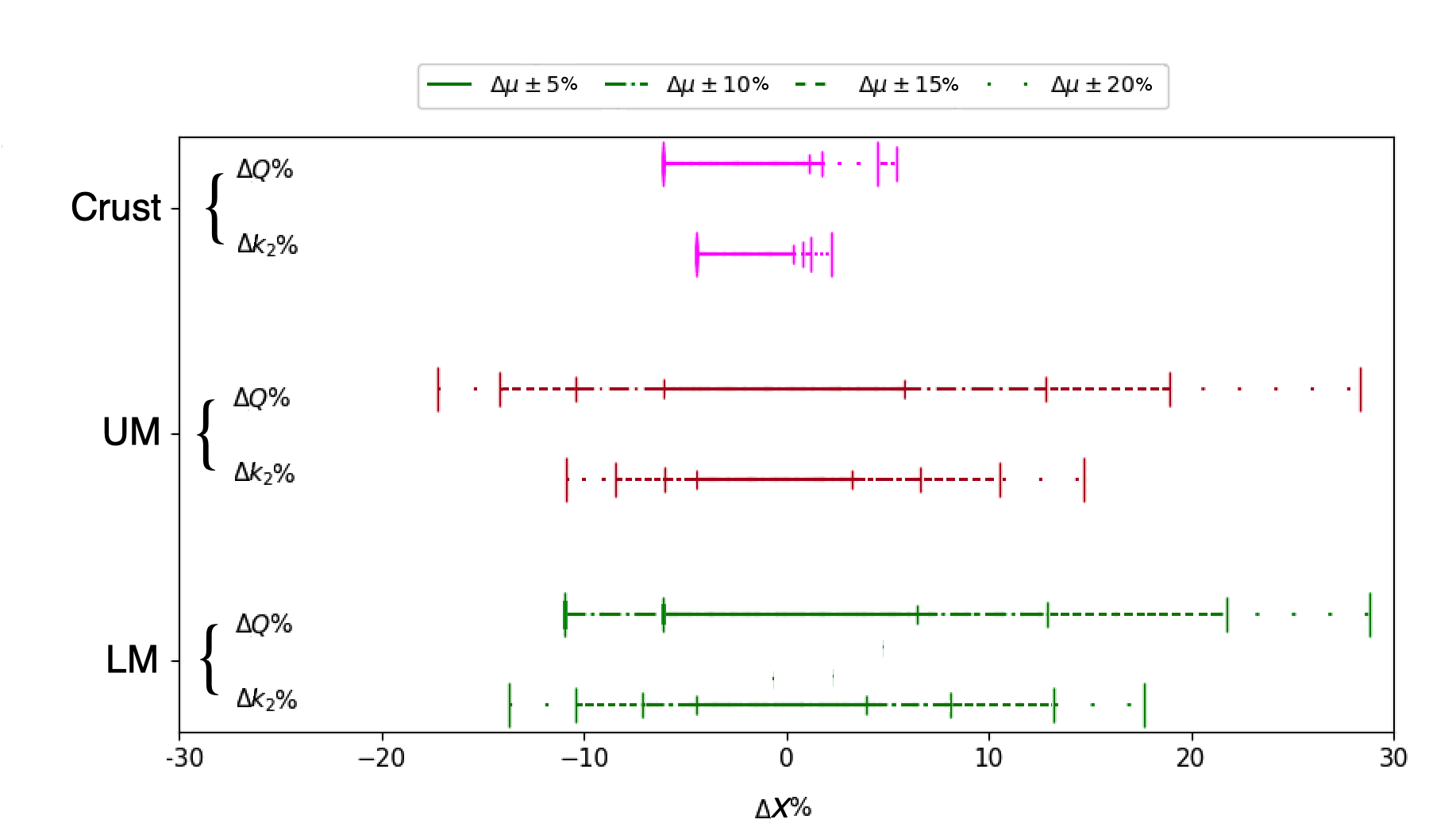}
    \caption{The difference in percentage for the real part of $k_2$ and $Q$ between the new results and the original results after varying the rigidities ($\mu$). The x-axis is the difference in percentage either to $k_2$ or to $Q$, therefore $X$ being either $k_2$ or $Q$.}
\label{Fig:Different_rigidities}
\end{figure}

We select random subsets of the original $65000$ models of \textbf{Class 1} and filter them with the MoI, $k_2$ and $Q$ filters used in this work. Fig. \ref{Fig:Different_rigidities} illustrates the percentage of the filtered models with respect to the number of models in each random subset. Fig. \ref{Fig:Subsets_percentage} shows the percentage of filtered models (y-axis) as a function of the number of models (x-axis) of the subsets. Fig. \ref{Fig:Subsets_percentage} shows that for the $65000$ originally simulated models, the MoI, $k_2$ and $Q$ filters preserve $7.2\%$ of the models (or $4703$ models). This value is approached after $10000$ simulated models. Therefore simulating more models does not provide a higher percentage of models after filtration. The same conclusion is valid for the other two filters: the MoI and $k_2$ filters applied together and the MoI filter solely applied.

We test the effect of the rigidity variation on a subset of $10000$ models of the original $65000$ original 4-layer models of \textbf{Class 1}. For each of the models from this subset we vary the rigidity of only one layer and then we calculate the TLN $k_2$ and quality factor $Q$ for the new models. The core is considered to be an inviscid fluid therefore $\mu_{\mathrm{Core}} = 0$ Pa and therefore it does not vary. Therefore the layers rigidities that are tested are of the lower mantle, the upper mantle and the crust. The rigidity of each layer is varied each from the original values (see Table \ref{Tab:PRIOR}) either by $\pm 5\%$, $\pm 10\%$, $\pm 15\%$ or $\pm 20\%$.

We denote by $O_{X}$ and $N_{X}$ the original and new parameters respectively with $X = k_2$ or $X = Q$. Fig. \ref{Fig:Different_rigidities} illustrates
$\Delta k_2 = \displaystyle{\frac{N_{k_2}-O_{k_2}}{O_{k_2}}} \times 100$ and $\Delta Q = \displaystyle{\frac{N_{Q}-O_{Q}}{O_{Q}}} \times 100$ after we vary the rigidity of one layer. The y-axis and the legend indicate which layer rigidity has varied from the original models and the percentage of variation respectively. Fig. \ref{Fig:Different_rigidities} shows that the rigidity of the crust has the least effect on $k_2$ and $Q$ probably due to its smaller size. The effect of the rigidity of the other layers (lower mantle and upper mantle) has almost the same significance. The maximum effect caused by the variation of the lower mantle or the upper mantle rigidities by $20\%$ increases $Q$ by almost $30 \%$ and $k_2$ by less than $20\%$. The estimated $k_2$ from Table \ref{Tab:TableVenus} has an uncertainty with $2$-$\sigma$ of $22.37\%$, therefore it is of the magnitude of the difference in percentage a different rigidity of one layer causes to the resulting $k_2$ and $Q$.

\newpage

\section{Sensitivity to creep parameter $\alpha$}
\label{sec:Appendix_alpha}
\renewcommand{\thefigure}{C\arabic{figure}}
\setcounter{figure}{0}
\renewcommand{\thetable}{C\arabic{table}}
\setcounter{table}{0}

\begin{figure}[h!]
    \centering
    \includegraphics[width=9cm]{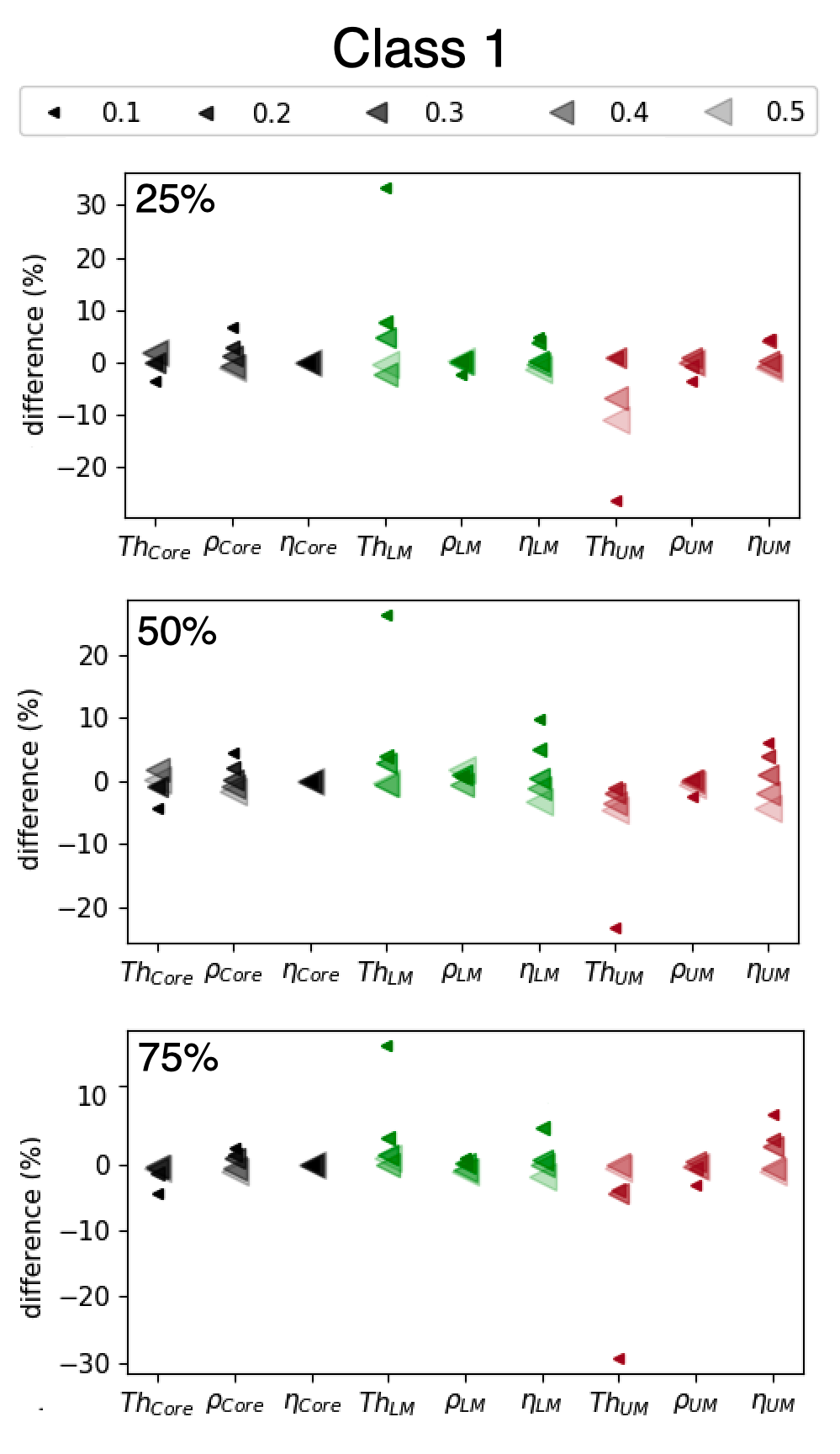}
    \caption{The difference in percentage (\%) for \textbf{Class 1} between the new results with $\alpha$ between $0.1$ and $0.5$ and $\alpha = 1/3$. The x-axis corresponds to the layer parameters and the y-axis is the percentage difference. From top to bottom, the supblots correspond to the first quartile (25\%), second quartile or median (50\%) and third quartile (75\%).}
\label{Fig:Plot_alpha1}
\end{figure}

\begin{figure}[h!]
    \centering
    \includegraphics[width=9cm]{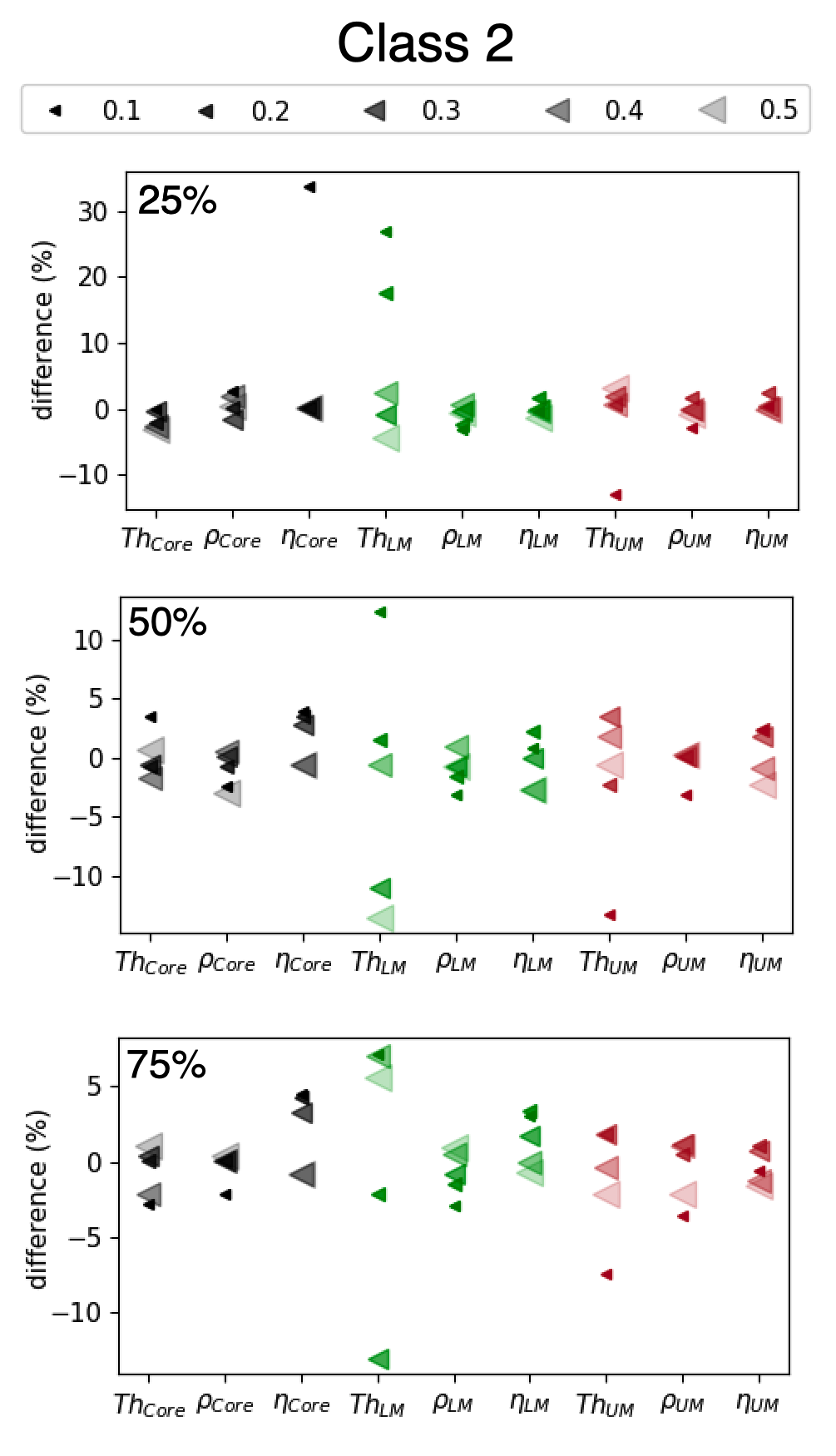}
    \caption{The difference in percentage (\%) for \textbf{Class 2} between the new results with $\alpha$ between $0.1$ and $0.5$ and $\alpha = 1/3$. The x-axis corresponds to the layer parameters and the y-axis is the percentage difference. From top to bottom, the supblots correspond to the first quartile (25\%), second quartile or median (50\%) and third quartile (75\%).}
\label{Fig:Plot_alpha2}
\end{figure}

\begin{figure}[h!]
    \centering
    \includegraphics[width=9cm]{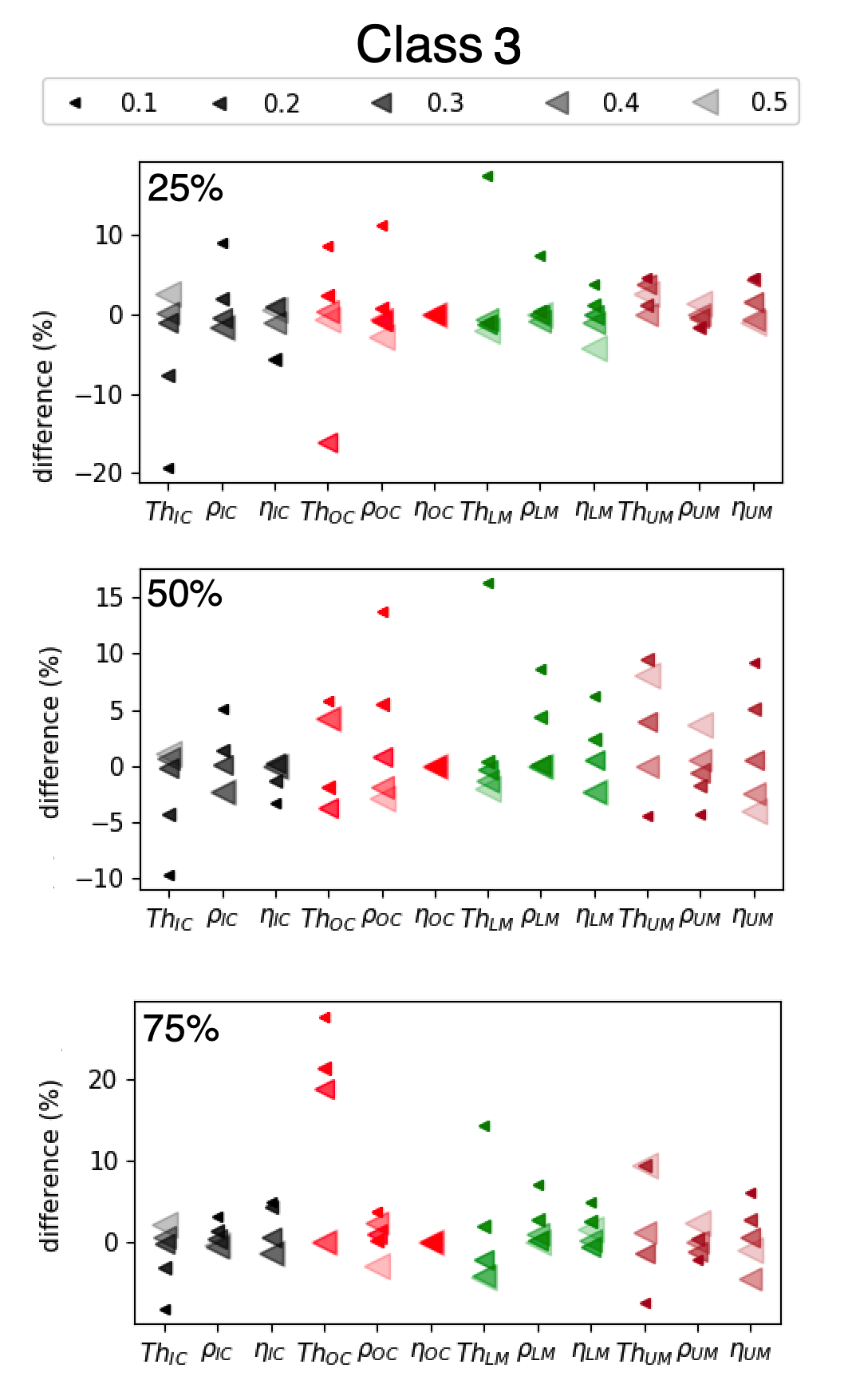}
    \caption{The difference in percentage (\%) for \textbf{Class 3} between the new results with $\alpha$ between $0.1$ and $0.5$ and $\alpha = 1/3$. The x-axis corresponds to the layer parameters and the y-axis is the percentage difference. From top to bottom, the supblots correspond to the first quartile (25\%), second quartile or median (50\%) and third quartile (75\%).}
\label{Fig:Plot_alpha3}
\end{figure}

In this section we investigate which layer parameters are the most sensitive to the $\alpha$ parameter of the Andrade rheology. The experimental parameter $\alpha$ is still not very well constrained, the value used in this work is $\alpha=1/3$ \citep{Louchet2009Andrade}. We test the effect of $\alpha$ on a subset of the $65000$ original models. We randomly select $5000$ models and fix $\alpha$ values between $0.1$ and $0.5$ \cite{CastilloRogez2011} with a step of $0.1$. For each value of $\alpha$ and for $\alpha=1/3$, we calculate the TLN $k_2$ and quality factor $Q$ for the subset of models. Afterwards we apply the same filters (mass, MoI, $k_2$ and $Q$) to each case. We obtain new results for the $5000$ models subset for each class and different values of $\alpha$. Finally we study the difference in the first, second and third quartiles obtained with the original and new $\alpha$ values. Figs. \ref{Fig:Plot_alpha1}, \ref{Fig:Plot_alpha2} and \ref{Fig:Plot_alpha3} represent the difference in percentage for the quartiles of each layer parameters. The viscosities of each layer are studied with a unit of $\log10$ Pa.s as in Table \ref{tab:results}.

Fig. \ref{Fig:Plot_alpha1} shows that the variations between the original and new results of each of the three quartiles are between $+33\%$ and $-30\%$ for \textbf{Class 1}. These two furthermost values correspond respectively to the $Th_{\mathrm{LM}}$ and $Th_{\mathrm{UM}}$ and for $\alpha=0.1$. For an approximate lower mantle and upper mantle thicknesses of $1800\ \mathrm{km}$ and $960\ \mathrm{km}$ (see Table \ref{tab:results}), a change of $+33\%$ and $-30\%$ amounts to a variation of $594\ \mathrm{km}$ and of $-288\ \mathrm{km}$, respectively. The other parameters quartiles vary between $\pm 10 \%$ depending on the value of $\alpha$. For an approximate lower mantle viscosity of $10^{20.78}$ Pa.s, a $\pm 10\%$ variation results in new viscosity values of $10^{18.7}$ Pa.s and $10^{22.86}$ Pa.s. For an approximate value of an upper mantle density of $3700\ \mathrm{kg.m^{-3}}$, a change of $\pm 5\%$ amounts to a variation of $\pm 185\ \mathrm{kg.m^{-3}}$ of the density. \textbf{Classes 2} and \textbf{3} have the similar patterns as \textbf{Class 1}. Fig. \ref{Fig:Plot_alpha2} shows that the \textbf{Class 2} quartiles vary between $-13 \%$ and $33 \%$ depending on the layer parameters and the values of $\alpha$. The first quartile varies $\approx$ by maximum of $-12.8\%$ and $33.7\%$ corresponding to $\alpha=0.1$ for $\rho_{\mathrm{Core}}$ and $Th_{\mathrm{UM}}$ respectively. The other layer parameters first quartile vary between $\pm 10\%$ except $Th_{\mathrm{LM}}$ which varies by $26.9\%$ and $17.5\%$ for $\alpha=0.1$ and $\alpha=0.2$ respectively. The second (respectively third) quartile vary in between $-13.4\%$ and $12.36\%$ (respectively $-13\%$ and $7.2\%$). These maximum variations correspond to $Th_{\mathrm{LM}}$. The other layer parameters vary between $\pm 5\%$ except $Th_{\mathrm{UM}}$ for $\alpha=0.1$ which varies by $-13.2\%$ and $-7.5\%$ for the second and third quartiles respectively. Fig. \ref{Fig:Plot_alpha3} shows that the maximum variation of the first and second quartiles of \textbf{Class 3} correspond to $\alpha=0.1$ and to $Th_{\mathrm{IC}}$ and $Th_{\mathrm{LM}}$. The first quartile of the parameters of these 2 layers varies by $-19 \%$ and $+17 \%$, respectively and their second quartile varies by $-9.6\%$ and $16\%$, respectively. The first quartile of the other parameters vary between $\pm 10\%$ except $Th_{\mathrm{OC}}$ and $\rho_{\mathrm{OC}}$ which vary by $-16\%$ and $11\%$ for $\alpha=1/3$ and $\alpha=0.1$ respectively. The second quartile of the other parameters varies between $\pm 10\%$ except $\rho_{\mathrm{OC}}$ and $Th_{\mathrm{LM}}$ which vary by $13.6\%$ and $16\%$, respectively, both for $\alpha=0.1$. The third quartile of \textbf{Class 3} varies between $-8\%$ and $27.6\%$ and these furthermost values correspond to $\alpha=0.1$. The most affected parameters are $Th_{\mathrm{OC}}$ (between $18.8\%$ and $27.6\%$ for $\alpha$ of $0.1$, $0.2$ and $0.3$) and $Th_{\mathrm{LM}}$ ($14\%$ for $\alpha = 0.1$). The other layer parameters vary between $\pm 10\%$. Therefore the $\alpha$ parameter affects $k_2$ and $Q$ for each of the classes, hence it affects the distribution of the layer parameters after filtering with the ranges of $k_2$ and $Q$. Nonetheless the effect is not big enough to affect the original general study.

\end{document}